\documentclass[]{rsos}
\usepackage[version=4]{mhchem}
\usepackage{comment}
\usepackage[numbers]{natbib}
\usepackage{xcolor}
\usepackage{float}
\usepackage{booktabs}
\usepackage{multirow}
\usepackage{multicol}
\usepackage{pdflscape}
\usepackage{rotating}
\usepackage{tabularx}
\usepackage{pdflscape}
\usepackage{rotating}
\usepackage{tabularx}
\usepackage{arydshln}

\begin{document}

\title{The development of the ozone layer under an evolving Sun}

\title{Simulations of the Evolving Ozone Layer: Implications for the Early Earth and Exoplanets}

\title{Simulations of the Evolving Ozone Layer: Implications for the Early Earth, Exoplanets, and the Search For Life}

\author{
A. Kumar$^{1, 2}$, G.J. Cooke$^{1}$}

\address{$^{1}$Institute of Astronomy, University of Cambridge, Cambridge, CB3 0HA, UK\\
$^{2}$Department of Earth and Planetary Sciences, Harvard University, Cambridge, MA 02138, USA}

\subject{Atmospheric science/atmospheric chemistry}


\corres{G.J. Cooke\\
\email{gjc53@cam.ac.uk}}

\begin{abstract}
Over 4 billion years, Earth's atmosphere has been transformed from a composition devoid of molecular oxygen (\ce{O2}) to one where \ce{O2} is now 21\% of our modern atmosphere by volume. With high quantities of \ce{O2}, an ozone (\ce{O3}) layer can form and partially shield the surface from biologically harmful ultraviolet radiation (UV). The evolution of the \ce{O3} layer has previously been simulated under the modern Sun by independent groups with separate models. Here, we present simulations of Earth's atmosphere from four open-source one-dimensional (1D) photochemical models and a 3D chemistry-climate model, focussing on \ce{O3}. We find that the \ce{O3} layer thickness exhibits a clear deviation between the 1D models and the 3D model due to several factors, including treatments of \ce{O2} photolysis, chemical networks, \ce{H2O} microphysics and 3D transport. We find that 1D models introduce a bias for unobserved climate states and show how these biases may propagate to calculations of biological UV damage and surface habitability. We show that when applied to specific exoplanet atmospheres, these discrepancies can become more pronounced, propagating through to observational predictions for the next generation of telescopes, such as HWO and LIFE, that will search for habitable and inhabited exoplanets. 
\end{abstract}

\maketitle

\section{Introduction}
\label{Introduction}

Over the course of Earth's history, its atmosphere has undergone a profound transition, from an anoxic (O$_2$-poor) state during the Archean eon (4.031 -- 2.500 Ga, where Ga is billions of years ago), to the O$_2$-rich (21$\%$ by volume) atmosphere animals inhabit today \citep{catling2020archean}. A possible influencing factor in Earth's biological history is the formation and thickness of the ozone (\ce{O3}) layer, which is partly dependent on the atmospheric \ce{O2} concentration. \ce{O3} molecules absorb biologically harmful ultraviolet (UV) radiation, thus shielding life on the surface. While terrestrial organisms experience the benefits of the modern \ce{O3} layer, this protective screen has not always been present throughout Earth's past.

\begin{figure*}[h!]
    \centering
    \includegraphics[width = 1\textwidth]{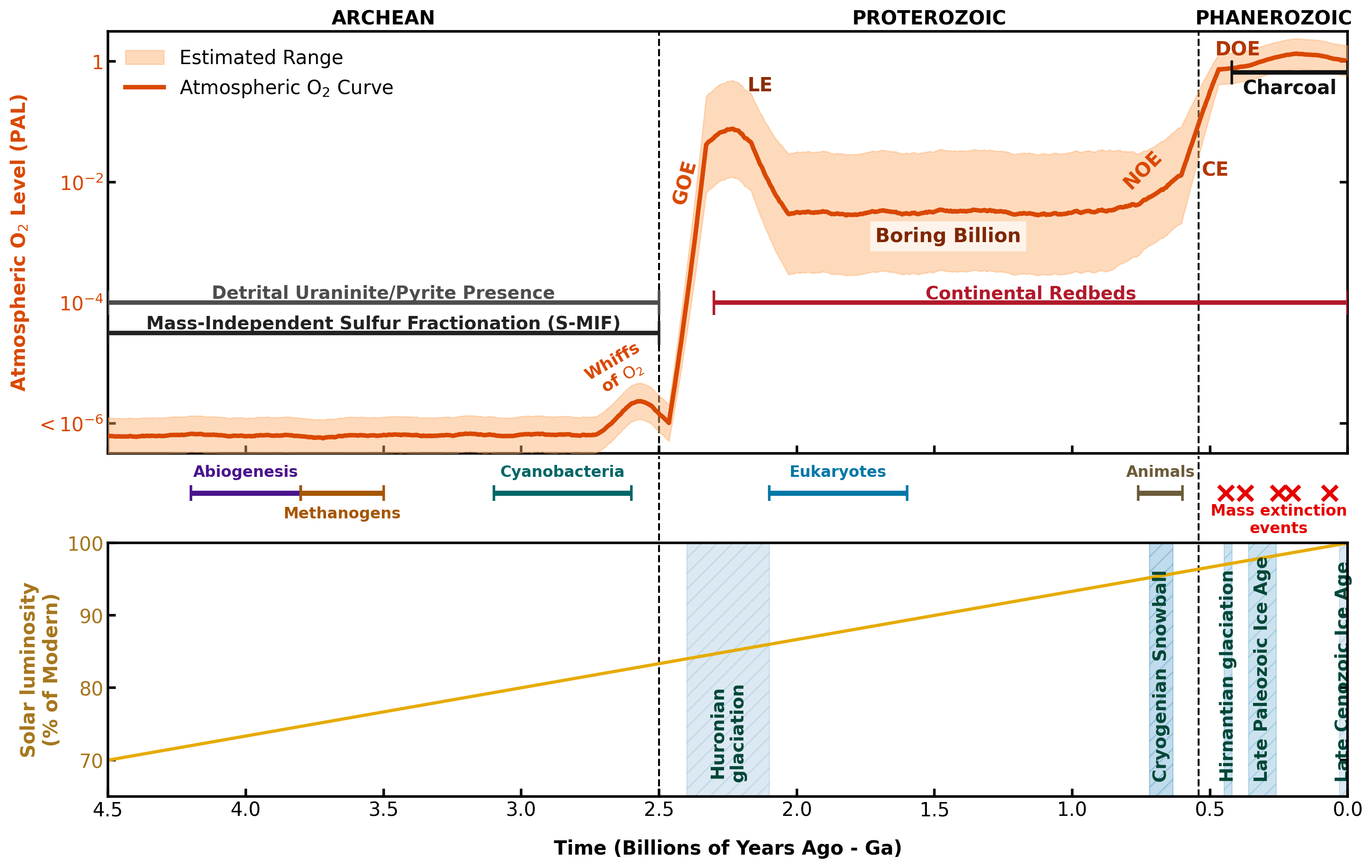}
    \caption{The schematic diagram shows the \ce{O2} concentration in Earth's atmosphere over the last 4 billion years, alongside other major astrophysical and climatic events which have occurred alongside, and affected, the evolution of life on Earth. \textbf{Top panel:} An approximate guide to the history of oxygen is shown, with the shaded region showing possible ranges \citep{holland2006oxygenation, bekker2012oxygen, lyons2014rise, lyons2021oxygenation}. The data is not exact and should be seen as an illustration of changes rather than a recreation of the past oxygen record. We direct readers to the following papers, which discuss the history of \ce{O2} \citep{2005AREPS..33....1C, lyons2021oxygenation, lyons2014rise}. The following acronyms are present: NOE = Neoproterozoic Oxidation Event, DOE = Devonian Oxidation Event, GOE = Great Oxidation Event, LE = Lomagundi Excursion, CE = Cambrian Explosion. Smaller increases in \ce{O2} during the end of the Archean are shown as oxygen ``whiffs'' \citep{anbar2007whiff}. The ``Boring Billion'' is also marked \citep{Buick1995_Boring}. Evidence for the absence of \ce{O2} and presence of \ce{O2}, including detrital uraninite \citep{johnson2014o2}, the S-MIF signal that disappears near the GOE \citep{farquhar2000atmospheric}, and the charcoal record during the Phanerozoic, are marked \citep{scott2006diversification}. \textbf{Middle section:} error bars show approximate timelines for abiogenesis (the origin of life), the origin of methanogens, the origin of cyanobacteria, the origin of Eukaryotes, and animals \citep{betts2018integrated, dos2015uncertainty}. Marked by red crosses are the timings of the five major mass extinction events that occurred during the Phanerozoic Eon \citep{raup1982mass, algeo2024theory}. \textbf{Bottom panel:} The increasing luminosity of the Sun is plotted against the same time axis as the top panel, starting 4.5 billion years ago at 70\% the modern luminosity \citep{feulner2012faint}. Several glaciations are marked by the light blue zones with their corresponding names labelled \citep{kopp2005paleoproterozoic, hoffman2017snowball, finnegan2011magnitude, montanez2013late, zachos2001trends}.}
    \label{Oxygenation history figure}
\end{figure*}

Although localised \ce{O2} production and accumulation likely occurred during the Archean \citep{doi:10.1126/sciadv.abj0108}, the widespread oxygenation of Earth's surface and atmospheric oxygenation occurred in several key stages. The first of these was at the end of the Archean, during one of the most significant environmental shifts in our planet's history - the Great Oxidation Event \citep[GOE;][]{lyons2014rise}. This $\sim200$ Myr transition started around 2.4 Ga \citep{poulton2021200} and resulted in atmospheric O$_2$ levels rising from less than 10$^{-6}$ of the Present Atmospheric Level (PAL) to between 1\% and 10\% PAL \citep{holland2006oxygenation, bekker2004dating}. Evidence for the GOE arises from a myriad of sources \citep{catling2020archean}, including: the disappearance of mass-independent fractionation of sulphur isotopes \citep{farquhar2000atmospheric}; the cessation of atmospheric xenon fractionation \citep{zahnle2019strange}; the appearance of continental red beds \citep{shawwa2024earth}; the termination of widespread Banded Iron Formations (BIFs) \citep{bekker2010iron, yin2023origin}; the disappearance of minerals like detrital uraninite and pyrite in sediments \citep{johnson2014o2, johnson2019experimental}; and the presence of oxidised palaeosols post-GOE \citep{babechuk2019pervasively}. Following the GOE, atmospheric O$_2$ dropped \citep{canfield2013oxygen, bekker2012oxygen, lyons2014rise} at the start of the Proterozoic eon (2.500 -- 0.539 Ma Ga). Much of the Proterozoic saw a prolonged period of apparent environmental stability, known as the ``Boring Billion'' \citep{Buick1995_Boring}. Atmospheric \ce{O2} levels during this period remained relatively stable at around 10$^{-3}$ to 10$^{-1}$ PAL \citep{lyons2021oxygenation, lyons2014rise}, with other work suggesting periods with slightly higher levels \citep{steadman2020evidence}. While the Mesoproterozoic may be characterised by a stasis when measured in terms of atmospheric redox, carbon isotope excursions, supercontinent cycles \citep{westall2024precambrian}, and a constant day length \citep{Mitchell_Day_Length}, this era was hugely important for biological evolution \citep{westall2024precambrian}. The late Proterozoic likely underwent a second rise in atmospheric \ce{O2}, in an event dubbed the Neoproterozoic Oxidation Event \citep[NOE;][]{och2012neoproterozoic}, evidenced by shifts in carbon isotope values \citep{krissansen2015statistical} and other redox proxies \citep{turner2016thick, pogge2015selenium, lu2017iodine}. The atmospheric \ce{O2} level following the NOE is still debated, but estimates range from between 10 and 50$\%$ PAL \citep{yeung2017low, blamey2016paradigm, steadman2020evidence}. In the subsequent Phanerozoic eon (0.541 Ga -- present), \ce{O2} levels are understood to have fluctuated between 10$\%$ and 150$\%$ PAL, first reaching approximately modern day concentrations during the Paleozoic \citep{lyons2021oxygenation, wallace2017oxygenation}. The broader geological, atmospheric, and biological history of the Earth is summarised in Fig.~\ref{Oxygenation history figure}.

Parallel to this dramatic evolution of our atmosphere, the Sun has also evolved significantly over Earth's history. Models of stellar evolution predict that the young Sun 4 billion years ago was only 70--75$\%$ as bright as the modern Sun \citep{gough1981solar, newman1977implications}. The Faint Young Sun Paradox \citep{feulner2012faint} describes the discrepancy between the early Sun's reduced luminosity and the geological evidence for liquid water on Earth's surface. A fainter Sun leads to greater ice cover, resulting in increased reflection of incoming radiation and lower temperatures. This "ice-albedo feedback'' loop could theoretically result in a runaway glaciation and global ice cover, leading to a "Snowball Earth" state, where the entire planet was covered in ice \citep{kirschvink1992late}. However, this scenario is in conflict with evidence of liquid water, including 4.4 Ga old zircon grains \citep{mojzsis2001oxygen, 2001Natur.409..175W}, 3.4--3.5 Ga old pillow basalts and sedimentary rocks \citep{lowe1980archean, furnes2004early}, and claimed evidence of microbial life from 2.5--3.5 Ga old microfossils and stromatolites \citep{schopf2006fossil, schopf2007evidence, noffke2013microbially}. While climatic models have largely reconciled this paradox for the Archean \citep{Charnay_FYS_2020, Goldblatt_Clouds_2021}, the mechanisms sustaining ice-free conditions throughout the Proterozoic are not fully resolved, though they often hinge on variations in atmospheric composition and an amplified greenhouse effect \citep{Olson_2016_CH4, Fiorella_2017, cooke2022revised, 2023RSOS...1030056J}. 

Many of the aforementioned astrophysical, geological, and biological factors are likely to have influenced the thickness of the \ce{O3} layer. The thickness is also referred to as the "column depth" (given in terms of molecules per area in a vertical column above the surface), and varies with season and position above the Earth's surface due to several factors - including atmospheric chemistry, temperature differences, and the Brewer-Dobson circulation \citep[BDC;][]{2014_Butchart}. A thorough understanding of the O$_3$ layer's temporal variation and its influence on biological habitats has historically been derived from one-dimensional (1D) atmosphere modelling studies \citep{kasting1980evolution, Kasting_1985_Oxidant}. Only a few pior works \citep{cooke2022revised, 2023RSOS...1030056J, Jaziri_2022_GOE} have compared these 1D models to 3D coupled chemistry-climate models. Once considered a `solved problem' \citep{kasting_2025_unsolved}, the co-evolution of the \ce{O3} layer with atmospheric \ce{O2} concentrations is now quantitatively unclear because no single model (1D or 3D) accounts for all the different influencing parameters together \citep{cooke2022revised, Jaziri_2022_GOE, 2023RSOS...1030056J, Ji_2024JGRD..Correlated_K, kasting_2025_unsolved}.

Investigating discrepancies between various 1D and 3D models is crucial for several scientific pursuits. 1D photochemical models are widely used to study the habitability of early Earth-like and Earth-like exoplanets \citep{segura2003ozone, rugheimer2018spectra, rugheimer2015uv, wunderlich2020distinguishing, meadows2023feasibility, sneed2026ultraviolet}. Identifying why these predictive disparities arise is muddled by differences in assumptions, model setups, chemical reaction rates, vertical mixing, temperature-pressure profiles, and boundary conditions, which are not always thoroughly reported in various works, making it difficult to unravel where the fundamental disagreements emerge from.

In this work, we aim to disentangle the underlying principles at work behind the atmospheric results from the simulations. We also characterize their broader consequences for the atmospheric modelling of the early-Earth and Earth-like exoplanets. We use four open-source photochemical models, three of which are widely used by the exoplanet community, and run hundreds of simulations of the Earth's atmosphere under a range of conditions pertinent to its geological history. Focussing on the development of the \ce{O3} layer, we systematically vary the atmospheric \ce{O2} mixing ratio, the largely unknown pressure-temperature profiles of the post-GOE Earth, and change the incoming stellar spectrum to account for an evolving Sun. We explore model sensitivities to a range of additional input parameters to explore how various assumptions affect atmospheric concentrations of key species. We compare results from these models to each other, and to 3D chemistry-climate simulations. Subsequently, we comment on how ``tuning'' models to reproduce the modern observed atmosphere of Earth can propagate through as a fundamental error to paleo-atmosphere and exoplanet atmosphere calculations. Furthermore, we quantify the effects of differing modelling predictions in our theoretical understanding of surface habitability. Finally, given that several of these models are employed to simulate potentially inhabited exoplanets to determine the spectral fingerprints of living worlds, we explore the implications of our work in the search for life beyond our planet.
\section{Computational Atmospheric Modelling}

Numerical models of planetary atmospheres, surfaces, and oceans, are useful tools for understanding and predicting the chemical composition, evolution, and climate of planets. Applications range from Earth's modern anthropogenic climate change \citep{tokarska2020past, Danabasoglu_CESM2_2020} to investigating the environments of exotic exoplanets \citep{Shields_2019_Climates, 2025PSJ.....6..256W}. This section provides an overview of 1D photochemical models, 3D atmospheric chemistry-climate models, and introduces the computational tools used in this work. We introduce the four open-source 1D photochemical models used here, \texttt{Atmos}, \texttt{VULCAN}, \texttt{Photochem}, and the \texttt{Kasting} model, along with the open-source 3D chemistry-climate model \texttt{WACCM6}. Furthermore, we describe two supplementary models: the \texttt{Planetary Spectrum Generator} (PSG), used to assess how model discrepancies propagate into observable exoplanet transmission spectra, and the \texttt{Tropospheric Ultraviolet and Visible Radiation Model} (TUV), employed to evaluate how model discrepancies alter predictions of surface UV environments and planetary habitability.

1D photochemical models construct the atmosphere as a single vertical column, predicting the variation of chemical species with altitude (or pressure). These models can be coupled to a climate model, where the temperature profile is adjusted based on the chemical composition, surface albedo and the radiation field. Because 1D models are computationally efficient, they can incorporate highly complex chemical networks involving hundreds of species and thousands of reactions. They are useful for exploring fundamental chemical pathways, calculating global mean states, and conducting long-term evolutionary studies, such as the rise of oxygen on the early Earth or the abundance of trace gases in planetary atmospheres at different epochs.

3D atmospheric chemistry-climate models expand on 1D models to include horizontal transport and spatial heterogeneity, where each atmospheric grid cell has a solar zenith angle (SZA) that depends on the planetary system's orbital parameters and the time in the simulation. Unlike 1D models, 3D models can capture the influence of atmospheric circulation, such as winds and atmospheric waves (e.g. Kelvin waves and gravity waves), and the day-to-night (diurnal) cycle, which are essential for predicting spatial and temporal variations, including seasonal effects on Earth. Therefore, 3D models have the potential to provide a more realistic representation of an atmosphere. Unfortunately, they are significantly more computationally expensive and cannot always include as many chemical processes as 1D models whilst simultaneously remaining feasible for research and forecasting purposes.

A core component bridging both 1D and 3D models is the radiative transfer scheme, which calculates the propagation of stellar and planetary radiation through an atmospheric column. Radiative transfer governs both climate (via heating and cooling fluxes) and photochemistry (via wavelength dependent actinic fluxes that drive photolysis). While these models provide robust frameworks for simulating planetary atmospheres, their accuracy is limited by the mathematical parametrisation of radiative transfer. 
A particular concern motivating this work is the treatment of \ce{O2} photolysis in the Schumann--Runge (SR) bands (175 -- 205 nm), where strong temperature dependence and overlapping continua make accurate parametrisation difficult, possibly leading to significant errors when modelling oxygenated atmospheres \citep{2023RSOS...1030056J}. Since high-resolution line-by-line calculations are computationally expensive, models often rely on parametrisation schemes to evaluate UV transmission in the SR bands. Methods for calculating transmission include using polynomial representations of temperature and SZA dependent cross sections to calculate dissociation rates \citep{allen1982effective}, using rapid parametrisations derived directly from line-by line models \citep{1993JGR....9810543M}, and using algorithms based on local temperature and the \ce{O2} column along the line of sight \citep{1996AnGeo..14...68K}. Recently, to address the limitations of older exponential sum fits in capturing explicit temperature dependence and multiple Rayleigh scattering at low \ce{O2} abundances, the \texttt{Kasting} model adopted a new correlated-$k$ parametrisation \citep{Ji_2024JGRD..Correlated_K} based on updated line-by-line calculations \citep{Fernandez_2007_O2, 2022JQSRT.27707949G, minschwaner1992polynomial}. This technique calculates cumulative absorption probabilities from high resolution line-by-line spectra across multiple temperature grids. Of the four 1D photochemical models considered here, only the \texttt{Kasting} model has been updated with the correlated-$k$ scheme \citep{Ji_2024JGRD..Correlated_K, kasting2025atmospheric}. However, it overpredicts the amount of \ce{O3} for the modern atmosphere at 330 Dobson Units versus 292 DU \citep{braesicke2019update}. Since \texttt{Photochem}, \texttt{Atmos} and \texttt{VULCAN} do not employ the up-to-date accurate parametrisation of \ce{O2}-photolysis in the SR bands and \ce{O2} cross sections differ (see Fig. ~\ref{stellar flux}), we suspect that they likely predict different \ce{O3} columns even when the model planetary setup is identical, thereby forming the premise of our work.

\subsection{1D Atmospheric Models}

Here, we detail the various open-source 1D models used in this work. All four models discretise the atmosphere into plane-parallel vertical layers in hydrostatic equilibrium and compute the vertical concentration profiles of atmospheric gases and aerosols by solving the coupled mass continuity equations. These equations account for chemical production and loss alongside vertical transport, which is parametrized through an eddy diffusion coefficient ($K_{\text{zz}}$) and molecular diffusion across discrete atmospheric layers.
A stellar spectrum is input along with a user inputted SZA, diurnal averaging factor (DAF) and surface albedo. To maintain computational stability and efficiently converge to a steady state, the models rely on the evaluation of a jacobian matrix. Finally, they allow user-specified upper and lower boundary conditions, which includes options for fixed mixing ratios, constant surface fluxes, and deposition velocities at the lower boundary, and diffusion-limited escape at the upper boundary.

\subsubsection{Atmos}

\texttt{Atmos}\footnote{Atmos: \href{https://github.com/VirtualPlanetaryLaboratory/atmos}{https://github.com/VirtualPlanetaryLaboratory/atmos}} is an open source one-dimensional coupled photochemistry and climate code \citep{arney2016pale, 2021ascl.soft06039V}.  The photochemical model in \texttt{Atmos} \footnote{https://github.com/VirtualPlanetaryLaboratory/atmos/tree/master/PHOTOCHEM} (referred to internally as \texttt{Photo} or \texttt{Photochem}) \footnote{not to be confused with from the standalone \texttt{Photochem} model described in subsection \ref{Photochem_section}} was developed by \citet{kasting1979oxygen}, updated by \citet{zahnle2006loss}, and is described in detail in \citet{arney2016pale} and \citet{meadows2018habitability}. For photochemical simulations, we use the \texttt{"Modern Earth Complex"} template described in \citet{Lincowski2018}, which includes a high resolution wavelength grid with 309 reactions and 73 species. The model solves the continuity equations using a fully implicit reverse Euler method. It also implements a centered finite volume scheme for particle transport and molecular diffusion, and uses the $\delta$-2 stream method for radiative transfer \citep{Toon_1989_Rapid}. The model includes the effect of lightning indirectly using a user defined column-integrated photochemical production rate of Nitric Oxide (NO).

The climate portion of the model \footnote{https://github.com/VirtualPlanetaryLaboratory/atmos/tree/master/CLIMA} (referred to internally as \texttt{Clima}) draws from early models \citep{kasting1984response, kasting1986climatic, kasting1983loss, kasting1984effects} and was updated by \citet{kopparapu2013habitable} and \citet{arney2016pale}. It also uses the $\delta$-2 stream approximation for radiative transfer, and uses the correlated-k coefficients to calculate the spectral absorption of \ce{O2}, \ce{O3}, \ce{CO2}, \ce{H2O}, \ce{CH4}, and \ce{C2H6} across 38 spectral intervals. The model includes the effects of pressure broadening of gases like \ce{N2}. Infrared fluxes are calculated with the RRTM routine \citep{1997JGR...10216663M}, employing 16-term sums in each spectral band for calculated k-coefficients. The tropospheric temperature profile is derived by tracing a moist adiabatic lapse rate up to the altitude where the stratospheric temperature is attained \citep{kasting1988runaway}, unless the atmosphere is desiccated, in which case a dry adiabat is used instead. Both \texttt{Photo} and \texttt{Clima} can be coupled by running them back and forth, with the photochemical model providing species profiles to the climate model, and the climate model providing the pressure-temperature profile to the photochemical model. 

\subsubsection{Photochem}

\label{Photochem_section}

\texttt{Photochem}
\footnote{https://github.com/Nicholaswogan/photochem} is an open source one-dimensional coupled climate–photochemistry model that has been used to study the atmospheres of a wide range of planets, exoplanets and brown dwarfs \citep{2025PSJ.....6..256W, wogan2024jwst, benneke2024jwst, cooke2024considerations, mukherjee2025effects, schmidt2025comprehensive, beiler2024tale, wogan2025sonora, wogan2023origin}. While heritage concepts were drawn from \texttt{Atmos}, the model was developed from the ground up with several differences, discussed in detail in \citet{2025PSJ.....6..256W}. Building on networks in \citet{2016DPS....4821201Z}, \citet{wogan2023origin} and \citet{wogan2024jwst}, the updated reaction network for the Earth \footnote{https://github.com/Nicholaswogan/UofAPhotochem \textunderscore Workshop/blob/main/3 \textunderscore ModernEarthPhotochemistry/zahnle \textunderscore earth.yaml} contains 98 gas-phase species composed of H, O, C, N, S, Cl and He, and incorporates 610 forward reactions alongside 95 photolysis reactions. The principle of microscopic reversibility is used to reverse non-photolysis reactions \citep{2011ApJ...738...72V}. The network also includes 16 condensed-phase species. Hydrocarbon haze particles are formed from gas-phase reactions, while aerosols like \ce{H2O} and \ce{NH3} are produced and destroyed via evaporation and condensation. 
The model discretises the atmosphere using a finite volume method. To maintain numerical stability, particularly when advection dominates (e.g., heavy molecules in light background gases), it employs a first-order upwind scheme for advective transport terms, alongside a second-order centred scheme for diffusive terms. The resulting system of stiff ordinary differential equations is solved using the CVODE integrator \citep{hindmarsh2005sundials}, which implements backward differentiation formulas and provides more stability compared to backward Euler methods. The model computes the jacobian matrix using forward-mode automatic differentiation via the custom open-source library Differentia \citep{2025zndo..15678369W}, further increasing solver stability. Photolysis rates are computed using a quadrature two-stream radiative transfer scheme based on \citet{Toon_1989_Rapid}. 

\subsubsection{VULCAN}

\texttt{VULCAN}\footnote{\texttt{VULCAN}: \url{https://github.com/shami-EEG/VULCAN}} is an open source one-dimensional photochemical kinetics code for simulating planetary atmospheres \citep{tsai2017vulcan}. It has been benchmarked against several models, including TEA \citep{blecic2016tea} and ARGO \citep{2016ApJS..224....9R}, and has been used to simulate the atmospheres of planets, exoplanets and brown dwarfs \citep{tsai2017vulcan, tsai2021comparative, tsai2024biogenic, dai2024investigation, gibbs2025simulations}. We use two VULCAN photochemical networks to run sensitivity tests within \texttt{VULCAN}. Both of these include Chapman, HO\textsubscript{x}, and NO\textsubscript{x} \ce{O3} chemistry. The less complex network is the N–C–H–O scheme\footnote{N–C–H–O Earth network: \url{https://github.com/shami-EEG/VULCAN/blob/master/thermo/NCHO\textunderscore earth\textunderscore photo\textunderscore network.txt}}, with 64 gaseous species, 312 forward reactions, and 47 photodissociation branches. The more complex network is the 2025 full S–N–C–H–O–Ar scheme for oxidizing atmospheres\footnote{S–N–C–H–O–Ar network: \url{https://github.com/shami-EEG/VULCAN/blob/master/thermo/SNCHOAr\textunderscore full\textunderscore photo\textunderscore network\textunderscore 2025.txt}}, with 100 gaseous species, 589 forward reactions, and 70 photodissociation branches. Relative to the N–C–H–O network it adds sulphur chemistry (including H$_2$S, SO, SO$_2$, SO$_3$, H$_2$SO$_4$, OCS, CS$_2$, and S$_8$). The two networks are not a strict subset-superset though: the smaller scheme includes CH$_3$NH, which is absent from the 2025 full sulphur network. Unless otherwise noted, we use the S–N–C–H–O–Ar network in the presented simulations. Reverse rates for the thermal reactions are computed from NASA-9 Gibbs energies \citep{2002ntrs.rept85330M}. Users are able to initialise simulations with FastChem \citep{stock2018fastchem, stock2022fastchem}, from constant mixing ratios, from a lookup table, or from a previous run. We initialise with constant mixing ratios for some species such as \ce{O2}, \ce{CO2}, and \ce{N2}. The transport equation is mathematically discretized on a staggered grid, employing a first-order upwind scheme for advection and a second-order central difference scheme for diffusion. To handle the stiffness of the resulting system, \texttt{VULCAN} employs a semi-implicit Rosenbrock integration method. 

\subsubsection{Kasting model}

The Kasting-group 1D photochemical model, a widely used one-dimensional framework for simulating Earth and exoplanet atmospheres  \citep{kasting1979oxygen, kasting1980evolution, segura2003ozone, 2023RSOS...1030056J, Ji_2024JGRD..Correlated_K} has been used and updated over 4 decades. For photochemical simulations, we use the updated \texttt{Kasting} model from \citet{Ji_2024JGRD..Correlated_K}\footnote{Updated \texttt{Kasting} model - \url{https://zenodo.org/records/10384004}}. The reaction network includes 55 chemical and aerosol species composed of H, O, C, N, S, and Cl and includes 222 chemical reactions. The relative humidity distribution is from \citet{manabe1967thermal}. In addition to standard reactions, tropospheric nitric oxide (NO) production by lightning is explicitly parametrised using earlier methods by \citet{chameides1981rates}, modified by \citet{kasting1979oxygen} and \citet{Kasting_1985_Oxidant}. NO production is scaled to lower \ce{O2} levels assuming a high temperature thermodynamic equilibrium at a freeze-out temperature of 3500K in the shock wave surrounding the lightning bolt. The equilibrium \ce{O2} mole fraction is resolved via a Newton-Raphson iterative root finding algorithm, which evaluates \ce{N2}-\ce{O2}-\ce{CO2}-\ce{H2O} composition to a convergence tolerance. The derived theoretical column production is then distributed proportional to the rainout rate from \citet{giorgi1985rainout}. 

Photolysis rates may be calculated using two treatments of SZA: a user inputted single SZA, or an 8-point Gaussian quadrature method \citep{2023RSOS...1030056J}, which evaluates photolysis across eight discrete zenith angles (84.55\textdegree, 73.64\textdegree, 62.74\textdegree , 51.84\textdegree , 40.94\textdegree , 30.05\textdegree , 19.17\textdegree , and 8.35\textdegree) along with a DAF of 0.5 to represent the sunlit hemisphere. Photolysis of \ce{O2} in the SR bands was originally parametrised following \citet{allen1982effective}, who divided the SR region into 17 wavelength intervals based on quantum mechanical calculations \citep{Frederick_1980_O2, Frederick_1980_O2_water} and laboratory measurements \citep{hudson1968bandwidth}. The original implementation used a band model valid for present-day \ce{O2} levels, with absorption occurring primarily in the upper stratosphere and mesosphere. To treat multiple scattering in low-\ce{O2} atmospheres, the model adopted the \citet{Toon_1989_Rapid} $\delta$–2-stream method and replaced the older \citep{Hudson1972Photo} parametrization with the \citet{allen1982effective} band model. Exponential sum fits were derived from \citet{allen1982effective} to couple with the scattering treatment, though these lacked explicit temperature dependence and thus performed poorly under low-\ce{O2} conditions. As mentioned earlier, the model's \ce{O2} photolysis scheme has been recently updated to include a correlated-k parametrisation based on line-by-line calculations, accounting for the strong temperature dependence of SR absorption and overlapping continua (the SR continuum and the Herzberg continuum)\citep{Fernandez_2007_O2, Ji_2024JGRD..Correlated_K}. This approach performs reliably across a wide range of \ce{O2} levels. Comparisons with laboratory data \citep{Yoshino1992} and the \texttt{WACCM6} parametrisation show that the correlated-k method improves accuracy for both modern and low-\ce{O2} atmospheres when multiple scattering is included.

\subsection{The Whole Atmosphere Community Climate Model (WACCM)}
\label{WACCM6}

The Whole Atmosphere Community Climate Model (WACCM6) is a three-dimensional chemistry-climate model configured from the Community Earth System Model (CESM) \citep{gettelman2019whole}. The current version and older versions have been used to study the climates and atmospheres of the early Earth and exoplanets \citep{Chen2019_Hab, Chen2021_Flare, 2023MNRAS.524.1491L, luo2023coupled, cooke2022revised,  cooke2025oxygen}. Here, we build on simulations from \citet{cooke2022revised} and use \texttt{WACCM6} within CESM version 2.1.3 \footnote{CESM2.1.3: \url{https://www.cesm.ucar.edu/models/cesm2}}. \texttt{WACCM6} integrates fully interactive sub-models for the atmosphere, ocean, land, land-ice and sea-ice. The model is an upper atmospheric extension to the Community Atmosphere Model version 6 (CAM6). \texttt{WACCM6} includes non-orographic gravity wave drag parametrisations for both convective and frontal sources, which are essential for driving the meridional overturning circulation of the stratosphere. The model includes water vapour microphysics (e.g. evaporation, rainout, clouds). \texttt{WACCM6} also includes atmosphere-ocean interactions, producing a more realistic measure of dynamic responses to changes in environmental conditions (like temperature). The chemical network includes 98 species and 298 reactions and is described in \citet{emmons2020chemistry}. An implicit numerical scheme is employed to calculate the concentrations of 75 species, while the concentrations of 22 long-lived species were calculated using an explicit numerical scheme \citep{Brasseur_Solomon_2005, sandu1997benchmarking}. The model uses the \citet{price1992simple} parametrisation for lightning. Radiative transfer is calculated using Rapid Radiative Transfer Model for GCMs (RRTMG) \citep{2008JGRD..11313103I, 1997JGR...10216663M}. 

WACCM6 reproduces many major observed features of atmospheric \ce{O3} and its historical evolution \citep{gettelman2019whole, emmons2020chemistry}. In the stratosphere, WACCM6 captures the broad evolution of total \ce{O3} column and the development of Antarctic \ce{O3} depletion over the twentieth century \citep{gettelman2019whole}, with some improvements relative to WACCM4 \citep{Stone2021}. In the troposphere, CESM2-WACCM6 shows generally good agreement with global ozonesonde observations and with aircraft observations of tropospheric composition, showing biases up to $\sim 10$ ppbv at the surface \citep{emmons2020chemistry}. Additionally, when compared with the ERA-Interim, WACCM6 captures the Brewer-Dobson circulation well \citep{Fu_2020LGM, Abalos2021_BDC}. All of this is not to say WACCM6 does not have its biases: for instance, the model can overestimate stratospheric \ce{O3} loss at high southern latitudes \citep{gettelman2019whole, Davis2023}. Whilst WACCM6 is not a benchmark for paeloclimate or exoplanet atmospheres, it is one of the best performing CMIP6 models for \ce{O3} simulations \citep{Fiore2022}. Further benefits and limitations of WACCM6 for paleoclimate and exoplanet simulations will be discussed in Section \ref{Discussion section}. For the original simulations in \citet{cooke2022revised}, \texttt{WACCM6} did not include \ce{CO2} and \ce{H2O} absorption in the SR bands, or scattering at wavelengths $\leq$ 200 nm. The absorption of \ce{CO2} and \ce{H2O} was included in \citet{2023RSOS...1030056J} and is included in the \texttt{WACCM6} simulations used here, published alongside this work in \citet{cooke_kumar_2026}.

\subsection{Planetary Spectrum Generator (PSG)}

While 1D and 3D models simulate the physical and chemical state of an atmosphere, translating these atmospheric profiles into observables requires a radiative transfer and instrument simulation tool. We input our atmospheric outputs into the \texttt{Planetary Spectrum Generator}\footnote{Planetary Spectrum Generator: \url{https://psg.gsfc.nasa.gov}} \citep[PSG;][]{2018JQSRT.217...86V, 2022fpsg.book.....V}. \texttt{PSG} is an online, line-by-line and correlated-k radiative transfer suite designed to model planetary spectra across a broad range of wavelengths for a wide variety of planetary bodies, including comets, solar system planets, and exoplanets \citep{2026arXiv260325694Z, noonan2026emerging, 2024AAS...24324902K}. It generates simulated observations for a wide range of instruments using the Planetary and Universal Model of Atmospheric Scattering \citep[PUMAS;][]{2015Sci...348..218V}) as well as spectroscopic databases \citep[e.g, HITRAN;][]{2017JQSRT.203....3G}. Here, we produce idealised transmission spectra with no associated noise component.

\subsection{The Tropospheric Ultraviolet and Visible (TUV) Radiation Model} 
\label{TUV section}

To translate atmospheric profiles from our 1D models into quantifiable impacts on surface habitability, we use the \texttt{Tropospheric Ultraviolet and Visible radiation} \footnote{TUV, v5.4: \url{https://www2.acom.ucar.edu/modeling/tuv-download}} \citep[TUV;][]{madronich1997theoretical, madronich1992implications} model . \texttt{TUV} is an open source radiative transfer model widely used for calculating photolysis rates and computing surface UV radiation environments for a wide range of input parameters and planetary environments \citep{2010E&PSL.299..263P, 2015AsBio..15..207T, 2018AsBio..18..481T}. By inputting the specific vertical profiles of temperature and \ce{O3} from \texttt{WACCM6}, \texttt{Atmos}, \texttt{Photochem}, \texttt{VULCAN}, and the \texttt{Kasting} models, we use \texttt{TUV} to calculate the actinic flux and spectral irradiance reaching the planetary surface.  We explore the biological effects of calculated surface fluxes in \texttt{TUV} using empirical Biological Weighting Functions (BWFs). Since organic molecules and biological processes exhibit wavelength dependent sensitivities to UV radiation, assessing surface habitability requires calculating the Biologically Effective Dose (BED) rate:

\begin{equation}
    \text{BED} = \int E(\lambda) \cdot W(\lambda) \, d\lambda,
\end{equation}

\noindent where $E(\lambda)$ is the spectral irradiance at the surface and $W(\lambda)$ is the BWF for a specific biological response at wavelength $\lambda$. We calculate the in vitro DNA damage BWF \citep{setlow1974wavelengths}, acting as a proxy for the rate of potentially lethal pyrimidine dimer formation in unshielded surface-dwelling microbes. 

\subsection{Simulations performed}
\label{Simulations performed}

In this section, we detail our modelling choices and outline the specific setups used here. A summary of simulations performed is shown in table \ref{simulation summary table}.

\begin{table}[hbt!]
\centering
\caption{Summary of simulations performed across the 3D chemistry-climate model (\texttt{WACCM6}) and the four 1D photochemical models (\texttt{Photochem}, \texttt{Atmos}, \texttt{VULCAN} and the \texttt{Kasting} model). The baseline parameter space explores Earth-like atmospheres with variable \ce{O2} levels ranging from 0.1\% to 150\% of the Present Atmospheric Level (PAL). Baseline simulations utilised fixed mixing ratios at the lower boundary, while additional fixed flux simulations for \ce{O2} $\leq$ 100\% PAL were run in the Kasting model. Additional sensitivity analyses assess the impact of varying SZAs, surface albedos, and methane boundary conditions. Simulations also extend to paleoclimate and astrobiological contexts by simulating conditions under the stellar spectra of a young sun and the atmosphere of exoplanet Proxima Centauri b.}
\label{simulation summary table}
\begin{tabularx}{\textwidth}{@{}
    >{\hsize=2.0\hsize}X 
    >{\hsize=0.8\hsize}X 
    >{\hsize=0.8\hsize}X 
    >{\hsize=0.8\hsize}X 
    >{\hsize=0.8\hsize}X 
    >{\hsize=0.8\hsize}X 
@{}}
\hline
\textbf{Simulation} & \textbf{WACCM6} & \textbf{Photochem} & \textbf{Atmos} & \textbf{\texttt{VULCAN}} & \textbf{Kasting} \\ 
\hline
Earth's variable \ce{O2} (0.1\%--150\% PAL) & \checkmark & \checkmark & \checkmark & \checkmark & \checkmark \\ 
Fixed \ce{O2} flux ($<$100\% PAL) & No & No & No & No & \checkmark \\
SZA Sweeps (45$^\circ$--60$^\circ$) & N/A & \checkmark & \checkmark & \checkmark & \checkmark \\ 
Surface Albedo (0.06--0.3) & N/A & \checkmark & \checkmark & N/A & 0.25 only \\ 
Methane Perturbations (0.1x--10x) & No & \checkmark & \checkmark & No & No \\ 
YS Spectra (0.0--2.4 Ga) & No & \checkmark & No & No & No \\ 
Coupled YS Climate-Photochemistry (0.0,2.4 Ga) & No & No & \checkmark & No & No \\ 
Proxima Centauri b & \checkmark & \checkmark & \checkmark & \checkmark & No \\ 
\hline
\end{tabularx}
\end{table}

We run 8 \texttt{WACCM6}  simulations of the Earth published alongside this article in \citet{cooke_kumar_2026}, utilising the configuration described in \citet{cooke2022revised}, with the addition of absorption of \ce{H2O} and \ce{CO2} in the SR Bands. To account for uncertainties in estimates of both the magnitude and timing of the rise in atmospheric \ce{O2}, our simulations cover the entire range of estimates over earth's history, with atmospheric \ce{O2} varying between 150\% PAL, 100\% PAL (pre-industrial, PI), 50\% PAL, 10\% PAL, 5\% PAL, 1\% PAL, 0.5\% PAL, and 0.1\% PAL. In all cases, the background gas (\ce{N2}) was changed to maintain atmospheric pressure. The PI simulation is representative of the Earth's atmosphere circa 1850, along with modern day ocean and land settings. Surface boundary conditions were maintained while \ce{O2} was altered, which as noted by \citet{cooke2022revised} is unrealistic, but isolates the effect of changes in oxygenation to atmospheric chemistry and dynamics. These simulations incorporated minor changes to the initial conditions in CESM \footnote{ExoCESM Github: \url{https://github.com/exo-cesm/CESM2.1.3/tree/main/O2\textunderscore Earth\textunderscore analogues}}, which are detailed in \citet{cooke2022revised}. The spatial resolution was set at $1.875^\circ$ by $2.5^\circ$ (96 latitude and 144 longitude points), with the atmosphere containing 70 vertical atmospheric layers from 1000 hPa to $4.5 \times 10$\textsuperscript{-6} hPa (approximately 150 km in altitude). Simulations used an atmospheric time step $\Delta t$ of 30 minutes. These runs did not include tropospheric \ce{O3} production due to volatile organic compounds. For each scenario, the simulation was run for 30 to 90 years until the trend in total hydrogen atoms stabilized in the middle atmosphere (12–100 km), with the outputs averaged over the last 4 simulated years. 

We set up our 1D models as closely as possible to the 3D chemistry-climate runs to ensure that divergences in simulation outputs are largely driven by fundamental differences between 1D and 3D models, or by intrinsic variations amongst the individual models themselves (such as distinct chemical networks and numerical solvers). At the lower-boundary, we use the same fixed mixing ratios boundary conditions \footnote{\texttt{WACCM6}  fixed boundary conditions : \url{https://github.com/exo-cesm/CESM2.1.3/blob/main/O2_Earth_analogues/README.md}} and surface fluxes \footnote{CESM CMIP6 emissions : \url{https://github.com/lkemmons/cesm-cmip6-emissions}} as \citet{cooke2022revised} where possible (shown in Table \ref{WACCM6_boundary_conditions}). For species not explicitly defined in the \texttt{WACCM6}  runs, we retain default surface fluxes and deposition velocities in all models. Available species differ in different models, and we assume that default boundary conditions for these species have been individually chosen for each model. Therefore, while we harmonise species with boundary conditions in \texttt{WACCM6} (Table \ref{WACCM6_boundary_conditions}), we retain model defaults for all other species. We test the effect of this decision in our preliminary runs and simulate the atmosphere using \texttt{Photochem}, but with the default boundary conditions for \ce{NOx} species
from \texttt{VULCAN}, \texttt{Atmos} and \texttt{Kasting}. We find that the \ce{O3} column only varies by 0.01\%, 0.52\% and 0.29\% respectively in these runs. While lower boundary fluxes in our \texttt{WACCM6}  simulations incorporated temporally and spatially varying CESM CMIP6 emissions from the year 1850 cycled over a 12 month period, our 1D runs incorporate static global averages to represent annual mean fluxes, with the total sum used for species having multiple sources (e.g., biomass burning, oceans, soils etc). For models that do not support altitude-dependent volcanic injection of SO$_2$, we apply the aggregate flux in Table \ref{WACCM6_boundary_conditions} as a single lower-boundary condition. The lower boundary conditions used in each model are listed in Appendix \ref{boundary conditions appendix}.

\begin{table}[h]
\centering
\caption{\texttt{WACCM6}  boundary conditions adopted for our 1D simulations. Values represent fixed boundary conditions and CESM2 emission fluxes.}
\label{WACCM6_boundary_conditions}
\begin{tabular}{l c c}
\hline
Species & Mixing Ratio & Flux (molecules cm$^{-2}$ s$^{-1}$) \\
\hline
\multicolumn{3}{c}{\textit{Fixed Mixing Ratios (\citet{cooke2022revised})}} \\
\hline
N$_2$ & 0.78 & - \\
O$_2$ & 0.21 & - \\
CH$_4$ & $0.808 \times 10^{-6}$ & - \\
CO$_2$ & $284 \times 10^{-6}$ & - \\
H$_2$ & $5.00 \times 10^{-7}$ & - \\
OCS & $3.37 \times 10^{-10}$ & - \\
N$_2$O & $2.73 \times 10^{-7}$ & - \\
CH$_3$Cl & $4.57\times 10^{-10}$ & - \\
\hline
\multicolumn{3}{c}{\textit{Emission Fluxes (CESM2 CMIP6, also used in \citet{cooke2022revised})}} \\
\hline
CO & - & $4.98 \times 10^{10}$ \\
NO & - & $2.37 \times 10^{9}$ \\
CH$_2$O & - & $5.16 \times 10^{8}$ \\
SO$_2$ (surface) & - & $1.2 \times 10^{8}$ \\
SO$_2$ (volcanic) $^{+}$ & - & $1.31 \times 10^{9}$ \\
C$_2$H$_6$S (DMS)$^{*}$ & - & $1.62 \times 10^{9}$ \\
\hline
\multicolumn{3}{l}{\scriptsize{$^*$ DMS is only included in \texttt{Atmos}}} \\
\multicolumn{3}{l}{\scriptsize{$^+$ Volcanic fluxes are distributed between an altitude range of 0.24 and 5.45km in the WACCM runs.}} \\
\end{tabular}
\end{table}

\begin{figure}[b!]
    \centering
    \includegraphics[width = 1 \linewidth]{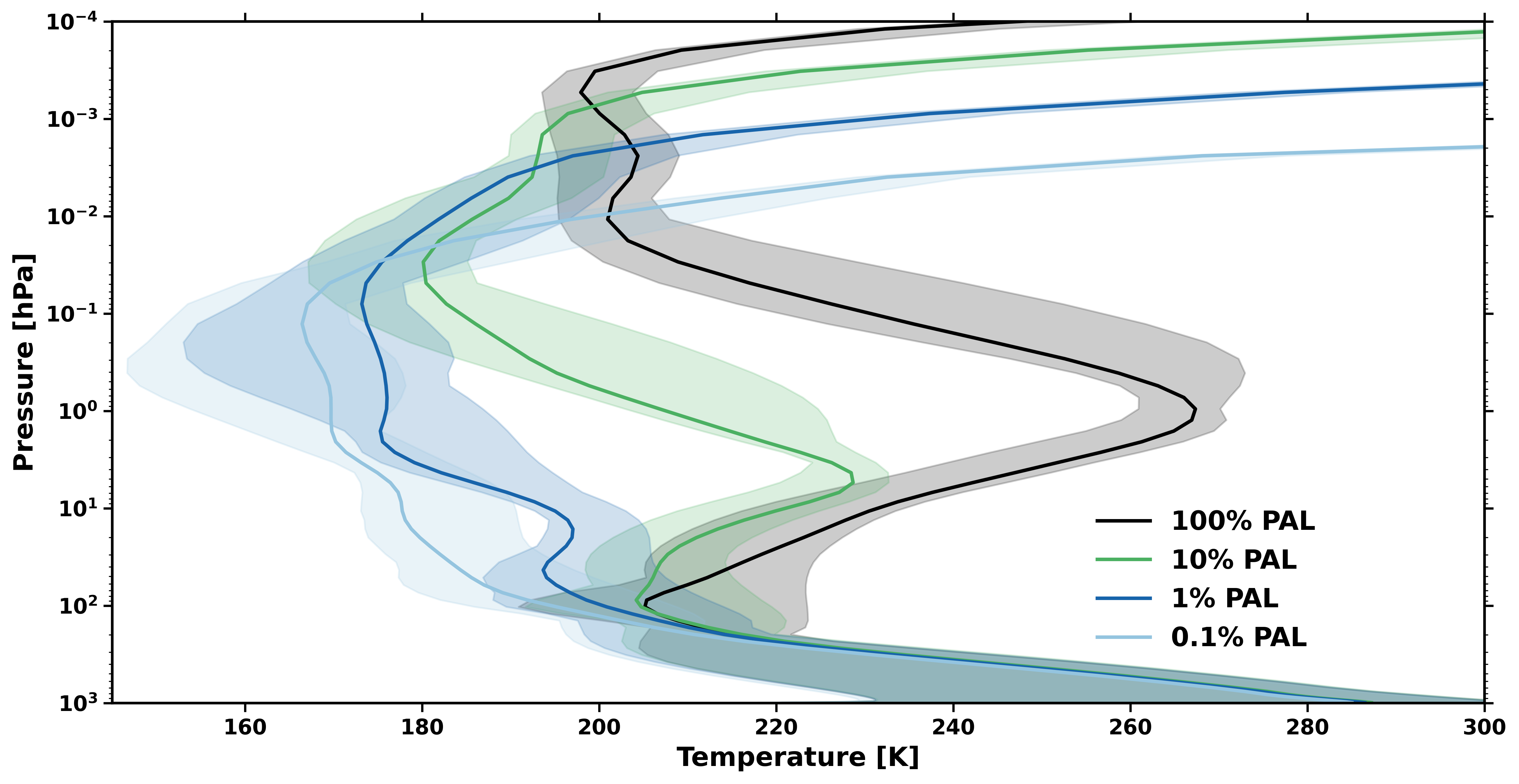}
    \caption{Temperature profiles (in kelvin) varying with pressure and shown for four \texttt{WACCM6}  simulations. The global mean temperature profiles from the 100\% PAL (black), 10\% PAL (green), 1\% PAL (blue), and 0.1\% PAL cases (light blue) are shown as the solid lines, with the shaded region representing the total range of temperature (time-averaged).}
    \label{P-T WACCM6}
\end{figure}

To capture the broad parameter space covered by our WACCM runs, we vary three key inputs in our 1D simulations: the atmospheric oxygenation state, the pressure-temperature (P-T) profile, and SZA. Again, we vary atmospheric \ce{O2} between 150$\%$, 100$\%$, 50$\%$, 10$\%$, 5$\%$, 1$\%$, 0.5\%, and 0.1$\%$ PAL. For each of these oxygenation states, we use P-T profiles generated from the \texttt{WACCM6}  simulations and input the vertical global mean profile into the 1D models (shown in Fig.~\ref{P-T WACCM6}). Minor variations in P-T profiles are expected in the 1D model runs due to inherent differences in how these are implemented in each model. 

Our 3D simulations calculated photolysis rates using specific angles which vary in each grid cell with each atmospheric time step. Translating this to a 1D model is complicated by the fact that they utilise different default SZAs and DAFs. The specific default DAF used in the 1D models is model dependent, ranging from simple fractional multipliers (\texttt{Atmos}, \texttt{Photochem}, \texttt{VULCAN}) to more complex 8-point Gaussian quadrature (\texttt{Kasting}). These default settings are summarised in Table \ref{sza_diurnal}. To explore the sensitivity of atmospheric composition to insolation geometry and bridge the gap between these differing defaults, we ran our baseline simulations using an SZA of 48.2$^\circ$ alongside a DAF of 3/8 (0.375)  as suggested by \citet{2014JAtS...71.2994C}. Furthermore, we conducted parameter sweeps varying the SZA between 45$^\circ$ and 60$^\circ$ (paired with a DAF of 0.354 and 0.5 respectively) across the models. This captures the range most commonly used in 1D photochemical modelling literature (\citep{moses2011disequilibrium, hu2012photochemistry, 2014JAtS...71.2994C}, also see Appendix C in \citet{tsai2021comparative}). 

\begin{table}[b!]
\centering
\caption{Summary of default SZAs and diurnal averaging factors native to the 1D models, alongside the default surface albedo in each model.}
\label{sza_diurnal}
\begin{tabular}{l c l c}
\hline
\textbf{Model} & \textbf{Default SZA} & \textbf{Diurnal Integration Scheme} & \textbf{Default Surface Albedo} \\
\hline
\textbf{Photochem} & $60^\circ$ & Diurnal factor of 0.5 & 0.2\\
\textbf{Atmos} & $60^\circ$ & Diurnal factor of 0.5 & 0.25 \\
\textbf{\texttt{VULCAN}} & $58^\circ$ & Diurnal factor of 0.5 & N/A \\
\textbf{Kasting} & $48.2^\circ$ & Diurnal factor of 0.375 \textit{or} & 0.25 \\
 & & 8-point Gaussian quadrature & \\
\hline
\end{tabular}
\end{table}

Our 1D models also diverge in their treatment of albedo. Earths surface reflectivity is highly varied, ranging from highly absorptive open oceans and vegetated landmasses to highly reflective regions of sea ice and fresh snow  \citep{2015RvGeo..53..141S, hartmann1994global}. Over its history, it is reasonable to assume that the surface albedo of the Earth has varied \citep{rosing2010no}. In our 3D simulations, albedo is calculated from aerosols and surface properties, while in our 1D simulations, we retain the default surface albedo configurations inherent to each model (listed in Table \ref{sza_diurnal}). We also vary surface albedo in \texttt{Photochem} and \texttt{Atmos} between 0.06, 0.1, 0.15, 0.2, 0.25 and 0.3 to explore model sensitivities to this parameter.

\subsubsection{Photochem}
\label{Photochem simulation description}
We modify the default initial atmospheric composition in the \texttt{Photochem} documentation \footnote{"Modern Earth Tutorial":  
https://github.com/Nicholaswogan/UofA\textunderscore Photochem\textunderscore Workshop/tree/main/3\textunderscore ModernEarthPhotochemistry} and the default settings from \citet{2025PSJ.....6..256W}, as this represents the latest publicly available configuration for simulations of the Earth. Similar to \citet{cooke2022revised} and \citet{cooke2023variability}, we simultaneously adjust \ce{N_2} surface boundary conditions whenever \ce{O_2} boundary conditions were modified, ensuring the total atmospheric pressure remained constant. We vary the oxygen content of the atmosphere using by a fixed mixing ratio lower boundary condition, and use \texttt{WACCM6}  generated P-T profiles interpolated onto the altitude grid of \texttt{Photochem}. Following \citet{2025PSJ.....6..256W}, we adopt a constant relative humidity of 40$\%$, which serves as a tropospheric average between Earth's modern ranges of 77$\%$ and 20$\%$ \citep{manabe1967thermal}. We retain the wet deposition scheme for soluble species (e.g., H$_2$SO$_4$, soluble gases), assuming a modern Earth rainfall rate of 1.1 $\times$ 10$^{17}$ molecules cm$^{-2}$ s$^{-1}$ following the \citet{giorgi1985rainout} parametrisation. The tropopause altitude was set to 1.1 $\times$ 10$^6$ cm. Upper boundary conditions are uniformly zero escape flux for all species except hydrogen-bearing species (H and H$_2$), which undergo diffusion-limited escape. We use lower boundary conditions listed in Table \ref{WACCM6_boundary_conditions} and retain default settings from \citet{2025PSJ.....6..256W} for all other species (see in Table \ref{boundary conditions table} in Appendix \ref{boundary conditions appendix}). 

We identified a discrepancy regarding the default lower boundary conditions for \ce{O3} and various HO\textsubscript{x} species (including H, OH, \ce{HO2}, and \ce{H2O2}). While explicit lower boundary conditions for these species are included in the \texttt{Photochem} tutorial configuration file, they were omitted from the configuration file in \citet{2025PSJ.....6..256W}. During our preliminary simulations, we found that omitting these conditions, specifically the dry deposition of \ce{O3}, caused a notable effect in the predicted \ce{O3} column under the Proxima Centauri stellar spectrum in our exoplanet simulations. In modern Earth environments, \ce{O3} dry deposition velocities vary widely depending on the surface type, ranging from 0.1 to 1.2 cms$^{-1}$ \citep{neirynck2018variability, clifton2020dry, silva2018investigating, lyp2025dry, garland1980mechanism}. For these simulations, we adopted a moderate deposition velocity of 0.1 cms$^{-1}$, maintaining consistency with the default listed \ce{O3} deposition velocity in the \texttt{Kasting} model. Finally, for completeness and alignment with other models in this work, we also reinstated lower boundary conditions for the aforementioned HO\textsubscript{x} species using default values from the \texttt{Photochem} tutorial configuration file. 

\subsubsection{Atmos}
\label{atmos simulations performed}

We modify the default initial atmospheric composition and boundary conditions in the \texttt{Modern Earth Complex} photochemical template \footnote{Modern Earth Complex Photochemical Template : \url{https://github.com/VirtualPlanetaryLaboratory/atmos/tree/master/PHOTOCHEM/INPUTFILES/TEMPLATES/ModernEarthComplex}}. Using the initial atmospheric composition and the input temperature profile, \texttt{Atmos} constructs a vertical grid of 200 layers with a uniform thickness of 0.5 km, and then internally computes the pressure and number density profiles via hydrostatic integration. We interpolate the \texttt{WACCM6}  generated temperature profile onto the altitude grid of \texttt{Atmos} and allow the model to internally compute the pressure and number density profiles. Upper boundary conditions were uniformly zero except for H and \ce{H2}, which undergo diffusion limited escape. Lower boundary conditions were adapted to mirror the pre-industrial control simulations in table \ref{WACCM6_boundary_conditions} where possible, retaining default boundary conditions for all other species (shown in table \ref{boundary conditions table} in Appendix \ref{boundary conditions appendix}). 

\subsubsection{VULCAN}
We modify the default Earth template \footnote{Earth config file: \url{https://github.com/shami-EEG/VULCAN/blob/master/cfg_examples/cfg_Earth.txt}} with default boundary conditions \footnote{Earth boundary conditions: \url{https://github.com/shami-EEG/VULCAN/blob/master/atm/BC_bot_Earth.txt}} and utilise the full S-N-C-H-O reaction network \footnote{Full S-N-C-H-O reaction network: \url{https://github.com/shami-EEG/VULCAN/blob/master/thermo/SNCHO_full_photo_network.txt}} for oxidised atmospheres as well as the N-C-H-O reaction network. We used 120 atmospheric layers between 1 bar and $5\times10^{-8}$ bar (1000 hPa -- $5\times10^{-5}$ hPa) in pressure, specifying \ce{N2} as the background gas, a rocky base, and a gravity of 9.81 ms\textsuperscript{-2}. We used the stellar input spectrum labelled ``Gueymard\textunderscore solar.txt''\footnote{Gueymard Solar file: \url{https://github.com/shami-EEG/VULCAN/blob/master/atm/stellar_flux/Gueymard_solar.txt}} from \citet{Gueymard2004}. At the upper boundary, we assume diffusion-limited escape for H and \ce{H2}. We use the fixed surface mixing ratios and surface fluxes from the \texttt{WACCM6} control case in table \ref{WACCM6_boundary_conditions}, and retain default boundary conditions for all other species, as shown in table \ref{boundary conditions table} in Appendix \ref{boundary conditions appendix}. For water vapour, we adopt a constant relative humidity of 40\%.

\subsubsection{Kasting}

We modify the updated Kasting model for the Earth from \citet{Ji_2024JGRD..Correlated_K} with the new correlated-k parametrization for \ce{O2} photolysis \footnote{Updated \texttt{Kasting} model - \url{https://zenodo.org/records/10384004}}. Using the initial atmospheric composition and an input temperature profile, the model constructs a vertical grid of 100 layers with a uniform thickness of 1km. We interpolate the \texttt{WACCM6} temperature profile onto the altitude grid and allow the model to internally determine the resulting P-T structure. Upper boundary conditions included effusion velocities calculated to simulate the diffusion limited escape of light atmospheric species, including atomic hydrogen (H), molecular hydrogen (\ce{H2}), methane (\ce{CH4}), and water vapour (\ce{H2O}). For heavier species such as \ce{O2} and \ce{CO2}, effusion velocities were determined directly by their respective photolysis rates at the uppermost grid layer. To maintain mass balance and account for this high-altitude photolytic destruction, corresponding constant downward fluxes were applied to the photolysis products (\ce{CO}  and \ce{O}). Again, where possible, lower boundary conditions were adapted to mirror the fixed mixing ratios and surface fluxes from the pre-industrial \texttt{WACCM6}  simulations in table \ref{WACCM6_boundary_conditions}. For all other species, we retain default fluxes and deposition velocities (see Table \ref{boundary conditions table} in Appendix \ref{boundary conditions appendix}). 

To assess the impact of insolation geometry, photolysis rates were computed in two methods. First, we use a single insolation-weighted average SZA of 48.2 degrees coupled with a diurnal averaging factor of 0.375, as suggested by \citet{2014JAtS...71.2994C} and used in \citet{Ji_2024JGRD..Correlated_K}. Secondly, following \citet{2023RSOS...1030056J}, we applied an 8-point Gaussian quadrature to integrate incoming sunlight over the sunlit hemisphere. Furthermore, low-\ce{O2} simulations in \citet{Ji_2024JGRD..Correlated_K} utilised fixed flux boundary conditions at the lower boundary instead of fixed mixing ratios (Aoshuang Ji, Private Communication). Consequently, we also run fixed flux simulations for low-\ce{O2} scenarios (\ce{O2} less than 100\% PAL) with our \texttt{WACCM6}  P-T profiles for comparison. 

\subsubsection{Young Sun Simulations}

To model atmospheres under a young sun, we use parametrized solar fluxes from \citet{claire2012evolution} as model inputs. These spectra were manually used as inputs in \texttt{Photochem}, while \texttt{Atmos} has an inbuilt function to calculate altered spectra using methods in \citet{claire2012evolution} for varying solar ages when coupled with \texttt{Clima}. In our coupled runs in \texttt{Atmos}, the photochemical and climate models were iterated back and forth until we note that the change in the \ce{O3} column between successive runs is less than 0.5 DU.  In Clima, we use the "Modern Earth Complex" climate template \footnote{Modern Earth Complex climate template : \url{https://github.com/VirtualPlanetaryLaboratory/atmos/blob/master/CLIMA/IO/TEMPLATES/ModernEarthComplex/input_clima.dat}}. To ensure consistency between models, we align our relative humidity assumptions with our \texttt{Photochem} simulations. In \texttt{Photochem}, we adopted the constant 40\% relative humidity approximation from \citet{2025PSJ.....6..256W}. As \citet{2025PSJ.....6..256W} note, this value represents a tropospheric average of the \citet{manabe1967thermal} profile, which varies from 77\% near the surface to 20\% at the tropopause. Correspondingly, we utilize the built-in Manabe-Wetherald parametrization in \texttt{Atmos} and initialise the surface relative humidity at 77\% (RH$_{surf}$= 0.77). We acknowledge that this diverges from the 70\% value recommended by \citet{Lincowski2018} to capture globally averaged Earth data, but we adopt it to facilitate a direct comparison. 

\begin{figure}[t!]
    \centering
    \includegraphics[width = 0.75 \linewidth]{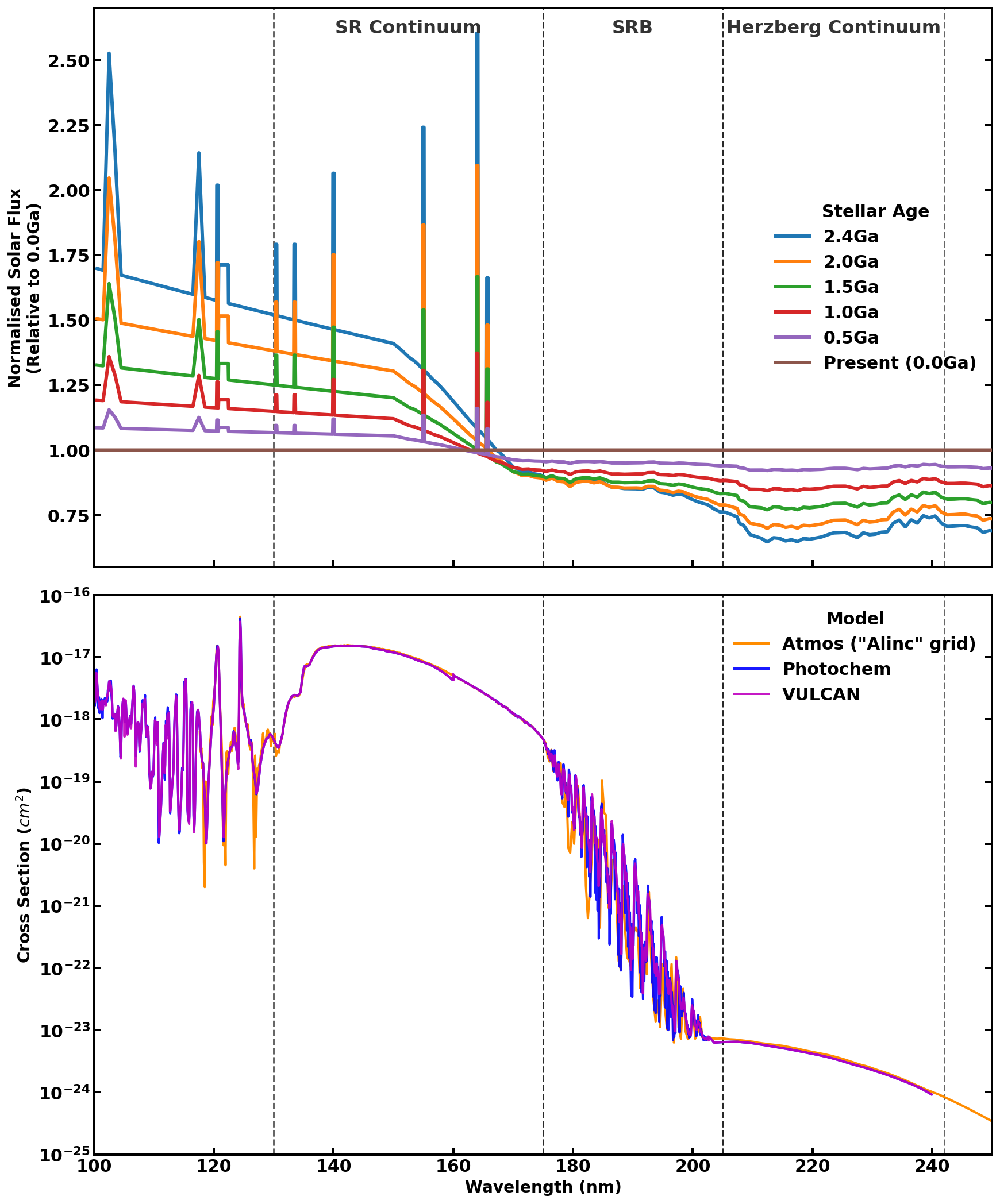}
    \centering
    \caption{Top: Evolution of solar UV fluxes relative to the present-day Sun (0 Ga) for 0.5–2.4 Ga. Marked regions include the Schumann–Runge continuum (130- 175 nm), Schumann–Runge bands (175–205 nm), and the Herzberg continuum (200–242 nm). Bottom: \ce{O2} absorption cross sections from \texttt{Atmos}, \texttt{Photochem}, and \texttt{VULCAN} models.}
    \label{stellar flux}
\end{figure}

Fig.~\ref{stellar flux} shows stellar fluxes in bands relevant to \ce{O3} photochemistry (i.e, the Schumann-Runge Continuum, the Schumann-Runge Bands and the Herzberg Continuum) for the sun at stellar ages of 0.0 (Present), 0.5, 1, 1.5, 2 and 2.4 Ga. Also shown in Fig.~\ref{stellar flux} are the various \ce{O2} absorption cross sections in these regions for the \texttt{Atmos}, \texttt{Photochem}, and \texttt{VULCAN} models. We run our uncoupled simulations for all stellar ages (0.0 - 2.4 Ga) across the oxygenation grid (0.1\% PAL - 150\% PAL), and run coupled simulations for 0.0 Ga (present) and 2.4 Ga for atmospheric \ce{O2} of 0.1\% PAL and 100\% PAL. 

\subsubsection{DNA damage simulations}
We modelled the surface UV environment and biological DNA damage under our simulated atmospheres using \texttt{TUV}. For each of the four \ce{O2} scenarios, we extract the altitude, temperature and \ce{O3} profiles from the \texttt{Photochem}, \texttt{VULCAN}, \texttt{Kasting}, and \texttt{Atmos} simulations. For  \texttt{WACCM6}, we extract these from a gaussian latitude mean, as well as the equatorial column.  We input atmospheric chemistry from each model at photochemical equilibrium using our baseline (SZA 48.2 \textdegree) simulations. We then utilise an illumination geometry in TUV across an SZA range of 0\textdegree to 85\textdegree to calculate temporal variation of surface UV driven by the diurnal cycle. We input the \ce{O3} column, temperature structure, and altitude grids from our chemistry simulations into \texttt{TUV}, retaining all other default parameters. Calculations were performed over a range of 100 to 420 nm. We output spectral irradiance, DNA damage calculated using \citet{setlow1974wavelengths}, and plant damage using \citet{caldwell1971solar} and \citet{flint2003biological}.

\subsubsection{Exoplanet simulations}
\label{Exoplanet simulation section}
We simulated the atmosphere of habitable zone exoplanet Proxima Centauri b under a range of conditions using \texttt{Photochem}, \texttt{Atmos} and \texttt{VULCAN} to explore how model assumptions translate to predictions for exoplanet spectra, particularly exoplanets around M-dwarfs where the stellar SED is known. Proxima Centauri b is the closest exoplanet to our solar system and orbits the star Proxima Centauri 1.3 pc away, making it a promising exoplanet candidate in the search for life elsewhere \citep{2016A&A...595A...2G, 2023SPIE12680E..1UF}. We use 4 \texttt{WACCM6} simulations of Proxima Centauri b from \citet{Cooke_2024_Lethal}, who assumed an Earth-like atmosphere with \ce{N2} as the background gas and atmospheric \ce{O2} ranging from 100\% PAL, 10\% PAL, 1\% PAL and 0.1 \% PAL. These simulations included absorption by \ce{O3}, \ce{O2}, \ce{CO2} and \ce{H2O} in the SR bands. Consistent with previous works, Proxima Centauri b is assumed to have an insolation of 884 Wm$^{-2}$ \citep[0.65 $S_0$,][]{Ridgway2023_Flares, yates2020ozone, boutle2017exploring}. Simulations assumed a minimum  mass of 1.07 M$_{\oplus}$ and a radius of 1.05 R$_{\oplus}$. Furthermore \citet{Cooke_2024_Lethal} assumed a rotation rate of 11.2 days; and assumed the planet to be tidally locked. We simulate the atmosphere assuming an Earth-like composition with the same \ce{O2} range, and use the GJ 551 (Proxima Centauri) MUSCLES stellar spectrum \citep{youngblood2016muscles, france2016muscles, p2016muscles}. As mentioned in section 2 (\ref{Photochem simulation description}), initial \texttt{Photochem} simulations revealed that omitting the \ce{O3} deposition velocity boundary condition led to drastically inflated \ce{O3} columns. Consequently, we reinstated this parameter across all models. Furthermore, our \texttt{Atmos} simulations converged prematurely when using the default atmospheric composition files. To resolve this, we initialized these runs using the \ce{O2} column output from our Earth simulations as our model input.

\section{Results}

We first present the major results in terms of the ozone-oxygen (\ce{O2}-\ce{O3}) relationship, showing the predictions of each photochemical model in terms of the total \ce{O3} column depth at each \ce{O2} mixing ratio. After this, we analyse the results to determine why each model gets different final atmospheric compositions in each scenario, before presenting the results of several perturbation experiments. 

\subsection{Ozone abundance and total column density}
\label{Ozone columns section}

\begin{figure}[t!]
    \centering
    \includegraphics[width = 0.95\textwidth]{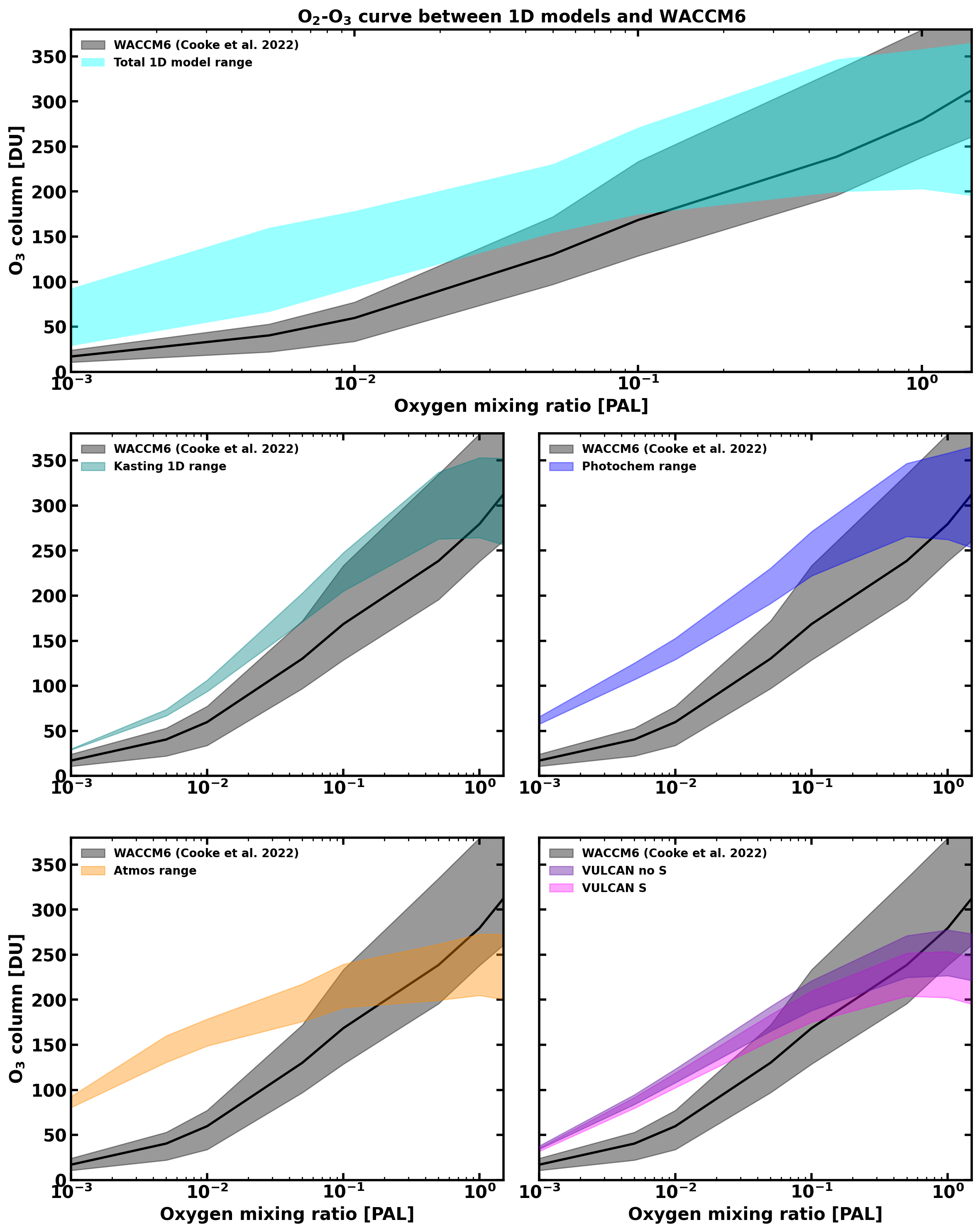}
    \centering
    \caption{The calculated total \ce{O3} column in Dobson Units (DU) from the atmospheric simulations conducted in this work is shown against the assumed atmospheric mixing ratio of \ce{O2}. In the top panel, the mean and range of the \texttt{WACCM6}  \ce{O3} columns is shown in black and grey, respectively, with the range of the 1D columns shown in blue. \ce{O3} columns from simulations using the \texttt{Kasting} model (teal; middle left panel), \texttt{Photochem} (blue; middle right panel), \texttt{Atmos} (orange; bottom left panel), \texttt{VULCAN} (bottom right panel; magenta for sulphur network and purple for the network without sulphur) are compared to \texttt{WACCM6}. The range for the 1D models shows simulations with a SZA between $45^\circ$ and $60^\circ$. The oxygen concentrations simulated for this figure were 0.1\% 0.5\%, 1\%, 5\%, 10\%, 50\%, 100\%, and 150\% PAL. For each \ce{O2} level, the column was calculated using the globally averaged vertical pressure-temperature profile from \texttt{WACCM6}  at that \ce{O2} concentration.}
    \label{ozone vs oxygen}
\end{figure}

Fig.~\ref{ozone vs oxygen} shows the \ce{O3} column in Dobson Units (DU) for various atmospheric \ce{O2} mixing ratios and all five models used in this work. The top panel compares the range in \texttt{WACCM6}  to the range in the four 1D models, and then the subsequent four panels compares \texttt{WACCM6}  to each individual 1D model. The boundary conditions applied (fluxes and mixing ratios) are set to be the same for each model where applicable. The thick line shows the mean \texttt{WACCM6}  columns, whilst the range is between the minimum and maximum time-averaged \ce{O3} columns. The range showed for the 1D models is accounting for SZAs of $45^\circ$ -- $60^\circ$, with the higher SZAs producing lower \ce{O3} columns. Generally speaking, the column depth of \ce{O3} increases with atmospheric oxygenation. There are some exceptions to this for high \ce{O2} atmospheres in the range of 50\% -- 150\% PAL of \ce{O2}: depending on the model and the zenith angle, the total \ce{O3} column can decrease with increasing \ce{O2}. 

At 100\% PAL, the \texttt{Kasting} model and \texttt{Photochem} show better agreement with \texttt{WACCM6} than \texttt{VULCAN} and \texttt{Atmos}. The discrepancy between the 1D models and \texttt{WACCM6}  is exacerbated at lower \ce{O2} mixing ratios. There is significant overlap with the \texttt{WACCM6}  range at \ce{O2} mixing ratios of $>50$\% PAL. As \ce{O2} reduces below 5\% PAL, \texttt{Photochem}, \texttt{Atmos}, and \texttt{VULCAN} all produce \ce{O3} columns above the range of the \texttt{WACCM6}  \ce{O3} columns. The \texttt{Kasting} model is much closer, and only diverges beyond the range at $<1\%$ PAL, with it producing the smallest \ce{O3} columns at 0.1\% PAL of the 1D models. Differences in predictions for \ce{O3} columns are critical for two pursuits: understanding the history of Earth, and interpreting observations of exoplanet atmospheres. Both will be explored here, but first we will explore why these \ce{O3} columns differ for Early Earth at all.

\begin{figure}[t!]
    \centering
    \includegraphics[width = 1 \textwidth]{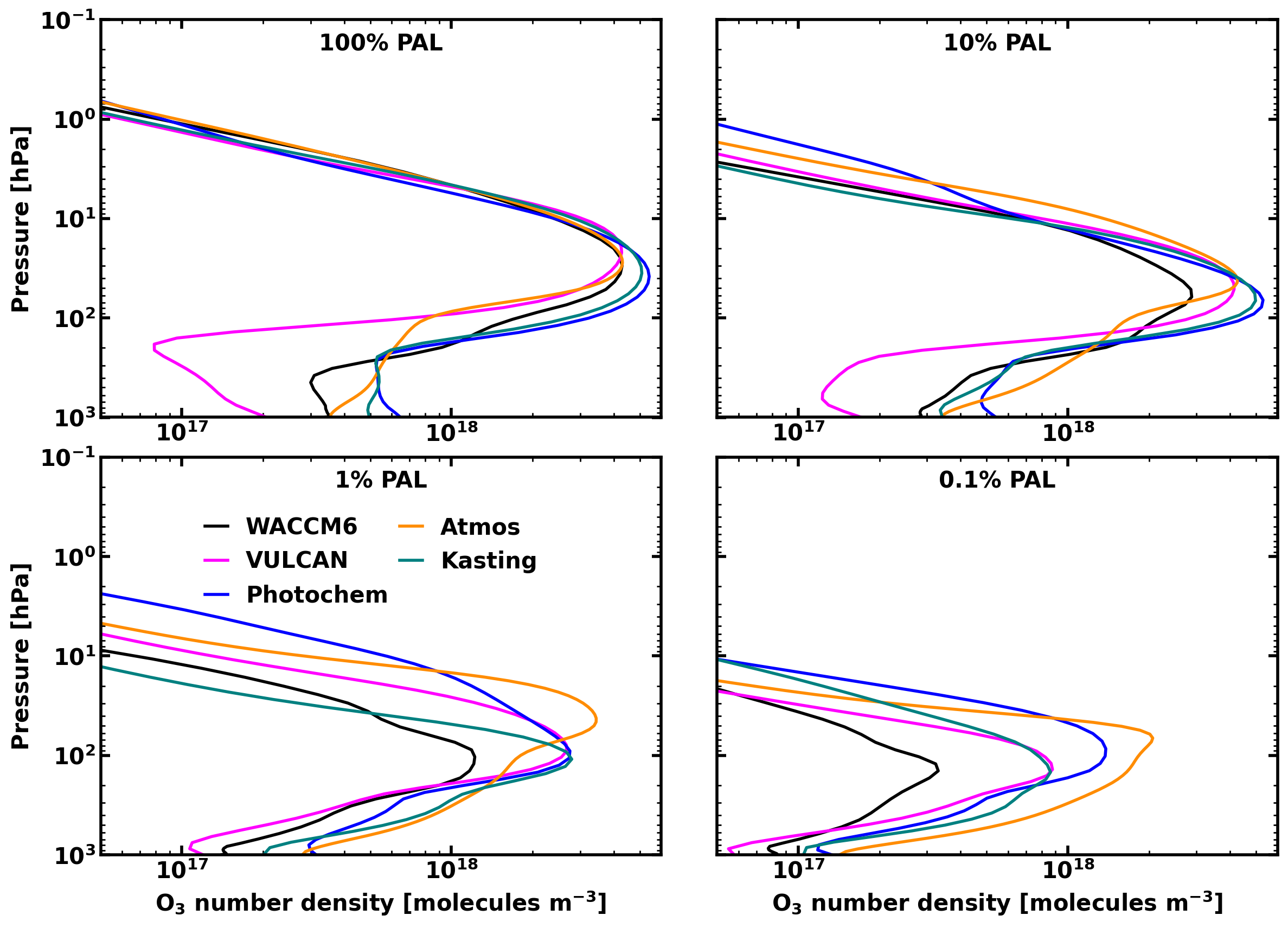}
    \caption{The number density of \ce{O3} molecules for all five models is plotted for the 100\% PAL (top left), 10\% PAL (top right), 1\% PAL (bottom left), and 0.1\% PAL (bottom right) atmospheres. \texttt{WACCM6}  is in black, the \texttt{Kasting} model in teal, \texttt{VULCAN} in magenta, \texttt{Photochem} in blue, and \texttt{Atmos} in orange. The \texttt{WACCM6}  profile is a globally averaged result.}
    \label{Ozone density}
\end{figure}

On the modern-day Earth, the majority of \ce{O3} resides in the stratosphere, roughly 15 to 35 km above the surface in the `ozone layer' \citep{Brasseur_Solomon_2005}. Fig.~\ref{Ozone density} displays the number density of \ce{O3} for all five models in the 100\% PAL, 10\% PAL, 1\% PAL, and 0.1\% PAL atmospheres. The figures show the pressures at which each model diverges in terms of the total \ce{O3} that is present. For modern-day \ce{O2}, the stratospheric \ce{O3} number density is in reasonable agreement between all five models, but they significantly diverge in the troposphere. As \ce{O2} is decreased, the discrepancy in the stratosphere is apparent: peak densities vary by a factor of 1.18, 1.85, 2.8, and 6.0 for each respective \ce{O2} concentration.

In 1D models, when the mixing ratio of \ce{O2} increases, the atmosphere ``self-shields'', and the peak altitude of \ce{O2} photolysis shifts upwards, attenuating the photon flux reaching the lower stratosphere, and starving the lower (and denser) atmospheric layers of atomic oxygen which is required for \ce{O3} production (see Section \ref{Production of Ox and O3} for further discussion on this). This occurs as a result of \ce{O2} and \ce{O3} absorption of ultraviolet radiation overlapping. When the peak altitude shifts higher, \ce{O3} formation can be slower because of lower pressures, and vice versa. Furthermore, these shifts affect radical production (e.g., OH) that can then change the final produced \ce{O3} abundance \citep{Kasting2003}. 

This mechanism also occurs somewhat in 3D models, but here the 3D nature of the planet, which includes atmospheric transport, affects the results. \citet{Jozefiak_2023} identified 3D transport as crucial to \ce{O3} calculations because the mean flow of the atmosphere moves \ce{O3} from where it is generated in the tropical atmosphere towards the extratropics. This does not necessarily occur in the same manner for all 3D models because accurately simulating the Brewer-Dobson circulation depends on simulation resolution,  orography, and gravity wave parametrisations \citep{Butchart2010Chem, Abalos2021_BDC}.

\begin{figure}[t!]
    \centering
    \includegraphics[width = 1\textwidth]{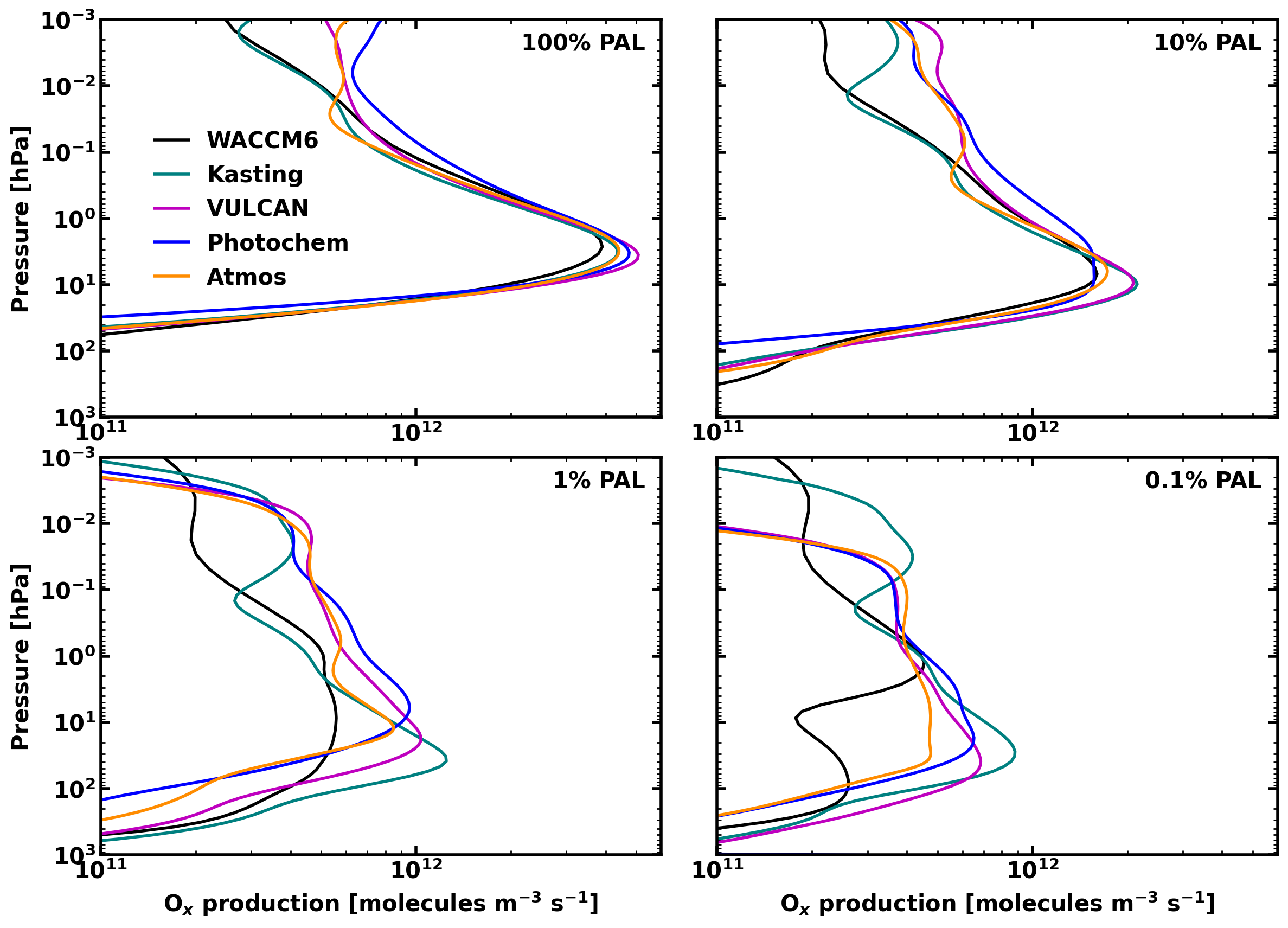}
    \centering
    \caption{The O\textsubscript{x} production rate from \ce{O2} photolysis is plotted for the 100\% PAL (top left), 10\% PAL (top right), 1\% PAL (bottom left), and 0.1\% PAL (bottom right) atmospheres. The colour scheme is the same as in Fig. ~\ref{Ozone density}.}
    \label{Ox production figure}
\end{figure}

A thorough understanding of the discrepant \ce{O3} abundances and total columns between 1D and 3D models is difficult due to the dichotomy in the inclusion of transport and a day-to-night cycle. An initial intercomparison was conducted by \citet{2023RSOS...1030056J}, who compared results from ROCKE-3D, \texttt{WACCM6}, and the \texttt{Kasting} model. Since then, \citet{Ji_2024JGRD..Correlated_K} updated the cross sections and parametrisations for \ce{O2} photolysis in the \texttt{Kasting} model and demonstrated their updated model produced \ce{O2} photolysis results much closer to the \texttt{WACCM6}  results. We also tested both an 8-point Gaussian quadrature and fixed fluxes, as described in \citet{2023RSOS...1030056J}, and the results follow a similar trend to the presented \texttt{Kasting} simulations. The following sections will further explore in more detail why the models diverge.

\subsection{Production of O\textsubscript{x} and \ce{O3}}
\label{Production of Ox and O3}

The production of O\textsubscript{x} is the first step in the Chapman mechanism \citep{chapman1930xxxv, Brasseur_Solomon_2005} for atmospheric \ce{O3} production, where \ce{O2} is photodissociated into atomic oxygen (O) in two main channels: 

\begin{align}
     \ce{O2 +} h\nu\ (175.9\ \textrm{nm} < \lambda < 242.4\ \textrm{nm}) &\ce{-> O + O}\label{O2 photolysis reaction 1},\\
      \ce{O2 +} h\nu\ (\lambda < 175.9\ \textrm{nm} ) &\ce{-> O(^1D) + O}, \label{O2 photolysis reaction 2}
\end{align}

\noindent where $h\nu$ represents a photon of frequency $\nu$, and $h$ is Planck's constant. The first electronically excited state of atomic oxygen is denoted as \ce{O(^1D)}, where the superscript `1' indicates the spin multiplicity and `D' represents the total orbital angular momentum. This state is energetically distinct from the ground state, which is characterized by the term symbol \ce{O^3P}, otherwise just written as O here by standard convention \citep{Brasseur_Solomon_2005}.

Across all modelling frameworks, the treatment of molecular \ce{O2} photolysis has the two aforementioned branching channels and calculates photodissociation across the Herzberg continuum, the highly structured SR bands, and the SR continuum. The major structural differences emerge in how these models resolve the complex absorption features of the SR region and account for thermal variations. This has been explored previously \citep{2023RSOS...1030056J}, such that the \texttt{Kasting} model now employs an updated parameterized approach as mentioned earlier \citep{Ji_2024JGRD..Correlated_K}. Conversely, \texttt{Photochem} and \texttt{VULCAN} ingest pre-tabulated, high-resolution cross-section and branching grids from unified data files, with the option for \texttt{VULCAN} to apply temperature dependence. \texttt{Photochem} uses high resolution \ce{O2} cross sections from \citet{2017A&A...602A.105H}, who source their high-resolution \ce{O2} cross sections in the SR bands from \citet{Yoshino1992}. \texttt{Atmos} offers both parameterized and high-resolution options, with the default Earth template using cross section data updated by \citet{Lincowski2018} and based on data from \citet{Yoshino1988, Yoshino1992, Lu2010}. However, the  high-resolution option does not work with the publicly available model, so we utilised the default grid. Finally, \texttt{WACCM6} calculates \ce{O2} photolysis rates in the SR bands using a parametrization from \citet{1996AnGeo..14...68K}. In the 3D model, the effective cross section is dependent on the temperature and slant-column. 

We display the \ce{O2} photolysis results in Fig.~\ref{Ox production figure} for all five models, which is essentially a measure of the production of `odd oxygen' (O\textsubscript{x}\ce{= O + O3}). It is the production of O\textsubscript{x} that initiates the Chapman mechanism, with \ce{O3} formation and destruction, as well as atmospheric dynamics, setting the overall concentration and distribution of \ce{O3}. Generally, 1D models calculate photolysis rates through a single SZA with a DAF to take into account the fact that half of the Earth is not illuminated by the Sun at any one time. The peak O\textsubscript{x} production rate occurs in the stratosphere in all the models at all the \ce{O2} concentrations, between 1 hPa to 100 hPa. The peak moves to higher pressures (lower altitudes) as \ce{O2} is reduced because more UV light can penetrate further as the optical depth of the atmosphere above is reduced.

In the 100\% PAL atmosphere, there is reasonably good agreement for \ce{O2} photolysis in the stratosphere between all 5 models between 0.1 hPa and 100 hPa: the peak photolysis rate difference at a pressure of $\sim 3$ hPa is roughly 1.3. These initial deviations are heightened when \ce{O2} is decreased. At 10\% PAL, the general shape and magnitude of the profiles are similar, with the photolysis peaking close to 10 hPa (all models are within a factor of 1.4 of each other) and then reducing before reaching the surface. The models start to diverge significantly more at 1\% PAL, with neither the shape nor magnitude (varying up to 2.1 in difference) consistent between the models. At this mixing ratio, \texttt{WACCM6} has the lowest peak production in the stratosphere and the \texttt{Kasting} model the largest. At 0.1\% PAL the peak rates differ by up to a factor of 1.9, with those peaks occurring between 1 hPa and 40 hPa. Generally, \texttt{WACCM6} has the lowest peak production, with \texttt{VULCAN} the highest for present day, and the \texttt{Kasting} model the highest for lower \ce{O2} mixing ratios. These results between the \texttt{Kasting} model and \texttt{WACCM6} are consistent with the findings reported in \citet{2023RSOS...1030056J}. Meanwhile, the total integrated photolysis throughout the entire atmosphere remains within 30\% at 100\% and 10\% PAL, within 40\% at 1\% PAL, and within 60\% at 0.1\% PAL. This result shows that the calculated mismatch in photolysis rates is more closely related to where the photolysis takes place rather than the total amount of photolysis. Importantly, this has consequences for the \ce{O3} layer because its formation and destruction significantly depends upon its atmospheric location.

Once \ce{O2} photolysis occurs, \ce{O2} and O can form \ce{O3} via the following 3 body reaction:

\begin{align}
\ce{O + O2 + M &-> O3 + M}, \label{O3 formation reaction}
\end{align}

\noindent where \ce{M} is any third body (usually \ce{N2} or \ce{O2} on Earth due to their abundance dominating the total amount of molecules in air). Reaction \ref{O3 formation reaction} proceeds faster at cooler temperatures and higher pressures. This is one reason why the temperatures were uniform between the presented 1D simulations.

We investigated the reaction rate coefficients for all five models in the various simulated atmospheres. Using the same pressure-temperature profile, at 100\% PAL, the rate coefficients between the models differ by a factor of 1.19 in the stratosphere (between 100 -- 10 hPa). This factor is 1.19, 1.22, and 1.27, for 10\%, 1\% , and 0.1\% PAL cases, respectively. \texttt{Photochem} has the smallest reaction rate, with the \texttt{Kasting} and \texttt{Atmos} models intermediate, and \texttt{VULCAN} having the largest rate. \texttt{WACCM6} is also intermediate and the rate depends on which part of the atmosphere is isolated for the analysis (e.g. tropics versus global reaction rate).

We also checked the incoming solar spectra between 100 nm and 400 nm. Differences in UV flux will bias the photolysis rates relative to other models. The Sun from \texttt{Atmos} and \texttt{Photochem} are from \citet{Thuillier2004}, the spectrum used in \texttt{VULCAN} is \citet{Gueymard2004}, and the one used in \texttt{WACCM6}  is from \citet{Matthes2017}. Based on comments in the code, we believe the solar data in the \texttt{Kasting} model comes from a World Meteorological Organization report on atmospheric \ce{O3} \citep{WMO1985AtmosphericOzone}. In the wavelength range between 180 -- 250 nm, the spectra vary between $\approx-15$\% and +40\% when compared to the \texttt{WACCM6}  solar spectra, with the mean integrated fractional deviation varying between 8\% and 11\%. We replaced the solar spectrum in \texttt{Photochem} with the \texttt{WACCM6} solar spectrum, and this reduced the \ce{O3} columns produced by \texttt{Photochem} by a factor of 1.03 (100\% PAL case) to 1.5 times lower (0.1\% PAL case) relative to the column depth when using the standard stellar file for the Sun in \texttt{Photochem}. This does not reduce the \ce{O3} columns enough to explain the overall discrepancies. Instead, it shows that solar data taken at different epochs affect simulation outcomes, serving as a warning against treating host star spectra as static inputs in photochemical modelling.

\texttt{VULCAN} has an option to include temperature dependent photodissociation cross sections for molecules, including \ce{O2}. This option is not considered in the default Earth configuration file. When we included the temperature dependence, for atmospheres at $\leq1$\% PAL \ce{O2} and 150\% PAL \ce{O2}, the \ce{O3} columns decreased, whereas between 5 -- 100\% PAL \ce{O2}, the \ce{O3} columns increased. However, we caution users not to use this option as currently implemented because the temperature dependent cross sections are on a much sparser wavelength grid compared to the default cross sections.

In summary, we have determined that the production of O\textsubscript{x} is substantially different between all five models. This determines the amount of \ce{O3} that can form, and the rate coefficient for the \ce{O3} formation reaction significantly differs between all five models too. The destruction of \ce{O3} will now be analysed in the following section.

\subsection{Destruction of \ce{O3}}
\label{Destruction of O3 section}

Both chemical reactions and photochemistry remove \ce{O3} from the atmosphere. Photolysis takes place in two branching channels:

\begin{align} 
    \ce{O3 +} h\nu\ (\lambda \geq 320\ \textrm{nm}) &\ce{-> O2(^3\Sigma^-_g) + O},\label{ozone photolysis 1} \\
    \ce{O3 +} h\nu\ (\lambda \leq 320\ \textrm{nm}) &\ce{-> O2(^1\Delta_g) + O(^1D)}. \label{ozone photolysis 2}
\end{align}

However, when \ce{O3} is destroyed through photolysis, the O and \ce{O2} produced can quickly react to form \ce{O3} again. Practically speaking, \ce{O3} is not removed from the denser layers of the atmosphere (mid-stratosphere -- surface) via its photochemical destruction. Rather, \ce{O3} is removed in other ways. For instance, it is destroyed when it reacts directly with atomic oxygen:

\begin{align} 
    \ce{O + O3 &-> 2 O2}.\label{ozone loss}
\end{align}

This direct recombination is relatively slow. On its own, it cannot account for the total observed \ce{O3} depletion in the stratosphere. Instead, most \ce{O3} destruction occurs via highly efficient catalytic cycles involving trace radical species. These radicals generally belong to the hydrogen (HO\textsubscript{x}\ce{=H + OH + HO2 + H2O2}), nitrogen (NO\textsubscript{x}\ce{= N + NO + NO2}), chlorine (\ce{ClOx}), and bromine (\ce{BrOx}) families. For a given catalyst, X, these reactions generally proceed in two consecutive steps:

\begin{align}
\ce{X + O3 &-> XO + O2}\label{ozone catalytic loss 1}, \\
\ce{XO + O &-> X + O2}\label{ozone catalytic loss 2}.
\end{align}

Since the catalyst X is consumed in step 1 and regenerated in step 2, the net result of this cycle is chemically identical to the direct \ce{O3} loss mechanism. Consequently, a single trace radical can cycle through these reactions, destroying thousands of \ce{O3} molecules before it is removed from the atmosphere by forming a more stable reservoir species. In the atmosphere, several species are involved in such a cycle, including OH, Cl, NO and \ce{NO2} \citep{1991oup..book.....W}. For \ce{HO}\textsubscript{x} species, the cycle can be written as :

\begin{align}
\ce{H + O3 &-> OH + O2} \label{H catalytic loss 1},\\
\ce{OH + O &-> H + O2} \label{H catalytic loss 2},
\end{align}

\noindent and, 

\begin{align}
\ce{OH + O3 &-> HO2 + O2} \label{OH catalytic loss 1},\\
\ce{HO2 + O &-> OH + O2} \label{OH catalytic loss 2}.
\end{align}

While these reactions give an indication of how \ce{O3} balance is maintained in the atmosphere, several catalytic cycles exist within the HO\textsubscript{x} family which do not fit into the general category described in reactions \ref{ozone catalytic loss 1} and \ref{ozone catalytic loss 2}. For instance the series of reactions: 

\begin{align}
\ce{OH + O &-> H + O2},\\
\ce{H + O2 + M &-> HO2 + M},\\
\ce{HO2 + O &-> OH + O2},
\end{align}

\noindent gives the net reaction \ce{O + O + M -> O2 + M}, destroying two odd oxygen species. This cycle is of importance above 40 km \citep{1991oup..book.....W} in the modern atmosphere. Similarly at altitudes ($\leq$ 30 km):

\begin{align}
\ce{OH + O3 &-> HO2 + O2},\label{OH catalytic loss no O 1} \\
\ce{HO2 + O3 &-> OH + O2 + O2}\label{OH catalytic loss no O 2}.
\end{align}

\begin{figure}[t!]
    \centering
    \includegraphics[width = 1 \textwidth]{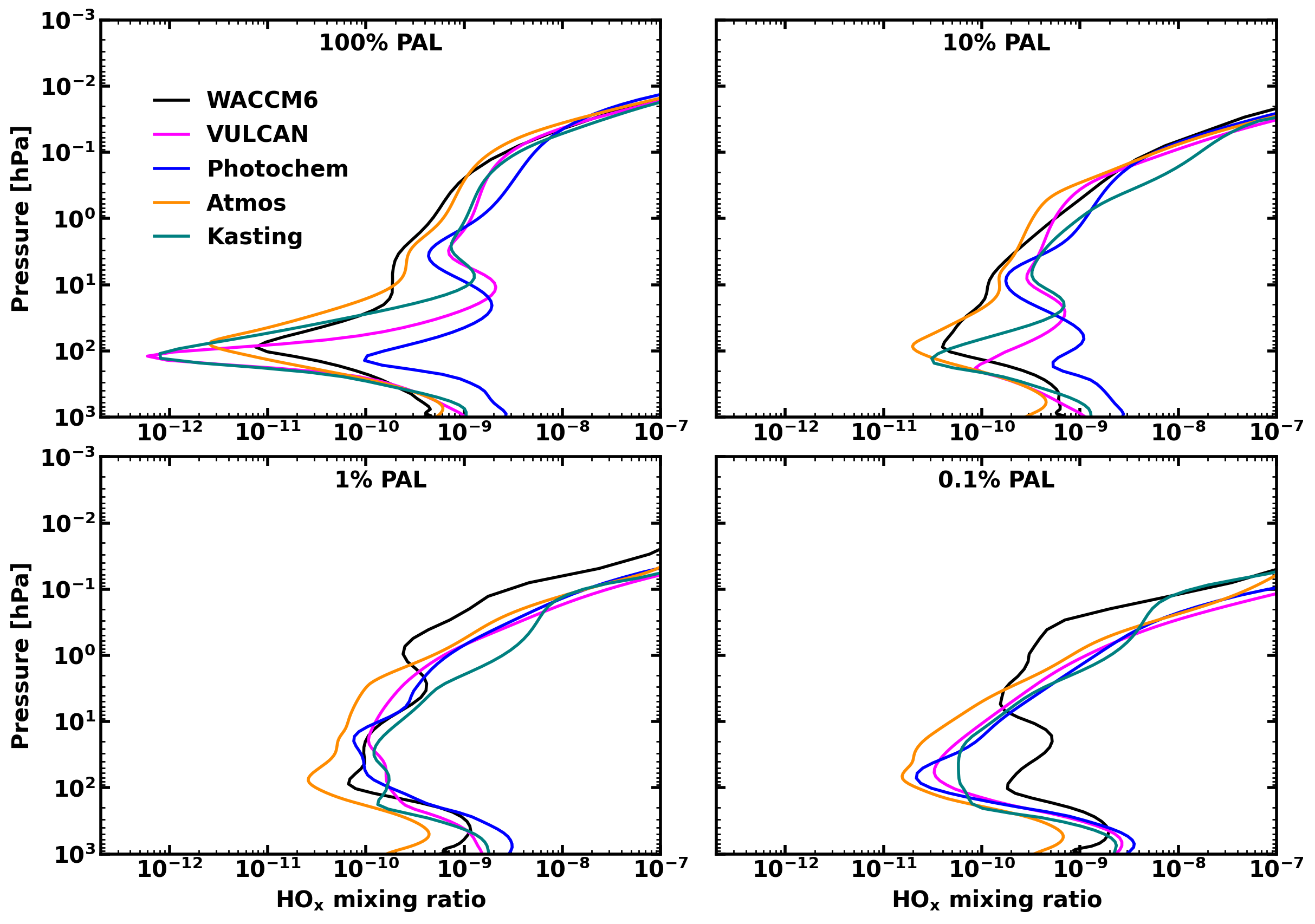}
    \caption{The mixing ratio of HO\textsubscript{x} (\ce{H + OH + HO2 + H2O2}) is shown against pressure in hPa for the 100\%, 10\%, 1\%, and 0.1\% PAL scenarios for all five photochemical models used in this work. The colours refer to the same models as in Fig.~\ref{Ox production figure}. The data from \texttt{WACCM6} is the globally averaged vertical profile.}
    \label{HOX mixing ratio figure}
\end{figure}

The net reaction becomes important since the concentration of atomic oxygen is low at those altitudes. Therefore, the effect of HO\textsubscript{x} species on the destruction of odd oxygen is driven by several species. In Fig.~\ref{HOX mixing ratio figure}, we present the HO\textsubscript{x} mixing ratio for each model and scenario. 

Similarly, for NO\textsubscript{x}, reactions \ref{ozone catalytic loss 1} and \ref{ozone catalytic loss 2} can be written as :

\begin{align}
\ce{NO + O3 &-> NO2 + O2} \label{NO catalytic loss 1},\\
\ce{NO2 + O &-> NO + O2} \label{NO catalytic loss 2}, 
\end{align}

giving the same net reaction as reaction \ref{ozone loss}. 

Whilst we have not compared the entire chemical network and all of the reaction rates, we compared several reaction rate coefficients relevant for \ce{O3} catalytic cycle destruction for each model. We found that there are clear differences for the \ce{NO + O3} and \ce{HO2 + O3} reactions.

In the modern atmosphere, concentrations of atomic O in the troposphere are very low because there is very little photolysis of \ce{O2}. Many of the HO\textsubscript{x}, NO\textsubscript{x}, and ClO\textsubscript{x} cycles require a free O atom to complete their cycle and remove \ce{O3}, but OH does not (see reaction \ref{OH catalytic loss no O 1} and \ref{OH catalytic loss no O 2}). In the stratosphere between 25 -- 40 km, NO\textsubscript{x} radicals dominate the loss of \ce{O3} \citep{Jucks1996}. Above this altitude, HO\textsubscript{x} radicals dominate the loss of odd oxygen \citep{nicolet1971aeronomic, Brasseur_Solomon_2005}.

HO\textsubscript{x} mixing ratios are determined in part from the \ce{H2O} concentrations, which we will explore in Section \ref{Water vapour profile section}. For HO\textsubscript{x} mixing ratios, the agreement is actually worst between all models for the 100\% PAL atmosphere, with lower stratospheric mixing ratios differing by up to a factor of $\sim 250$ (\texttt{Photochem} and \texttt{VULCAN} is the smallest) with \ce{H2O2} the dominant HO\textsubscript{x} molecule at this point. The surface HO\textsubscript{x} mixing ratio in all models and simulations is maintained between $1\times10^{-10}$ and $3\times10^{-9}$, with \texttt{Photochem} always having the most and \texttt{Atmos} generally having the least. As \ce{O2} is reduced, the predictions start to converge, with the best agreement at 1\% PAL. In the 0.1\% PAL case, \texttt{WACCM6}  has higher HO\textsubscript{x} than the 1D models in the lower stratosphere. The past agreement was worse until the \texttt{WACCM6} was updated to include absorption from \ce{CO2} and \ce{H2O} in the SR bands \citep{2023RSOS...1030056J}.

\begin{figure}[b!]
    \centering
    \includegraphics[width = 1 \textwidth]{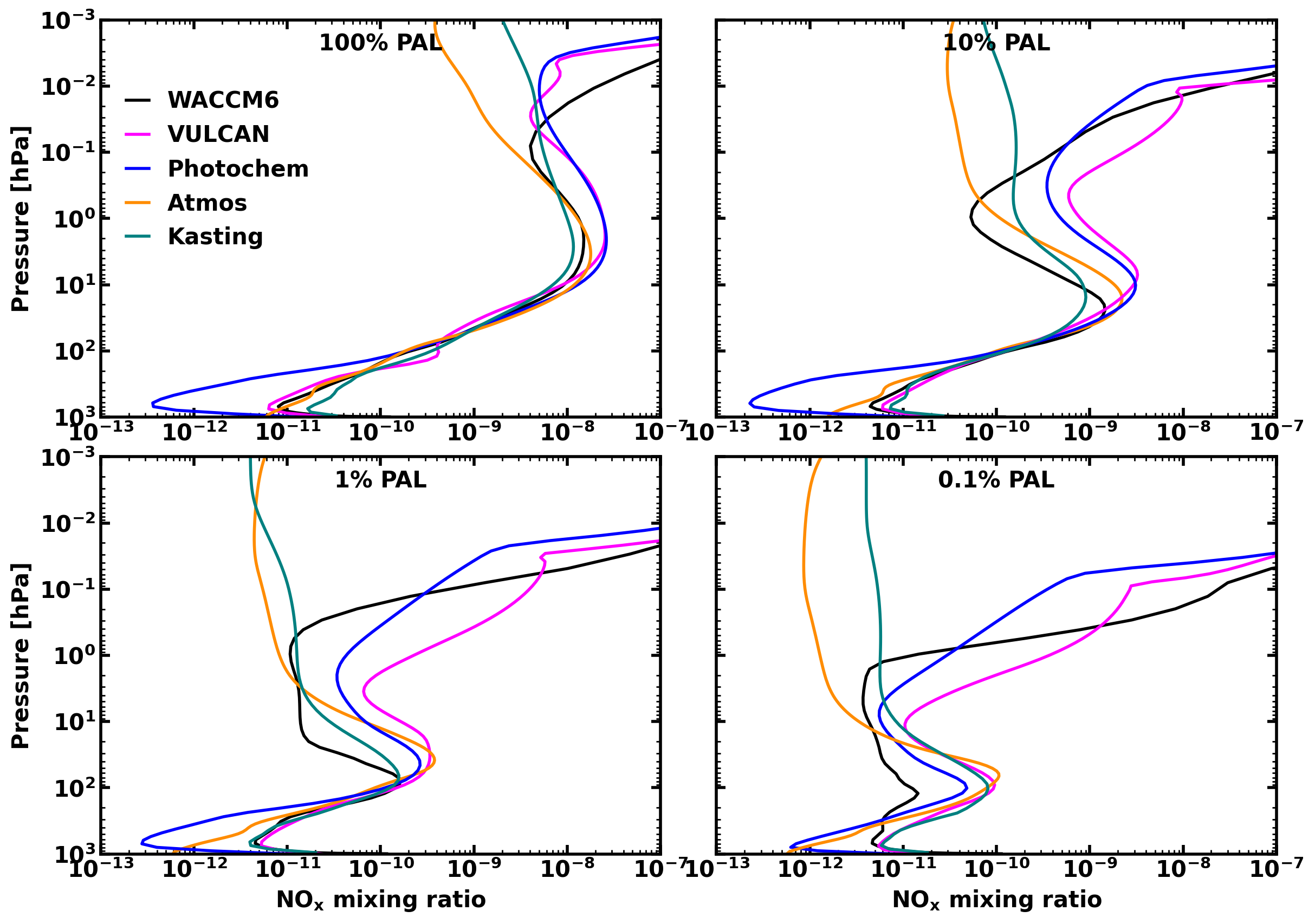}
    \caption{The mixing ratio of NO\textsubscript{x} (\ce{N + NO + NO2}) is shown against pressure in hPa for the 100\%, 10\%, 1\%, and 0.1\% PAL scenarios for all five photochemical models used in this work. Note that N is not included in the \texttt{Atmos} and \texttt{Kasting} model data. The colours refer to the same models as in Fig.~\ref{Ox production figure}. The data from \texttt{WACCM6}  is the globally averaged vertical profile.}
    \label{NOX mixing ratio figure}
\end{figure}

In Fig.~\ref{NOX mixing ratio figure} we present the NO\textsubscript{x} mixing ratios as a function of pressure. Between the surface and the upper stratosphere, \ce{NO2} and NO are both important components of NO\textsubscript{x}, with local concentrations depending on the altitude and time of day. \ce{NO2} photolysis during the daytime depletes \ce{NO2} and produces NO, with \ce{NO2} becoming more abundant at nighttime. For 100\% PAL, the stratospheric NO\textsubscript{x} mixing ratios are within a factor of 3 between all models, in stark contrast to the comparatively poor HO\textsubscript{x} agreement. However, when calculating partitioning between \ce{NO2} and NO, a dichotomy between the 1D models and \texttt{WACCM6}  emerges. \ce{NO2} is underpredicted in the 1D models relative to \texttt{WACCM6} when considering the region near the peak in \ce{O3} concentrations and at altitudes above the peak.

All models exhibit an increasing importance for \ce{O3} destruction from HO\textsubscript{x} in the stratosphere and troposphere compared to NO\textsubscript{x} as \ce{O2} decreases, apart from \texttt{Photochem} which does not have this trend in the troposphere. In the upper stratosphere at 100\% PAL ($<1$ hPa), HO\textsubscript{x} overtakes NO\textsubscript{x} and provides the majority of catalytic \ce{O3} destruction. This changeover moves to higher pressures and by 0.1\% PAL the switch between NO\textsubscript{x} to HO\textsubscript{x} destruction in the stratosphere is around $\sim 50$ hPa.  The exception to this is \texttt{WACCM6}, where HO\textsubscript{x} is the most important for destruction throughout the troposphere and stratosphere at 0.1\% PAL. This may be due to the fact that \texttt{WACCM6} does not include scattering in the SR bands, but it is also in part related to the 3D nature of the planet and the \ce{H2O} concentrations, as we will now discuss.

\subsection{Water vapour profile}
\label{Water vapour profile section}

In the Earth's troposphere, the concentration of \ce{H2O} is set by local temperatures and dynamics, including evaporation from rivers, lakes and oceans, as well as cloud formation and precipitation. Generally speaking, as air parcels rise, they lose heat and the air cannot hold as much water, so the \ce{H2O} mixing ratio reduces. Upon reaching the tropopause, atmospheric parcels reach the ice saturation mixing ratio minimum, also known as the `cold trap'. On Earth, this cold trap occurs in the tropical latitudes, specifically a region known as the Tropical Tropopause Layer (TTL) \citep{Fueglistaler2009}. \citet{cooke2025oxygen} recently explored this for the \texttt{WACCM6}  scenarios between 0.1\% PAL - 150\% PAL. In short, the cold trap sets the \ce{H2O} concentrations in the stratosphere, and this is in turn determined by the warmest temperature regions in the TTL. The global temperature profile and the extratropical temperatures do not determine the stratospheric \ce{H2O} temperature profile. Since all of the 1D models use global average temperatures from \texttt{WACCM6}, this distinction is not captured. Thus, 1D models cannot predict the \ce{H2O} profile accurately, and this biases the \ce{H2O} vertical profiles in the stratosphere \citep{cooke2025oxygen}.

\begin{figure}[t!]
    \centering
    \includegraphics[width = 1 \textwidth]{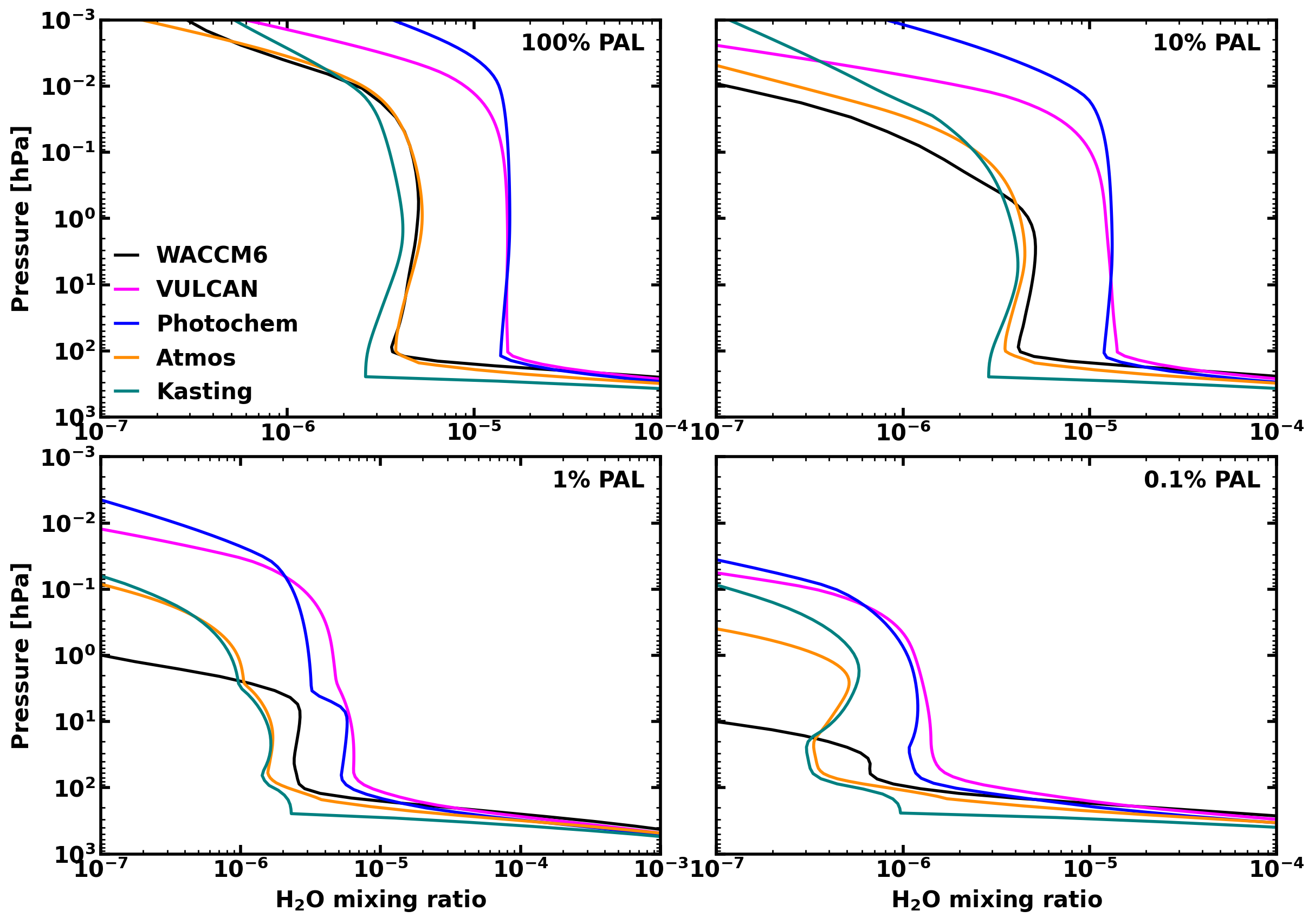}
    \caption{The mixing ratio of water vapour (\ce{H2O}) is shown against pressure in hPa for the 100\%, 10\%, 1\%, and 0.1\% PAL scenarios for all five photochemical models used in this work. The colours refer to the same models as in Fig.~\ref{Ox production figure}. The data from \texttt{WACCM6}  is the globally averaged vertical profile.}
    \label{Water vapour profile figure}
\end{figure}

The \ce{H2O} mixing ratio versus pressure for all 5 models are shown in Fig.~\ref{Water vapour profile figure}. In the 100\% PAL atmosphere, the \texttt{Atmos} and \texttt{Kasting} models best represent the global mean \ce{H2O} mixing ratio profile from \texttt{WACCM6}, with stratospheric entry values of 4 ppmv and 4.3 ppmv, respectively, compared to 3.7 ppmv in \texttt{WACCM6}. The peak stratospheric values are also within 1 ppmv of \texttt{WACCM6}. \texttt{VULCAN} and \texttt{Photochem}, on the other hand, have much higher stratospheric \ce{H2O}, with mixing ratios varying between 14--16.5 ppmv. The picture is similar at 10\% PAL but starts to change for 1\% PAL. \texttt{Atmos} and the \texttt{Kasting} model have lower stratospheric \ce{H2O} entry values than \texttt{WACCM6}. In all cases, \texttt{Atmos} and the \texttt{Kasting} model are the closest to \texttt{WACCM6}, while \texttt{VULCAN} and \texttt{Photochem} diverge. This is due to the fact that both \texttt{Atmos} and \texttt{Kasting} use the Manabe-Wetherald parametrization of atmospheric humidity, while \texttt{Photochem} and \texttt{VULCAN} use an average single altitude dependent relative humidity for the troposphere. Although \texttt{Photochem} includes an option for Manabe parametrization, we find that the humidity profile is unchanged when this setting is enabled.  Ultimately, the water vapour profiles in 1D models are biased to an unknown magnitude until comparison with a 3D model is made. Even then, the 3D model will have its own biases despite capturing more detailed water vapour microphysics.

Starting with photolysis of \ce{O3}, the concentration of \ce{H2O} alters the amount of \ce{HO}\textsubscript{x} in the local atmosphere.

\begin{equation}
    \text{O}_3 + h\nu \, (\lambda < 320\text{ nm}) \rightarrow \text{O}_2 + \text{O}(^1\text{D}),
\end{equation}

\noindent with the reactive $\text{O}(^1\text{D})$ atom reacting with \ce{H2O} to generate two hydroxyl radicals:

\begin{equation}
    \text{O}(^1\text{D}) + \text{H}_2\text{O} \rightarrow 2 \text{OH}.
\end{equation}

\noindent Overall,  this mechanism modifies the catalytic destruction of \ce{O3}. 

Furthermore, despite \texttt{WACCM6}  having intermediate stratospheric entry values, \ce{H2O} mixing ratios in the troposphere are higher for \texttt{WACCM6}  when compared to all of the 1D models in each scenario, which is further exacerbated in the moist air of the tropical atmosphere. On top of this, the \ce{H2O} $J$ rates are higher in \texttt{WACCM6}  than the 1D models. This combination naturally leads to larger photolysis rates of \ce{H2O}. In the low \ce{O2} cases production of OH from \ce{H2O} photolysis outweighs production of OH from the reaction between \ce{O^1D} and \ce{O3} \citep{2023RSOS...1030056J}. All OH mixing ratios in the troposphere are within an order of magnitude of each other when comparing the models.

Previously, \citet{2023RSOS...1030056J} found through sensitivity tests that larger amounts of tropospheric \ce{H2O} in the \texttt{Kasting} model reduced the \ce{O3} column. We increase the humidity in \texttt{VULCAN} from 0.4 to 0.6, which brings \ce{H2O} mixing ratios much closer to \texttt{WACCM6}  at pressures $>200$ hPa, but raises \ce{H2O} in the stratosphere to unrealistic values, simultaneously reducing the \ce{O3} column between 5--14\% for the $\leq1$ \%  PAL cases. 

\subsection{Atmospheric transport}
\label{3D effects section}
There is a clear dichotomy between the \texttt{WACCM6} 3D model and the four 1D models in terms of the total \ce{O3} columns they produce at various oxygen concentrations. However, it is possible that the discrepancy stems from photolysis treatments and the chemical network implemented for the atmosphere, rather than 3D effects. If this was the case, then the Brewer-Dobson circulation (BDC) would be effective at changing the latitudinal distribution of \ce{O3}, but not the globally averaged atmospheric column of \ce{O3}. To test this, we briefly analyse the atmospheric dynamics of the \texttt{WACCM6} simulations and perform two perturbation experiments.

\begin{figure}[t!]
    \centering
    \includegraphics[width = 1 \textwidth]{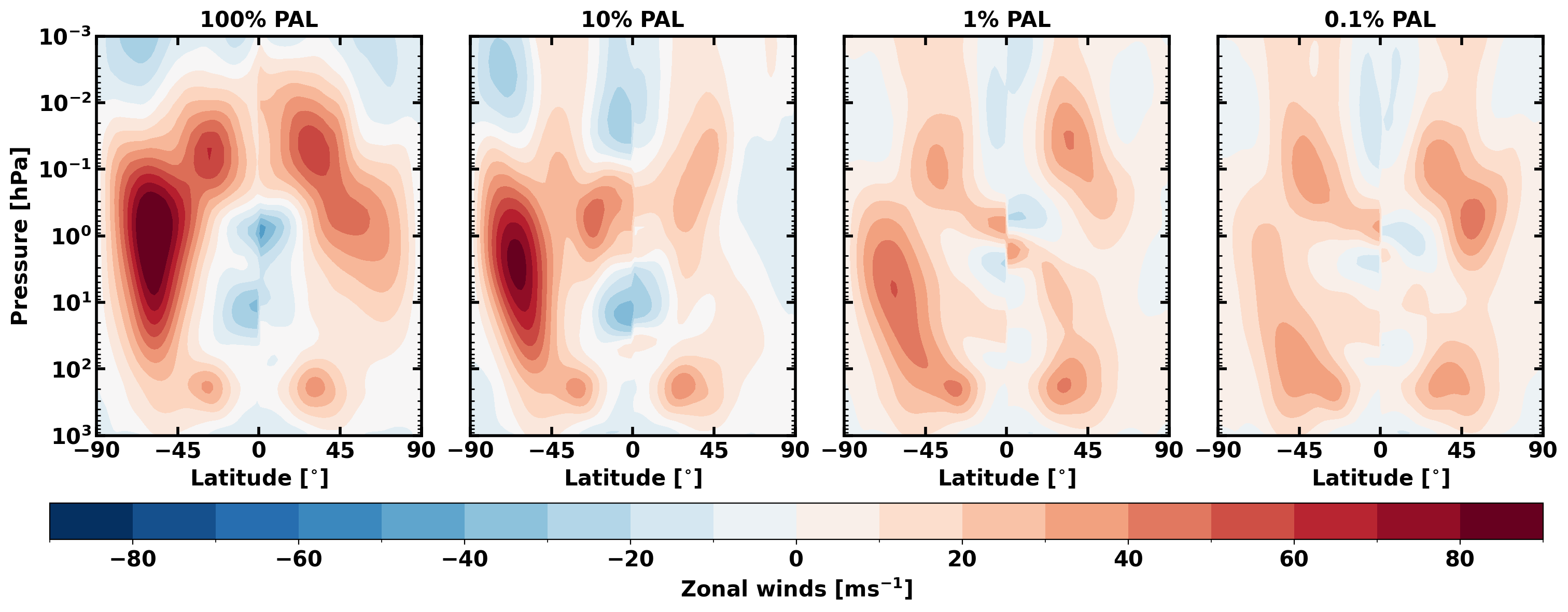}
    \caption{Zonal mean winds in m s\textsuperscript{-1} for four \ce{O2} levels (left to right: 100\%, 10\%, 1\%, and 0.1\% PAL). Each panel is a winter hemisphere composite: northern hemisphere latitudes use the DJF mean and southern hemisphere latitudes use the JJA mean of one model year. Red shows westerlies (positive) and blue shows easterlies (negative). Thus, the winter subtropical polar jet and polar-vortex structure is shown in each case.}
    \label{Zonal winds figure}
\end{figure}

We show two major changes to atmospheric circulation resulting from solely changing the atmospheric \ce{O2} mixing ratio. The zonal mean of the hemispheric winter zonal winds are shown in Fig.~\ref{Zonal winds figure}. This is a winter mean of December, January and February (DJF) for the northern hemisphere and June, July and August (JJA) for the southern hemisphere. The modern (100\% PAL) simulations produces the expected winter pattern of the polar jets and subtropical westerlies. At reduced \ce{O2}, as the stratosphere adjusts to a thinner \ce{O3} layer and thus reduced heating, the stratospheric zonal wind structure changes in conjunction: there is a major weakening of the southern hemisphere stratospheric jet as \ce{O2} is decreased, and accompanied more generally by the winter westerly winds reorganising in latitude and altitude. Whilst zonal winds do not measure the overturning BDC circulation directly, they mediate the atmospheric waves that drive the BDC. The zonal wind structure alters where gravity waves break \citep{Brasseur_Solomon_2005}, which act on the deep branch of the BDC.

\begin{figure}[b!]
    \centering
    \includegraphics[width = 1 \textwidth]{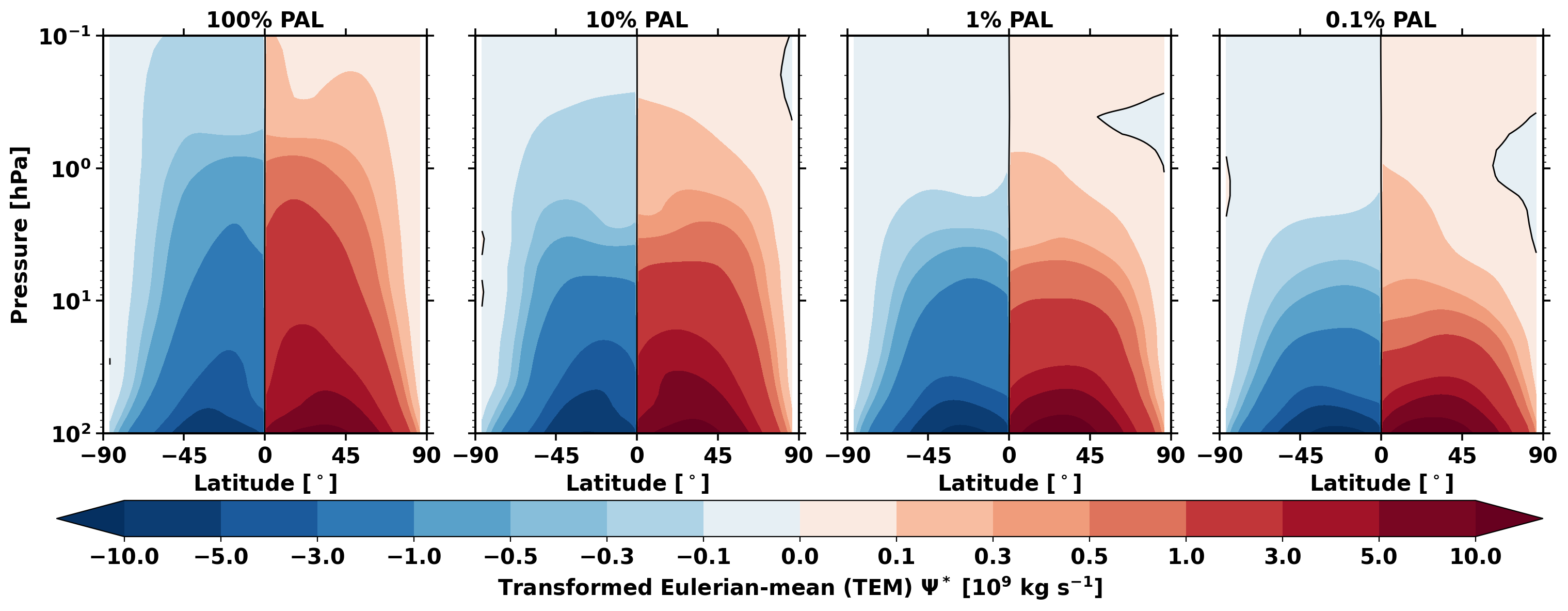}
    \caption{The residual TEM mass streamfunction ($10^9$ kg s\textsuperscript{-1}) for four \ce{O2} levels (left to right: 100\%, 10\%, 1\%, and 0.1\% PAL), using the same winter composite (NH = DJF, SH = JJA) as Fig.~\ref{Zonal winds figure}. TEM is derived from the diabatic residual circulation (radiative heating). Colour uses a symmetric logarithmic scale spanning approximately 0.1--$10\times10^9$ kg s\textsubscript{-1}, with the zero contour is shown in black. Closed cells indicate the meridional overturning of the Brewer–Dobson circulation, with the air rising in the tropics, travelling poleward, and then descending.}
    \label{TEM figure}
\end{figure}

In Fig.~\ref{TEM figure} we show the Transformed Eulerian Mean (TEM) streamfunction, which combines the mean fluid motion of the atmosphere with the wave-driven eddy component of atmospheric transport, showing the overall net meridional movement of mass in the atmosphere. For more details on TEM formulations we refer readers to \citet{andrews1976planetary} and \citet{Brasseur_Solomon_2005}.

The residual mass streamfunction $\Psi^*$ diagnoses the Brewer–Dobson overturning as a closed mass circulation in the meridional plane. Fig.~\ref{TEM figure} shows winter hemisphere composites of $\Psi^*$ (northern hemisphere DJF, southern hemisphere JJA) across four \ce{O2} concentrations. At 100\% PAL the streamfunction shows that overturning extends through the middle stratosphere. As \ce{O2} mixing ratio declines, this the lower-stratospheric (shallow branch) remains and intensifies near $\sim70$--100 hPa, whereas the deep branch that reaches $\sim1$ hPa weakens markedly by 0.1\% PAL.

These changes imply that extratropical \ce{O3} transport becomes increasingly restricted to the shallow branch at low \ce{O2}. Poleward transport via the deep branch is the pathway that pushes tropical \ce{O3} into the mid- and upper stratosphere under modern \ce{O2}. Therefore, as it weakens, the ability for the circulation to set the extratropical column reduces. Overall, these two circulation changes appear to be influenced by the temperature structure alterations and gravity wave dissipation. However, a full analysis of dynamical changes and their root causes is beyond the scope of this article.

\begin{figure}[b!]
    \centering
    \includegraphics[width = 1 \textwidth]{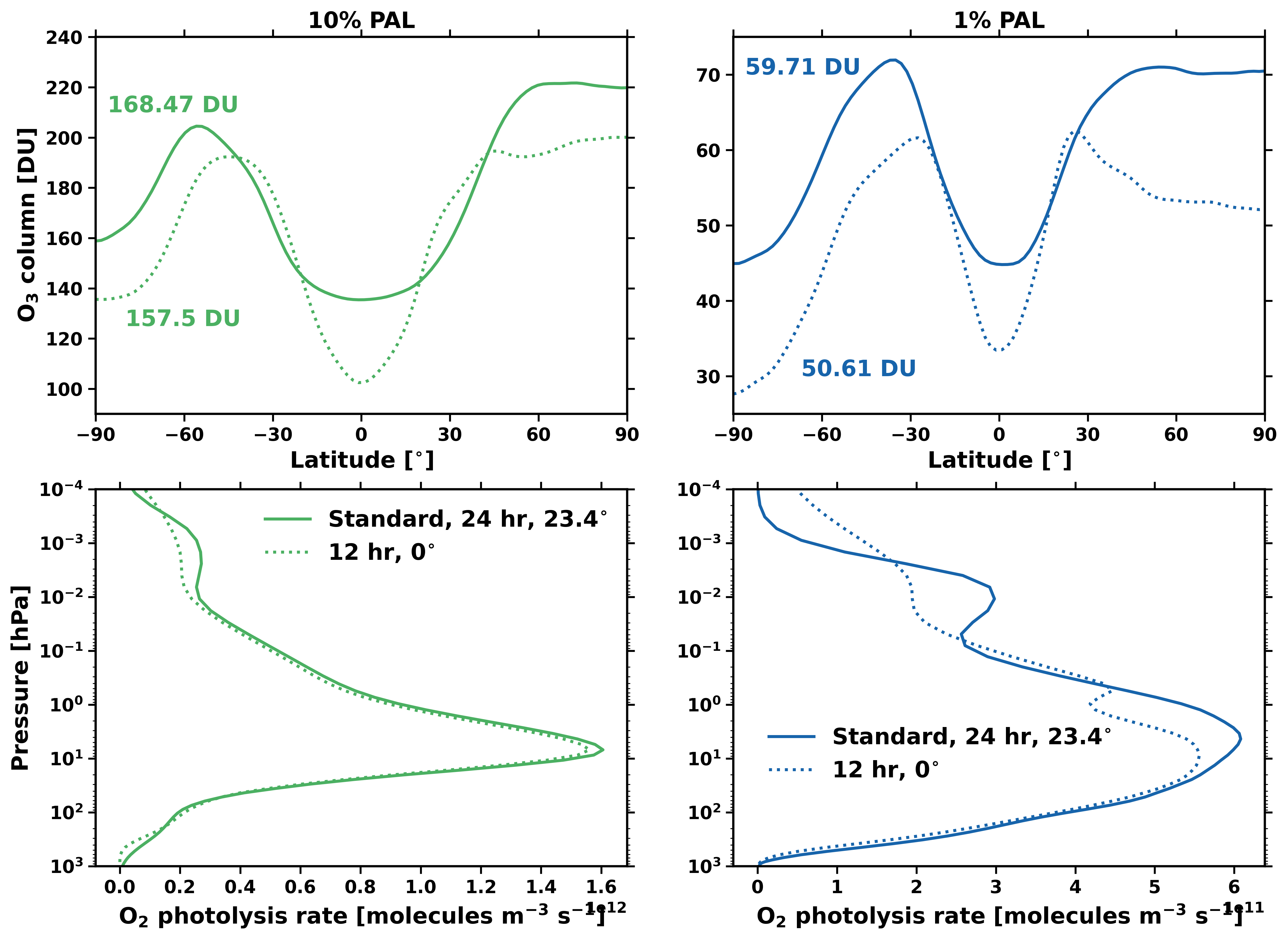}
    \caption{The top row shows how the ozone column varies with latitude for two 10\% PAL cases (left) and two 1\% PAL cases (right). The standard case is shown by the unbroken line, and the case with a 12 hour rotation rate and a $0^\circ$ obliquity is shown by the dotted line. The bottom row shows the global average \ce{O2} photolysis for the same simulations as the top panels.}
    \label{Transport experiment figure}
\end{figure}

Finally, we perform two perturbation experiments for the 10\% PAL and 1\% PAL simulations. Instead of a 24 hour rotation rate with a $23.4^\circ$ obliquity (standard modern Earth), we simulate the same cases with a 12 hour rotation rate and a $0^\circ$ obliquity. Fig.~\ref{Transport experiment figure} shows the two perturbation experiments compared to the standard 10\% PAL and 1\% PAL scenarios. The top row shows the \ce{O3} column with latitude, whilst the bottom row displays the \ce{O2} photolysis rate.

Whilst Earth likely had a faster rotation rate in the past \citep{Mitchell_Day_Length, 2014AGUFMPP53A1201B}, its axial tilt has likely been stable for several Gyr and may have hovered around 21--$25^\circ$ \citep{williams1993history}. Therefore, these experiments are purely for demonstrating the effect of 3D dynamics whilst holding the initial chemical conditions constant. Doubling the rotation rate increases the Coriolis force by a factor of 2, whilst reducing the obliquity to $0^\circ$ removes any significant seasons from the Earth simulations. In these cases, the extratropical peak of \ce{O3} sits closer to the equator in response to the perturbed atmospheric dynamics. Crucially, the total \ce{O3} column changes in both simulations, reducing from 168.85 DU to 157.5 DU in the 10\% PAL case (a 7\% reduction) and from 59.47 DU to 50.61 DU in the 1\% PAL case (a 15\% reduction). The bottom row shows the \ce{O2} photolysis rate against atmospheric pressure. Introducing the different dynamical setup affects the peak photolysis rate in the stratosphere by $< 3$\%. There are minor mixing ratio changes too for HO\textsubscript{x} and NO\textsubscript{x}, but these are much less significant than the HO\textsubscript{x} and NO\textsubscript{x} discrepancies between models or from changing \ce{O2} by an order of magnitude.

As a result of all our simulations, we have shown how atmospheric chemistry (the \ce{O2} concentration) affects atmospheric dynamics (Fig.~\ref{Zonal winds figure} -- \ref{TEM figure}), and how atmospheric dynamics affects chemistry (specifically \ce{O3}, Fig.~\ref{Transport experiment figure}), demonstrating that the response of \ce{O3} concentrations to a world with varying \ce{O2} concentrations is clearly affected by the 3D nature of the planet. Thus, alongside other effects, such as the propagation of water through the 3D tropopause, we have demonstrated that these effects do not exist in isolation.

\subsection{Additional Perturbation Experiments}

To investigate the effects of various parameters on predictions for atmospheric \ce{O3}, we conducted several sensitivity tests. These included varying surface albedo between 0.06 and 0.3, varying the amount of atmospheric methane between 0.1x and 10x of our pre-industrial baseline, and varying the solar spectrum to assess the influence of the young sun. Our results are presented in Appendix \ref{Perturbation experiments appendix} for brevity and summarised here. 

Albedo affects photochemistry since it dictates the fraction of solar radiation reflected back up through the atmosphere. While earth's surface albedo is highly varied (from oceans to land and snow), a key factor in its albedo is the reflectivity of water clouds. In 1D models, the effect of water clouds is captured by using slightly larger surface albedo than Earth's true value of 0.15 to effectively "paint" clouds on the surface \citep{2025PSJ.....6..256W}. We find that altering this parameter only results in minor changes to the overall \ce{O3} column, with column depths decreasing by roughly 2.9\%. In 3D chemistry-climate models (CCMs) on the other hand, it has been found that improvements in albedo treatments for ocean and snow-covered areas alter tropospheric OH and NO\textsubscript{x}, as well as photolysis rates of \ce{O3} \citep{laepple2005improved}. Online photolysis modules, such as the widely used Fast-J or Cloud-J schemes, compute photolysis rates dynamically by factoring in real-time surface reflectivity alongside cloud and aerosol scattering \citep{2023GMD....16.7433V, voulgarakis2009clouds}. Updating these albedo formulations can shift a model's simulated local tropospheric oxidizing capacity by $\pm20\%$, directly altering daily maximum surface \ce{O3} predictions and shifting regional \ce{O3} production bounds \citep{laepple2005improved, 2023GMD....16.7433V}. Furthermore, clouds themselves can change due to modulations in the \ce{O3} layer \citep{cooke2025oxygen}. We are unable to assess how clouds affect photolysis rates here since only \texttt{WACCM6} includes clouds. 

We also tested the effect of the young sun on atmospheric chemistry using simulations with varying stellar ages in \texttt{Photochem}. We also run coupled simulations in \texttt{Atmos}, where the climate and photochemical modules are iterated back and forth to generate self-consistent PT profiles that include the effects of both altered atmospheric \ce{O2} and incoming stellar flux. In our uncoupled simulations, we find that purely photochemical changes to the \ce{O3} layer are minor in both the high-O$_2$ (100$\%$ PAL) and low-O$_2$ (0.1$\%$ PAL) cases. When climate feedbacks are included in our coupled \texttt{Atmos} runs, the changes in column depth are minor for the 100\% PAL case, with the columns only decreasing from 264.8 to 263.7 DU (0.4\%). However, we find a significant increase in the \ce{O3} column in the coupled 0.1\% PAL atmosphere under a younger sun, with the column increasing from 44.3 DU to 93.5 DU. This is because the fainter young sun cools the troposphere, lowering \ce{H2O} and resulting in a reduced amount of HO\textsubscript{x} species, leading to a decrease in catalytic HO\textsubscript{x} driven \ce{O3} loss cycles. Our results suggest that climate-chemistry feedback loops are therefore important when assessing the atmospheric evolution of the early Earth, particularly for low-\ce{O2} states. This was also found to be true in \texttt{WACCM6} in \citet{cooke2022revised}.

The effect of changing \ce{CH4} concentrations was also explored in \texttt{Photochem}, \texttt{VULCAN}, and \texttt{Atmos}. In high \ce{O2} simulations (100\% PAL and 10\% PAL), \ce{O3} columns are flat or slightly increase with increased \ce{CH4}. In low \ce{O2} simulations on the other hand (1\% PAL  and 0.1\% PAL), the \ce{O3} column curve is flat or slightly decreases with increasing \ce{CH4}. Increased \ce{CH4} can remove HO\textsubscript{x} species in the atmosphere through reactions such as :
\begin{align}
\ce{CH4 +} h\nu \ce{&-> CH3 + H} \label{ch4_photolysis} \\
\ce{CH4 + OH &-> CH3 + H2O}
\end{align}

The thickness of the \ce{O3} column directly dictates \ce{CH4} photolysis rates. Consequently, varying the \ce{O2} levels leads to distinct alterations in the \ce{O3} columns. Furthermore, \ce{CH4} can alter NO\textsubscript{x} species in the atmosphere as well, through peroxy radicals produced through downstream reactions following the photolysis of \ce{CH4}. Reactions such as:

\begin{align}
\ce{HO2 + NO &-> OH + NO2} \\
\ce{CH3O2 + NO &-> CH3O + NO2},
\end{align}

modulate the NO\textsubscript{x} cycles as well. 

\subsection{Summary of results}

To summarise, we compared 5 computational atmospheric modelling tools (the 3D chemistry-climate model \texttt{WACCM6} and four 1D models - \texttt{Atmos}, \texttt{Photochem}, \texttt{VULCAN} and the \texttt{\texttt{Kasting}} model) across atmospheric \ce{O2} mixing ratios from 0.1\% to 150\% PAL. In the 1D models, we varied the SZA between $45^\circ$ and $60^\circ$, with the main reported results at a SZA of $48.2^\circ$. We find that the total \ce{O3} column generally increases with rising \ce{O2}, but several models show a modest decrease or plateau at 50-150\% PAL depending on SZA. At modern (100\% PAL) \ce{O2}, the models agree reasonably well and fall within the range of \texttt{WACCM6}, but they diverge at lower \ce{O2} levels from the maximum \texttt{WACCM6} columns. At low \ce{O2}, the \texttt{\texttt{Kasting}} model agrees the best with \texttt{WACCM6}, followed by \texttt{VULCAN}, \texttt{Photochem} and \texttt{Atmos}. These differences are a result of several factors, including: (1) model differences in O\textsubscript{x} production, \ce{O3} destruction, photolysis cross sections, reaction networks, and reaction rates; (2) 3D effects, including transport and the diurnal cycle (3) differing input stellar spectra; and (4) differing \ce{H2O} profile parametrisations. Additional sensitivity tests quantified secondary influences, with changing surface albedo, stellar spectra and methane concentrations all altering the \ce{O3} column to various extents. 

\section{Discussion}
\label{Discussion section}

We have presented results from several open source models on simulated atmospheres of the Earth with varying \ce{O2} mixing ratios, with the goal of determining how they differ in predictions of \ce{O3} densities and distributions throughout the atmosphere. In this section, we further discuss the reasons behind the discrepancies in calculated \ce{O3} columns between the models. Subsequently, we comment on possible implications for Earth's oxygenated history, including how lower simulated \ce{O3} columns could result in increased rates of damaging UV radiation reaching the surface. Finally, we show why our results are of critical importance for simulations of oxygenated terrestrial exoplanets orbiting different types of stars, and suggest future directions for the intertwined fields of paleo-Earth and habitable exoplanet research.

\subsection{3D effects for Earth's \ce{O3}}

Discrepancies between 1D and 3D models are naturally expected from several physical processes that a one-dimensional framework cannot represent. There is a vast literature on Earth's \ce{O3}, with several 3D effects already known to play significant roles in shaping its local and global concentration. As we have shown in section \ref{3D effects section}, these effects play a role in altering the atmospheric composition of the Earth. In this section, we present a non-exhaustive review of effects that are not captured by 1D models that may be relevant when simulating Earth's paleoatmosphere.  We summarise the 3D effects reviewed in this section in table~\ref{3D_effects_table}. We do not discuss local anthropogenic pollution, which plays an important role in \ce{O3} formation near the surface (smog), because it would not have been present on the early Earth. 

\begin{table}[b!]
\centering
\caption{Key 3D effects on the ozone layer are described here. Many of these effects are either not included or cannot be included in 1D photochemical models. References are grouped as follows: \textbf{(a)}: \citep{Brasseur_Solomon_2005}, \textbf{(b)}: \citep{Brasseur_Solomon_2005, monks2005gas, brown2012nighttime}, \textbf{(c)}:\citep{Proedrou_2016_3D_O3, Proedrou2016_SST, Chen_2018_Biosignature,Chen2019_Hab, Chen2021_Flare, Ridgway2023_Flares, 2023MNRAS.526..263B, Cooke2023Stellar, Cooke_2024_Lethal}, \textbf{(d)}: \citep{2014_Butchart, Graham2019_Snowball}, \textbf{(e)}: \citep{Schoeberl_Polar_O3_1991, Zhang_2018_ozone_vortex}, \textbf{(f)}: \citep{Chen2021_Flare, Ridgway2023_Flares}, \textbf{(g)}: \citep{Lelieveld1991_Clouds, Tie_2003_Clouds}, \textbf{(h)}: \citep{Solomon_O3_1986, Tritscher_2021_PSCs, Voigt2000_NAT}, \textbf{(i)}: \citep{cooke2025oxygen}, \textbf{(j)}: \citep{Sainsbury-Martinez_2025_Comets1, 2026ApJ..1001...57S}.}
\label{3D_effects_table}
\begin{tabularx}{\textwidth}{@{}
    >{\hsize=0.75\hsize}X 
    >{\hsize=1.85\hsize}X 
    >{\hsize=0.4\hsize}X 
@{}}
\toprule
\textbf{3D Effect} & \textbf{Importance for \ce{O3} Layer} & \textbf{References} \\ \midrule
Temperature structure & Controls temperature-dependent kinetic reaction rates as well as affecting photochemistry. 3D spatial and temporal gradients create localized variations in \ce{O3} production and destruction. & a \\
Diurnal Cycles & Changes heating rates between day and night. Results in dayside photolysis and therefore better captures the limited lifetimes of radical species like OH, ClO, and O. & b \\ \addlinespace
Horizontal, meridional, \& vertical transport & Redistributes species globally. Accounts for zonal wind and mixing which changes with seasons. Vertical transport affected by gravity waves which occur via topography and convective updrafts. & c \\ \addlinespace
Brewer-Dobson Circulation & Wave-driven poleward and subsequently downward flow. This distributes \ce{O3} towards higher latitudes. & d \\ \addlinespace
Polar Vortex & Acts as a physical barrier, isolating polar air and allowing for \ce{O3} depletion. Varies between hemispheres. & e \\ \addlinespace
Stellar flares & High-energy particles hit the planet's dayside, which drives asymmetric NO\textsubscript{x} production. & f \\ \addlinespace
Clouds & Affects stratospheric and tropospheric photolysis via scattering/albedo. & g \\ \addlinespace
Polar Stratospheric Clouds & Provides ice surfaces for heterogeneous chemistry and also temporal changes in the \ce{O3} layer from seasonal polar photolysis. & h \\ \addlinespace
Water vapour cold trap & Water vapour (\ce{H2O}) enters the stratosphere through the cold trap. The cold trap is not uniform and varies spatially and temporally depending upon the planetary regime. This affects stratospheric HO\textsubscript{x} and thus \ce{O3}. &  i \\ 
Impacts & Cometary impacts deliver large amounts of ice and heat upon ablation driving \ce{O3} changes. &  j\\ 

\bottomrule
\end{tabularx}
\end{table}

A primary driver of discrepancies between 1D and 3D models is the fact that 1D frameworks cannot inherently capture the spatial heterogeneity and horizontal transport on Earth. In a three-dimensional planetary environment, the distribution of atmospheric species like \ce{O3} is strongly affected by large scale atmospheric circulation. \ce{O3} is primarily made in the tropics where the solar irradiation is strongest. However, the Brewer-Dobson circulation redistributes \ce{O3} towards the poles \citep{2014_Butchart}. This circulation, driven primarily by Rossby waves and gravity waves \citep{2015EGUGA..17.4721G, 2020EGUGA..22.1418S}, creates significant seasonal and latitudinal variations in the \ce{O3} column depth. The northern hemisphere generally has more \ce{O3} than the southern hemisphere due to asymmetries in the land and ocean distribution, which causes a stronger and more persistent southern hemisphere stratospheric polar vortex \citep{waugh2010stratospheric, Waugh2017Vortex, 2021RvGeo..5900723M}, acting to isolate Antarctic air for longer than Arctic air. 

While 3D chemistry-climate models are able to capture the aforementioned dynamical features, 1D models are forced to parametrize transport purely in the vertical direction through molecular and eddy diffusion.  Consequently, the global mean state predictions are biased, especially for unobserved or paleo-climate states where dynamics could significantly diverge from the modern Earth.

The same circulation that redistributes \ce{O3} also governs the distribution of the species that destroy it.  Water vapour (\ce{H2O}), which impacts \ce{O3} destruction through odd hydrogen molecules (HO\textsubscript{x}) varies greatly in the troposphere. In the stratosphere, the \ce{H2O} mixing ratio is set by the temperatures in the tropical tropopause layer, as Earth's circulation directs its upwards trajectory through this region \citep{2015ApJ...813L...3K, Brasseur_Solomon_2005, Fueglistaler2009}. \citet{cooke2025oxygen} showed that for different oxygenation states (0.1\% PAL -- 150\% PAL) and various temperature profiles, a 1D representation can incorrectly predict the stratospheric \ce{H2O} mixing ratio.

Another critical 3D effect omitted by 1D models of the Earth's past is the evolution of Earth's rotation rate. Due to tidal dissipation caused by gravitational interactions with the Moon, Earth's rotation has gradually slowed down over time \citep{2024absc.conf99672C, 2014AGUFMPP53A1201B}. Planetary rotation rate is a strong driver of 3D atmospheric dynamics because it dictates the magnitude of the Coriolis force ($f$):

\begin{equation}
    f=2\Omega \sin \theta,
\end{equation}

\noindent where $\Omega$ is the angular frequency of the planetary rotation and $\theta$ is the latitude. In 3D models, a faster rotation rate strengthens the Coriolis effect, which dynamically restricts the meridional expansion of the Hadley cells \citep{1977JAtS...34..280S, 1980JAtS...37..515H}. Consequently, on a rapidly rotating early Earth, atmospheric circulation would have reorganized into narrower circulation cells, altering global climate and the distribution of atmospheric chemical species \citep{hunt1979effects}. Narrower Hadley cells are less efficient at transporting heat from the tropics to the poles, resulting in steeper temperature gradients, altered planetary albedo and distinct cloud distributions \citep{2017JCli...30.7465L, 2015ApJ...804...60K}. This in turn alters the Brewer-Dobson circulation, further changing the latitudinal distribution of \ce{O3}. 1D models cannot account for these rotation-driven changes in meridional transport and heat distributions, thereby biasing results when simulating the exact same planet with a faster or slower rotation rate. 1D models can in principle alter the photochemical diurnal period if the day length changes (although the four models assessed here do not have that option), but they cannot self-consistently simulate the meridional overturning or its seasonal absence. These limitations are especially acute for exoplanets where the rotation rate and obliquity cannot be assumed to be Earth-like.

Beyond large scale transport, 1D models struggle to capture localised temperatures and dynamics that are necessary for heterogeneous chemistry on Earth, like the formation and sustenance of polar stratospheric clouds (PSCs). In the modern atmosphere, PSCs form during the polar winter when stratospheric temperatures drop below critical thresholds \citep{2021EGUGA..23.6643T}. These clouds are composed of liquid supercooled ternary solution (STS; \ce{H2SO4-HNO3-H2O})  droplets, solid nitric acid trihydrate (NAT) particles, and water ice. PSC particles provide sites for heterogeneous reactions that convert stable, inert halogen reservoir species (e.g., HCl, \ce{ClONO2}) into active radicals that destroy \ce{O3} catalytically. Indeed, PSCs played a critical role in the formation of the Antarctic \ce{O3} hole \citep{Solomon_O3_1986}. 

Furthermore, 3D models incorporate realistic planetary geometry by computing specific SZAs for individual atmospheric grid cells based on the system's orbital parameters. This allows these models to naturally simulate the diurnal cycle and predict the resulting temporal and spatial photochemical variations. This is especially true for species which wholly depend on photochemistry to be present, such as OH and \ce{O(^1D)} \citep{2006Natur.442..184R, 1998acpa.conf.....S}. By contrast, 1D models tend to rely on a single, fixed SZA to approximate the planetary average or diurnal response. As demonstrated in our analyses, altering the zenith angle with standard boundaries changes the resulting \ce{O3} column. To achieve the average \ce{O3} column, a 1D model may require tuning of the fixed zenith angle. The predictive accuracy of these fine tuned zenith angle choices when applied to vastly different atmospheric states is currently unknown. Efforts to bypass the limitations of fixed SZAs were made using the 8-point Gaussian approach in the \texttt{\texttt{Kasting}} model \citep{2023RSOS...1030056J}, and similar methods need to be investigated in the future. 

The atmospheric response to stellar flares and coronal mass ejections is also three dimensional, which makes it difficult for 1D models to capture accurately. During these events, high energy extreme ultraviolet (EUV) and X-ray radiation irradiate the dayside of the planet, while energetic charged particles are funnelled by Earth's magnetic field into the high-latitude polar regions \citep{2008ACP.....8..765J, 2013ERL.....8d5010C}. In addition to the enhanced UV flux, these energetic particles significantly alter upper atmospheric chemistry by increasing the abundances of NO\textsubscript{x} and HO\textsubscript{x} species, which in turn alter the concentration of atmospheric \ce{O3} \citep{2005JGRA..110.9S43L, 2006GeoRL..3324811V}. 1D models typically simulate flare-induced particle precipitation by spreading it uniformly across the entire globe or hemisphere, and consequently predict strong and prolonged \ce{O3} depletion \citep{2019AsBio..19...64T, 2010AsBio..10..751S}. However, key 3D processes that occur over long timescales demonstrate the highly localised nature of the particle induced depletion, the subsequent downward advection of NO\textsubscript{x} into the middle stratosphere, or the gradual dilution and transport of these species. Furthermore, the absence of horizontal transport in 1D models misrepresents the balance between \ce{O3} destruction and production during the electromagnetic phase of a flare. Intense UV radiation enhances the photolysis of \ce{O2} on the irradiated dayside, generating atomic oxygen (O), which rapidly recombines to form \ce{O3} \citep{Ridgway2023_Flares}. In 1D models, this newly produced \ce{O3} remains confined to the dayside and is subsequently destroyed by the ongoing flare, resulting in net depletion. By contrast, 3D models show that atmospheric circulation can redistribute a significant portion of this newly produced \ce{O3} to the nightside, which can result in a net global \ce{O3} increase rather than depletion \citep{Ridgway2023_Flares}. Additionally, different types of FGK planets have various responses to flares, and \ce{O3} can also be depleted in 3D calculations \citep{Chen2021_Flare}.

Finally, while 1D models can approximate the basic photochemistry of impact delivered volatiles, they do not simulate the highly localised nature of asteroid or cometary impacts and the subsequent evolution of the atmosphere. 3D models are essential for resolving the dynamic horizontal mixing and transport of impact generated species. For example, shock heating during atmospheric entry or re-entry of ejecta produces nitrogen oxides (NO\textsubscript{x}) \citep{1997RvGeo..35...41T}, while oceanic impacts injects massive plumes of water vapour together with \ce{O3}-destroying halogens from seawater directly into the stratosphere \citep{2010E&PSL.299..263P}. 1D models also cannot accurately represent how zonal winds homogenize impact-delivered \ce{H2O} - a process that accelerates its UV photolysis and drives the subsequent production of ozone-destroying radicals (HO\textsubscript{x}) \citep{Sainsbury-Martinez_2025_Comets1}.

\subsection{Implications for early Earth}
\label{Implications for early Earth section}

Several works have explored the effect of the Earth's \ce{O3} layer on its habitability, primarily by its effects on the surface UV environment. In this section, we quantify how our results could alter our understanding of surface habitability using the \texttt{TUV} model, before considering the implications of altered \ce{O3} columns on palaeobiology and photogeochemistry of the early Earth. 

If 1D models incorrectly predict \ce{O3} columns, they may provide an overly optimistic assessment of Earth's UV shielding when considering a large portion of Earth's history from Fig.~\ref{Oxygenation history figure}. For example, a thinner early \ce{O3} layer would imply that a higher flux of biologically damaging UV radiation could have penetrated much deeper into the atmosphere and upper ocean, presenting an environmental stressor for life. While early hypotheses suggested that intense UV radiation could have prevented life from becoming terrestrial \citep{1965JAtS...22..225B}, this view has been disputed \citep{cockell2001history, cockell2011exposure, rambler1980bacterial}. Nevertheless, a consensus remains that the surface UV environment helped to shape the evolution of life on and close to the surface \citep{zedek2018holocentric, margulis1976reassessment, zedek2022holocentric}. A harsher surface radiation environment could have restricted the expansion of cyanobacteria and early complex multicellular life to deeper, shielded aquatic refugia, necessitating the rapid evolutionary development of heavy UV-screening compounds \citep{cockell1998biological, 2000OLEB...30..467C, mloszewska2018uv}. 

\begin{figure}[b!]
    \centering
    \includegraphics[width = 1\textwidth]{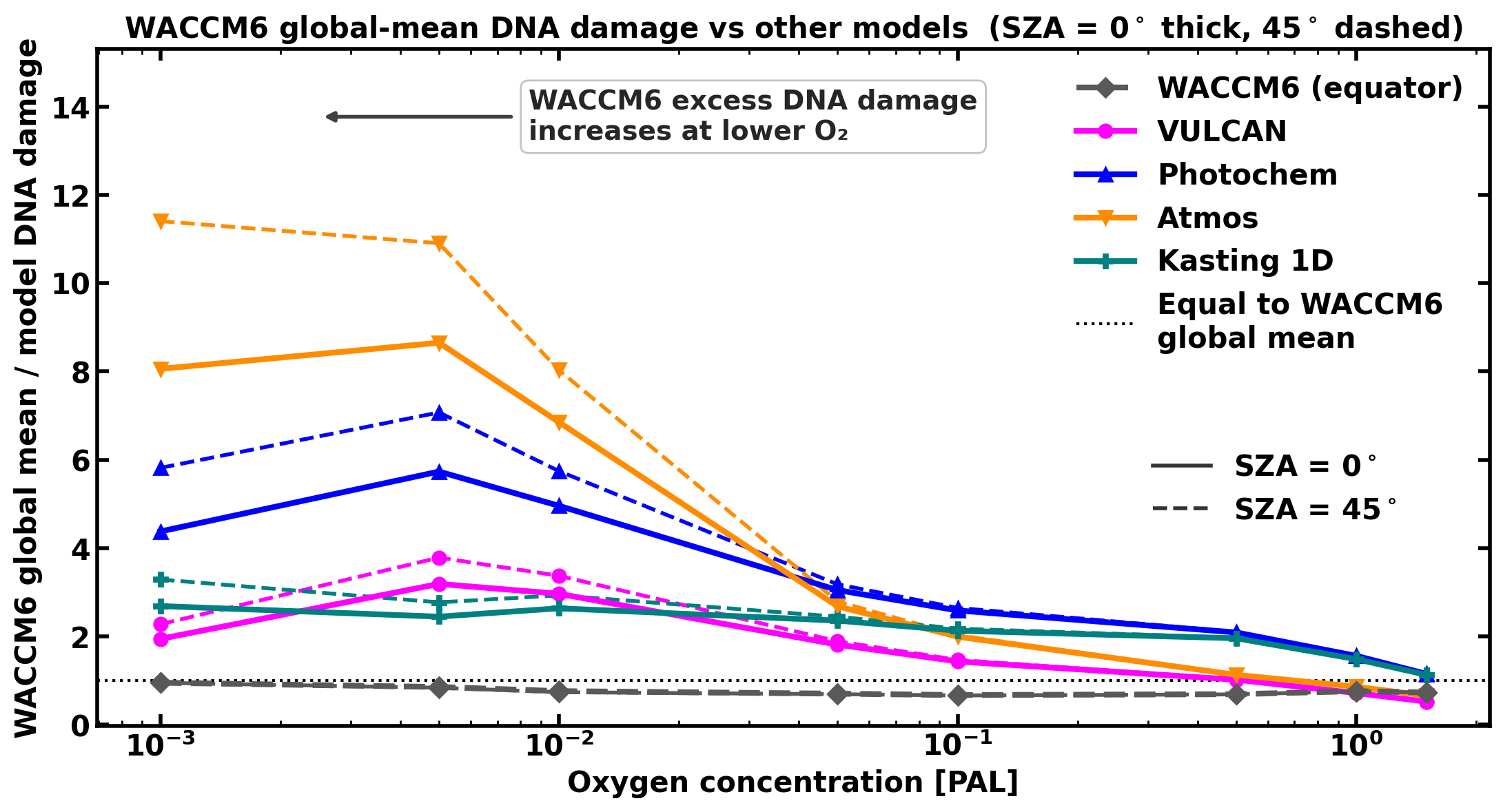}
    \caption{The \texttt{TUV} model was used to calculate the DNA damage at the surface for each model and for the simulations of the 0.1\%, 1\%, 10\% and 100\% PAL cases. Inputs included the pressure-temperature profile and the \ce{O3} profile and total \ce{O3} column. The figure shows the ratio between each simulation and the corresponding simulation in \texttt{WACCM6}, accounting for the global mean \texttt{WACCM6}  atmosphere. \texttt{VULCAN} is shown in magenta, \texttt{Photochem} in blue, \texttt{Atmos} in orange, and the \texttt{\texttt{Kasting}} model in teal. Additionally, we show the results in grey from profiles taken at the equator for \texttt{WACCM6}. The thick line shows the results for an overhead Sun with a solar zenith angle of $0^\circ$, whilst the dashed lines show a solar zenith angle of $45^\circ$. The black arrow shows how 1D models show decreasing rates of relative UV damage when compared with the WACCM6 model.}
    \label{DNA damage ratio figure}
\end{figure}

To quantify the effects of varied \ce{O3} column predictions on predictions for the surface habitability of the early Earth, we use the \texttt{TUV} model described in section 2 \ref{TUV section}. We input the atmospheric profiles from \texttt{WACCM6}  and the 1D models to compare the UV radiation hitting the surface in each case, and the biological UV damage that would occur as a result of an overhead Sun (a solar zenith angle of $0^\circ$) and a morning / afternoon Sun (a solar zenith angle of $45^\circ$). The results of DNA damage ratios are shown in Fig.~\ref{DNA damage ratio figure}. At 10\% PAL, where several of the 1D models start to produce higher \ce{O3} columns than \texttt{WACCM6}, the surface UV damage at $0^\circ$ SZA ranged from 1.5 -- 2.7 times lower than the surface UV damage under the global mean atmosphere predicted by \texttt{WACCM6}. This range is 2.7 -- 7.2 at 1\% PAL and  at 0.1\% PAL, the range is 1.8 -- 8.1. One should keep in mind that the equator receives the largest amount of radiation throughout the year. If we compare the \texttt{WACCM6} equatorial \ce{O3} column to that of the 1D models, then the DNA damage can actually increase again by up to a factor of 1.1 -- 1.5, with this effect most pronounced at 10\% PAL because of the range in \ce{O3} columns between the equator and extratropics. A $45^\circ$ SZA increases the relative UV damage because the radiation now has a longer path length to travel through before hitting the surface, but the absolute damage will be reduced when compared to $0^\circ$ SZA. 

It is pertinent to remind the reader that all the simulations performed here are not truly faithful to recreating all Proterozoic or Phanerozoic conditions, including their redox states. Indeed, the Proterozoic was the longest eon in Earths history, and both eons saw dramatic changes in a variety of environmental levers. Although the WACCM6 simulations here may not be fully representative of the Earth's climatic history, acknowledging modelling discrepancies between 1D and 3D models provides additional context for evaluating the role of UV radiation in past climatic and biological events, as well as the long-term evolution of life on Earth. During the Archean, harmful UV wavelengths may have been attenuated by iron-silica precipitates, yet the early expansion of cyanobacteria was potentially still limited by UV \citep{mloszewska2018uv}. It has also been suggested that rather than evolving strictly for photoprotection to overcome these constraints, ancient cyanobacterial pigments like scytonemins and mycosporine-like amino acids may have been selected to optimize thermodynamic photon dissipation instead \citep{simeonov2025cyanobacterial}. As life transitioned into the Proterozoic, a rapid increase in atmospheric \ce{O2} initiated a critical atmospheric transition. Even modest \ce{O3} columns of approximately 66~DU provided a dramatic, non-linear shielding effect by completely cutting off lethal UV-C radiation and multiplying organism survival rates forty-fold, though further increases beyond this thickness yielded smaller biological returns \citep{cockell2001history}. However, the \texttt{WACCM6}  calculations, if accurate, suggest that this protective \ce{O3} screen may have required a significantly higher $p\ce{O2}$—by a factor of 10 to 20 times estimates from previous 1-D model results \citep{cooke2022revised}. Later in Earth's history, during the Phanerozoic, disruptions to this established shield might have been the trigger for dramatic biotic crises. For instance, mass extinctions like the end-Devonian or end-Permian are hypothesized to involve severe \ce{O3} depletion and correspondingly high surface UV fluxes \citep{benca2018uv, marshall2020uv, 2017PPP...478....3B}. While an increased UV flux directly accelerates DNA damage and baseline mutation rates, long-term evolutionary outcomes suggest that organisms are able to adjust their error-prone repair mechanisms and metabolic investments to survive elevated mutagenic pressure \citep{bromham2024mutation}. Nonetheless, on short ecological timescales, sudden UV surges can still affect vulnerable populations as deleterious mutations accumulate faster than they can be purged, though this risk can be somewhat mitigated by rapid shifts in protective gene expression \citep{koski2022elevational}.

Beyond biology, an environment with higher surface UV fluxes could have implications for paleogeology, since it may alter our understanding of the photogeochemistry of the early Earth. The mass independent fractionation of oxygen isotopes (O-MIF, \ce{\Delta^{17}O}) for example originates in photochemical reactions involving \ce{O2}, \ce{O3} and \ce{CO2} in the upper atmosphere, and is subsequently transferred to minerals \citep{crockford2018triple}. The O-MIF signal has been interpreted to infer constraints on \ce{pO2} \citep{planavsky2020large}, but accurate interpretation may require robust photochemical modelling \citep{liu2021triple, gregory2021photochemical}. Similarly, the sulphur mass-independent fractionation (S-MIF) preserved in minerals is produced by UV photolysis of \ce{SO2} in a low-\ce{O2} atmosphere \citep{farquhar2000atmospheric, farquhar2001observation}. If the early Earth possessed a thinner \ce{O3} layer at a different transitional \ce{O2} level than previously calculated by 1D models, the thresholds and atmospheric constraints historically derived from them might need to be reassessed \citep{zahnle2006loss, pavlov2002mass}. Another implication for paleogeology relates to the origin of banded iron formations (BIFs). Abiotic photooxidation of dissolved \ce{Fe (II)} has been proposed as an alternative route to BIF formation \citep{braterman1983photo, franccois1986extensive, cairns1978precambrian} alongside oxygenic photosynthesis \citep{cloud1973paleoecological, yin2023origin, planavsky2014evidence} and anoxygenic photoferrotrophy \citep{widdel1993ferrous, kappler2005deposition, konhauser2002could}. This abiotic photo-oxidation relies on UV absorption by dissolved \ce{Fe (II)}, particularly \ce{FeOH+}. While BIFs in the Proterozoic show evidence of a mechanistic shift in comparison to Archean BIFs, the relative importance of photooxidation in the production of BIFs at different eras in Earth's past is still debated \citep{anbar1992photochemistry, konhauser2007decoupling, nie2017iron, tabata2021experimental, vergeli2025reassessing}. Assessing the viability and rate of these UV-driven pathways requires the estimation of the surface UV flux, which is sometimes adopted as a boundary condition from studies using 1D photochemical models (e.g, \citet{konhauser2007decoupling} used surface UV estimates from \citet{segura2003ozone}). To the extent that 1D and 3D models differ in their predictions of the \ce{O3} column, and in turn the surface UV environment, quantitative conclusions of downstream studies that rely on them may inherit the same biases. 
In the absence of a fully developed \ce{O3} layer, several abiotic mineral reactions could have also been driven by UV. For example, the photooxidation of siderite (\ce{FeCO3}) to ferric oxides and \ce{H2} in the presence of UV could have provided a key reductant for microbial metabolisms \citep{kim2013anoxic}. Another example is the photooxidation of rhodochrosite (\ce{MnCO3}) to manganite ($\gamma$-MnOOH) and \ce{H2} \citep{liu2020anoxic}. Manganite could efficiently oxidise Cr (III) to Cr (VI) under laboratory conditions \citep{johnson1991oxidation}. To our knowledge, whether a UV-driven Mn-oxide source ever operated at rates sufficient to leave a measurable Cr-isotope signal on the early Earth is an open question, but it is one worth noting alongside the use of Cr isotopes as an oxygenation proxy across the GOE \citep{frei2009fluctuations}. Consequently, a deeper understanding of the surface UV environment will aid in updating our understanding of the biogeochemical evolution of the early Earth. 

\subsection{Implications for exoplanets and the search for life elsewhere}
\label{Exoplanets discussion section}

Beyond Earth, our results have implications for the theoretical modelling of oxygenated exoplanet atmospheres, the use of \ce{O2} and \ce{O3} as  biosignatures, and the design of upcoming missions like the Habitable Worlds Observatory (HWO) \citep{Arney_HWO_2026} and the Large Interferometer For Exoplanets (LIFE) \citep{glauser2024large}.  In this section, we discuss the physical limitations of using 1D models to simulate exoplanet atmospheres. We demonstrate this by simulating the atmosphere of Proxima Centauri b and generate transmission spectra using \texttt{PSG}. We then illustrate why differing predictions for \ce{O3} complicate biosignature detection.

The prospect of finding signs of life on temperate Earth-like exoplanets with next generation telescopes has driven a growing interest in 3D chemistry-climate modelling of such planets, with an expanding body of recent work stressing its importance for these worlds. While 1D models have several advantages over 3D models, including their speed, flexibility, and the total number of chemical species and reactions that can be practically included, they have several disadvantages when simulating exoplanet atmospheres. These include the lack of a terminator and nightside region, 3D transport, and cloud processes, all of which can be crucial when predicting exoplanet spectra. Furthermore, the limitations of the fixed-zenith angle approximation in 1D models become particularly acute for exoplanets when moving beyond the 'fast-rotator' regime of the modern Earth. On Earth, the relatively rapid 24-hour rotation period allows for a quasi-global homogenization of chemical species whose lifetimes exceed the diurnal cycle. However, for slowly rotating or tidally locked exoplanets, the extreme temporal and spatial separation between the permanent day-side (a continuous source of photochemical radicals like HO\textsubscript{x}, NO\textsubscript{x}, and \ce{ClOx}) and the permanent night side (where species may be sequestered into long-lived reservoirs) creates gradients that a 1D global average cannot resolve. Modelling results depend upon the assumed rotation rate, whether the planet is tidally locked \citep{Chen_2018_Biosignature, Chen2019_Hab, Cooke2023Stellar, Cooke_2024_Lethal} or in a spin-orbit resonance (e.g., 3:2) \citep{2023MNRAS.526..263B}, and the incoming stellar energy distribution (SED). The question is, can 1D models successfully be applied to unobserved climate regimes that are outside of the solar system in order to extract useful information, or are the 3D effects too important? The answer is case-dependent and depends on which observable is in question. For \ce{O3}, at least, the 3D effects are extremely important, as this work and prior works have shown \citep{Chen_2018_Biosignature, Chen2019_Hab,  Chen2021_Flare, cooke2022revised, Cooke2023Stellar, Cooke_2024_Lethal, Jaziri_2022_GOE, 2023MNRAS.526..263B, Ridgway2023_Flares}. 

\begin{figure}[t!]
    \centering
    \includegraphics[width = 1\textwidth]{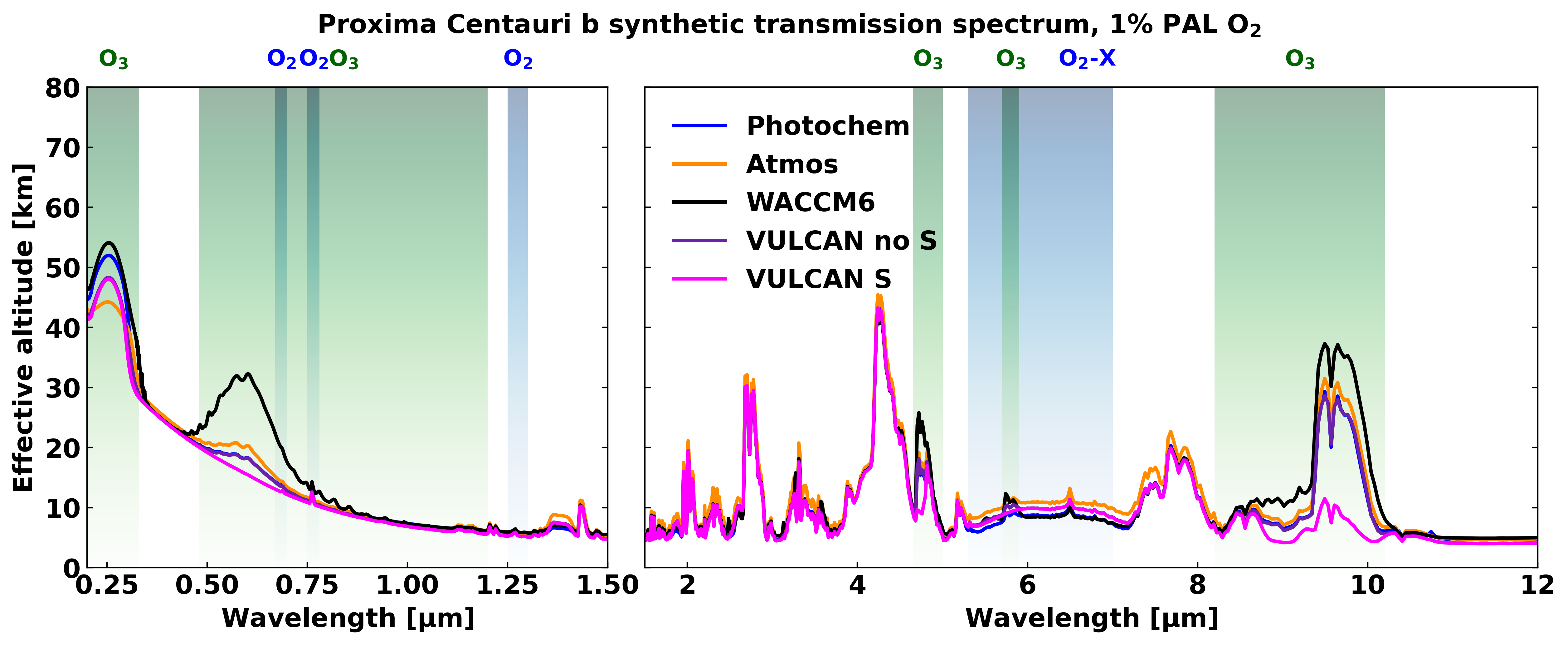}
    \caption{The synthetic transmission spectrum produced by \texttt{PSG} is shown between 0.2 -- 1.5 \textmu m (left panel) and 1.5 -- 12 \textmu m (right panel) for the 1\% PAL simulations of Proxima Centauri b for four models: \texttt{WACCM6}  (black), \texttt{Atmos} (dark orange), \texttt{Photochem} (blue), \texttt{VULCAN} without sulphur (purple) and \texttt{VULCAN} with sulphur (magenta). \ce{O3} features are highlighted in green and \ce{O2} features are highlighted in blue.}
    \label{Spectrum 2}
\end{figure}

To demonstrate that these 1D limitations matter in practice and that model choices affect observables, we extend our Earth setup to tidally locked exoplanet Proxima Centauri b. We use the same lower boundary conditions, P-T profile, and incoming SED from the 3D \texttt{WACCM6} simulations of Proxima Centauri b detailed in \citet{Cooke_2024_Lethal} and simulate its atmosphere with \texttt{Atmos}, \texttt{Photochem} and \texttt{VULCAN}. We use \texttt{PSG} to provide transmission spectra from the four atmospheric models over the 0.2 -- 12 \textmu m wavelength range, accounting for the range of different space-based telescopes. While these are not mission specific simulations, since instrument parameters of future missions are yet to be finalised, our simulated wavelength range encompasses the wavelength capabilities of the James Webb Space Telescope (JWST), HWO, and LIFE observatories. We show results for the 1\% PAL atmosphere in Fig.~\ref{Spectrum 2}. \texttt{VULCAN} simulations with and without sulphur are shown.

Between 0.2 -- 12 \textmu m, there are four key \ce{O3} features peaking at 0.25 \textmu m, 0.60 \textmu m, 4.7 \textmu m, and 9.65 \textmu m. The majority of the wavelengths have similar absorption depths, with the main deviations occurring for the aforementioned \ce{O3} features. For the 0.25 \textmu m UV feature, \texttt{WACCM6} has the largest absorption, followed by \texttt{Photochem}, both \texttt{VULCAN} simulations, and then \texttt{Atmos}. This differs to the features at 0.6 \textmu m and 9.65 \textmu m, where \texttt{WACCM6} has the largest absorption, followed by \texttt{Atmos}, \texttt{Photochem}, and \texttt{VULCAN} without sulphur and then with sulphur. While the 0.6 \textmu m feature shows a clear deviation between 1D and 3D models, the 9.65 \textmu m feature is closer in absorption strength for all models aside from \texttt{VULCAN} with sulphur. Similar statements can be made for the 100\% PAL, 10\% PAL, and 0.1\% PAL spectra that we have not shown for brevity. We summarise these results in Table~\ref{O3 feature table}.

In \texttt{WACCM6}, the \ce{O3} column counter-intuitively increases between 10\% PAL to 0.1\% PAL. \texttt{VULCAN} and \texttt{Atmos} do see a similar increase from 1\% PAL to 0.1\% PAL but do not replicate the large columns found in \texttt{WACCM6} at low \ce{O2}. The \ce{O3} accumulation is due to several interlinking factors. The pressure where \ce{O2} photolysis peaks is different with M dwarf stellar irradiation, and high levels of \ce{O2} prevent deep UV penetration. When the UV penetrates towards the lower stratosphere at reduced \ce{O2} mixing ratios, more \ce{O3} is produced and the atmospheric transport on a tidally locked world shuttles \ce{O3} around the planet to the nightside, where there is reduced catalytic cycle destruction and no photolysis. Thus the lifetime of \ce{O3} increases and can build up to high levels. Further discussion can be found in \citet{Cooke_2024_Lethal} and \citet{Cooke2023Stellar}.
 
It is not just the total \ce{O3} column which is important for determining the size of the absorption feature, but also where the \ce{O3} is distributed \citep{Cooke2023Stellar}. For instance, the 9.65 \textmu m feature in the 0.1\% PAL case is the same height for \texttt{WACCM6} and \texttt{Atmos}, despite \texttt{WACCM6} having an \ce{O3} column much greater than \texttt{Atmos} which can be observed when looking at the weaker feature around 9 \textmu m. The UV feature is present in all simulations but the visible feature is essentially non-existent for the \texttt{VULCAN} atmospheres, which sometimes have a weak mid-infrared signal.

\begin{table}[b!]
    \centering
    \caption{Summary of simulation results for Proxima Centauri b across varying \ce{O2} levels, assuming an Earth-like exoplanet, with \texttt{WACCM6}, \texttt{Atmos}, \texttt{Photochem}, and \texttt{VULCAN} with and without sulfur. Wavelength features ($\lambda$) and O$_3$ column depths are reported for each simulation percentage.}
    \label{O3 feature table}
    \begin{tabular}{l c ccc c}
        \toprule
        \multirow{2}{*}{\textbf{Model}} & \textbf{Simulation} & \multicolumn{3}{c}{\textbf{Wavelength feature height (km)}} & \textbf{O$_3$ Column Depth} \\
        \cmidrule(lr){3-5}
        & \textbf{Level} & 0.25 \textmu m & 0.60 \textmu m & 9.65 \textmu m & [DU] \\
        \midrule
        \multirow{4}{*}{\texttt{WACCM6}} 
        & 100\% & 67.7 & 28.2 & 36.4 & 229.2 \\
        & 10\%  & 59.3 & 26.2 & 35.7 & 156.1 \\
        & 1\%   & 54.0 & 32.2 & 37.1 & 310.0 \\
        & 0.1\% & 49.6 & 27.2 & 31.9 & 951.6 \\
        \addlinespace
        \multirow{4}{*}{\texttt{Atmos}} 
        & 100\% & 64.6 & 21.2 & 33.1 & 72.4 \\
        & 10\%  & 52.2 & 21.8 & 33.0 & 77.1 \\
        & 1\%   & 44.2 & 20.3 & 30.8 & 66.8 \\
        & 0.1\% & 42.6 & 20.9 & 31.4 & 71.0 \\
        \addlinespace
        \multirow{4}{*}{\texttt{Photochem}} 
        & 100\% & 66.6 & 27.5 & 36.5 & 244.8 \\
        & 10\%  & 59.7 & 21.7 & 32.9 & 110.8 \\
        & 1\%   & 51.9 & 18.3 & 28.6 & 50.7 \\
        & 0.1\% & 44.9 & 17.0 & 25.0 & 31.0 \\
        \addlinespace
        \multirow{4}{*}{\texttt{VULCAN no S}} 
        & 100\% & 61.4 & 28.6 & 36.4 & 239.6 \\
        & 10\%  & 54.4 & 23.0 & 33.0 & 118.5 \\
        & 1\%   & 48.3 & 18.3 & 28.0 & 47.8 \\
        & 0.1\% & 40.0 & 16.5 & 23.1 & 19.4 \\
        \addlinespace
        \multirow{4}{*}{\texttt{VULCAN S}} 
        & 100\% & 61.3 & 16.7 & 24.2 & 16.6 \\
        & 10\%  & 54.3 & 16.0 & 12.6 & 4.5 \\
        & 1\%   & 48.0 & 15.5 & 10.1 & 1.8 \\
        & 0.1\% & 39.8 & 16.0 & 16.7 & 5.8 \\
        \bottomrule
    \end{tabular}
\end{table}

The models achieve different atmospheric compositions for several reasons. For a fast rotator like modern Earth, the O density in a 1D model is close to the time-averaged O density in the 3D model \texttt{WACCM6}, diverging by a factor of $\sim1-10$. When simulating Proxima Centauri b, the globally averaged \texttt{WACCM6}  O density is within a factor of 2 of the dayside O density, with the nightside density a factor of $\sim100-1000$ times lower than the dayside density. If we take WACCM6 as a reference, then the 1D models approximate neither the nightside nor dayside density of O for Proxima Centauri b. The presence of O is essential for \ce{O3} production and loss, with \ce{O3} lifetimes on the nightside substantially extended for tidally locked exoplanets. Clearly, these modelling differences create significant degeneracies in observational predictions. For telescopes which do not have access to the entire 0.2 -- 11 \textmu m range (all three of the main \ce{O3} features), such degeneracies may be hard to resolve unless agreement can be found between future updated models.

For oxygenated atmospheres, we argue that 1D photochemical models provide varied predictions for \ce{O3} concentrations depending on which model is used and what input parameters are selected. For example, in our 100\% PAL Proxima Centauri b simulation in \texttt{Photochem}, omitting the \ce{O3} deposition velocity boundary condition (0.1 cms$^{-1}$) led to an increase in the \ce{O3} column from 244.8 DU to 8352.4 DU. As the deposition velocity of \ce{O3} is unknown for exoplanets, this introduces a potentially large uncertainty when modelling such atmospheres, given the apparent sensitivity of \ce{O3} to this parameter. A detailed exploration of this effect in the model is beyond the scope of this work. Even in our Earth simulations, depending on model choices and SZA, predicted \ce{O3} column depths vary by a factor of 1.6 at 100\% PAL and 2.5 at 0.1\% PAL. Our analyses were restricted to open-source models, but several proprietary 1D models are commonly used in the literature as well, including ExoPRIME \citep{payne2024oxygen, gomez2023search}, ARGO \citep{2016ApJS..224....9R, 2019Icar..329..124R, 2016ApJS..224....9R, 2021ApJ...922...44J}, EPACRIS \citep{hu2012photochemistry}, 1D Terra \citep{grenfell2025detectability}, and KINETICS \citep{1981JGR....86.3617A}. While these models may contain similar variations, their proprietary nature precluded their evaluation here. Future work should seek to include other models to ensure a more comprehensive cross-model comparison.

Understanding how to simulate the \ce{O3} layer accurately in low-\ce{O2} regimes is especially important since for much of Earth's history, its atmosphere resembled the lower-\ce{O2} Proterozoic rather than the modern, \ce{O2}-rich atmosphere. Earth's geological evolution has long served as an analogue for discussing possible biosignatures \citep{Leger_1993_O2_O3} and predicting Earth-like exoplanet spectra \citep{segura2003ozone, Meadows2006Diversity, Kaltenegger2007Spectral, arney2016pale, rugheimer2018spectra, cooke2023variability}. If many habitable exoplanets have similar low-\ce{O2} atmospheres, interpretations of their biological activity may rely on translating observed \ce{O3} features in their atmospheres to \ce{O2} sources and sinks. Indeed, it has been argued that \ce{O3} could act as a good tracer for \ce{O2} \citep{Leger_1993_O2_O3} (since \ce{O2} may not be directly detectable even with next generation telescopes). Consistent with other works \citep{Kozakis2022Proxy, cooke2022revised, Cooke2023Stellar}, our results suggest that directly tying an \ce{O3} abundance to \ce{O2} abundances may not be feasible due to model and observational uncertainties. Even if an atmospheric \ce{O2} level was somehow inferred, interpreting its biogenicity may still require using 1D models to estimate the underlying surface fluxes that could sustain it \citep{2026arXiv260421848W, 2018AJ....156..114K}. Our results indicate that estimates for such fluxes may be model dependent too. Additionally, focussing on detecting \ce{O3} as a proxy for \ce{O2} may not be fruitful if we do not know certain other relevant astrophysical variables \citep{Kozakis2022Proxy}, including the SED of the host star, as shown by \citet{Cooke2023Stellar} for oxygenated Earth-like scenarios for TRAPPIST-1 e, and more recently for Archean analogues by \citet{sneed2026ultraviolet}. Following this present work, we show that model degeneracies clearly exist too, even if the host star SED is known.

A further concern is that \ce{O3} need not be biological at all. Photochemical production of \ce{O2} and \ce{O3} from the photolysis of \ce{CO2} or \ce{H2O} can create false positives. Extensive literature has sought to quantify this abiotic build-up to help observers distinguish between biogenic signals from non-biological planetary processes using 1D models \citep{hu2012photochemistry, domagal2010abiotic, domagal2014abiotic, tian2014high, harman2015abiotic, Wordsworth_2014_H2O}. Our results suggest that 3D effects may alter such calculations, especially for tidally locked exoplanets around M dwarfs. To our knowledge, the theoretical thresholds to rule out false positives have not been thoroughly examined with 3D models. Although 3D models may be inaccurate too, it is clear that three-dimensional processes play major roles which 1D models cannot capture. The importance of understanding the dynamical regime has been confirmed independently for the LIFE mission by \citet{Braam2026}, who also warned against false positive scenarios for \ce{O2}-\ce{O3} in oxygenated scenarios for Proxima Centauri b. Consequently, we recommend that substantial work goes into improving the chemical schemes of 3D models, with 1D models as necessary guides for honing such updates. 

The search for life beyond our solar system continues, and we may require next generation telescopes to find it. While JWST could search for biosignatures on a handful of temperate, transiting exoplanets orbiting nearby M-dwarf stars, finding signs of life on these planets is contingent on overcoming several challenges (e.g, diminished absorption signals and stellar contamination \citep{2023ApJ...955L..22L}). Although biosignatures on Hycean worlds may be detected with JWST, these planets remain hypothetical \citep{madhusudhan2021habitability, 2024ESS.....550003M, werlen2025sub, 2023ApJ...947L..19R}, and their habitability remains an active area of investigation \citep{madhusudhan2023chemical, 2024ApJ...964L..19G, mitchell2025prospects, cooke2026ecological}. Detecting biosignatures on rocky planets around white dwarfs could be easier with JWST, and indeed  1D models have been used to predict \ce{O3} abundances and the spectra of such planets \citep{2020ApJ...901L...1K, 2020ApJ...894L...6K, 2018ApJ...862...69K}, but no habitable planet has been discovered around a white dwarf to date \citep{2025jwst.prop.7564L}. A large proportion of the astrophysics community has therefore remained focussed on Earth-like planets, and is designing future flagship missions like HWO and LIFE to search for signs of life on such nearby targets. Several works have used the open source models examined here in recent years to simulate the atmospheres of a range of planets and exoplanets \citep{2026arXiv260421848W, 2025PSJ.....6..256W,gialluca2026coupled, 2025AAS...24631305T, jaziri2026possible, kozakis2025ozone, Kozakis2022Proxy, sumida2025evaluating, wolf2025chemistry, young2024inferring}, with many focussing on HWO \citep{2026arXiv260426925G, 2025arXiv250714771K, pradhan2026detectability, moussa2026life} and LIFE \citep{Alei2022LIFEV, pradhan2026detectability, angerhausen2024large, rugheimer2026goldilocks}. Yet, we have shown that under a quasi-equalised set of assumptions, four open-source models currently used in the literature produce transmission spectra which have \ce{O3} features markedly different in absorption strength. Reflection spectra signals are likely to differ substantially too, but we have not included them due to the lack of clouds in the 1D models. Given the estimated costs of both HWO (\$11 billion USD) and LIFE (\$1 billion USD), resolving these underlying model discrepancies is of paramount importance for the success of these missions.

\subsection{Reported boundary conditions and reproducibility}

We have normalised all of the boundary and initial conditions between simulations where possible. If we were to normalise all of the 1D models in terms of their chemical networks and included reactions, we would likely get slightly different results to the ones we presented. However, to do so here defeats part of the purpose of our work. Generally, new users of a particular photochemical code undergo a tutorial, whether that is delivered by an expert user or through instructions followed online. This tutorial often involves running a simulation of the Earth because it is the atmosphere with the most observations upon which we can test our models. Users will typically run a ``default setup'' for modern Earth, involving a 1 bar surface pressure, 21\% \ce{O2} and 78\% \ce{N2} by volume for the composition, and a standard pressure-temperature profile. In some cases, such as \texttt{Atmos} or \texttt{Photochem}, there may exist default setups for other planets or exoplanets (e.g., Mars in \texttt{Atmos} and WASP 39 b in \texttt{Photochem}). When users change settings to simulate an exoplanet, they often change multiple settings including the SED and lower boundary conditions (chemical fluxes and mixing ratios). Some of these settings are not always reported in the literature, which makes reproducing the work from the papers alone difficult. Therefore, we recommend detailing these specific modelling choices as transparently as possible to ensure robust reproducibility across the modelling community. 

\subsection{Future work}
\label{Future work section}

In light of this work, we recommend that 1D photochemical models which simulate atmospheres that involve significant levels of \ce{O2} should be updated with the latest photolysis parametrisations \citep{Ji_2024JGRD..Correlated_K} and photodissociation cross sections, if they have not already. More work needs to go into understanding the limits of 1D models and their applicability to specific scenarios, such as early Earth and oxygenated Earth-like exoplanets. The relevance will depend on the science goals of the work. For example, some molecules which have atmospheric mixing ratio variations above the troposphere, such as \ce{CH4} and \ce{CO2}, barely affect the resulting transmission spectra in our simulated scenarios, unlike \ce{O3}. 

As for moving toward a deeper understanding of Earth's past, there are many other factors that should be investigated in the future. For 3D models, the influence of various continents (including supercontinents), varying amounts of greenhouse gases (\ce{CH4}, \ce{CO2}, and \ce{N2O}), and the faster rotation rate of early Earth \citep{2014AGUFMPP53A1201B}, should be considered. Future work should also explore simulations of the early Earth that capture Proterozoic and Phanerozoic conditions more accurately and account for overall redox balance.

Finally, 3D chemistry-climate model intercomparisions need to be performed in the future to determine the range of atmospheric and spectral predictions. For oxygenated exoplanets, this can include, for example: \texttt{WACCM6}, LMD-g, ROCKE-3D and the Unified Model. In Fig.~\ref{ozone vs oxygen} we could have showed the results from ROCKE-3D \citep{Way2017ResolvingPlanets} and LMD-generic \citep{Jaziri_2022_GOE}, two 3D models which simulated the \ce{O3} layer in response to oxygenation fluctuations. As discussed in \citet{2023RSOS...1030056J}, the \citet{Jaziri_2022_GOE} model does not include heterogeneous, halogen, or nitrogen chemistry, and additionally omits gravity waves. Furthermore, the setup in ROCKE-3D was not defined in full \citep{Way2017ResolvingPlanets}. Thus, direct comparison with these two results is not constructive here, but future systematic comparisons would be welcome. Notably, the \citet{Jaziri_2022_GOE} 3D model yielded lower \ce{O3} column depths than their 1D model with the same chemical scheme. On the other hand, WACCM6 does not currently include scattering in the SR bands and this should be updated. We recognise that dedicated intercomparison work is taking place for various types of exoplanet atmospheric models in the CUISINES framework \citep{2024PSJ.....5..175S}. Ideally, 1D and 3D models can be used in conjunction to build in flexibility for simulations sets and reduce the large computational costs associated with running 3D models. Machine learning protocols, which have recently been applied to exoplanet Global Climate Models \citep{stevenson2026thousandworlds}, could also be implemented for 3D chemistry-climate simulations, provided enough simulations are completed to produce a benchmark.

\section{Conclusion}
\label{Conclusions section}

Photochemical models are used to calculate the concentrations of gases on early Earth, solar system planets, and exoplanets. Therefore, whilst these models are used to predict atmospheric states for hundreds of real and hypothetical atmospheres, there are only a few examples upon which we can confidently validate them against. Any errors in these models are possibly exacerbated when extending to regimes where we cannot easily make detailed local measurements, such as the early Earth and exoplanet atmospheres. Thus, our work has important implications for both regimes, of which modellers and observers should be aware.

We simulated atmospheres over a range of \ce{O2} mixing ratios (0.1\% PAL - 150\% PAL) that are relevant from the Proterozoic Eon onwards (2.5 Ga to present day) for Earth using four open-source 1D photochemical models and a 3D chemistry-climate model. We investigated the influence of different lower boundary conditions on the final atmospheric composition, the assumed solar zenith angle, and the younger solar spectrum. We find that 1D models predict higher \ce{O3} columns at lower oxygenation states ($\leq 10$\% the present atmospheric level of \ce{O2}) compared to 3D models, and the discrepancy increases the more \ce{O2} is decreased. Whilst several effects are involved, the treatment of \ce{O2} photolysis, chemical networks and reaction rates, and the inclusion of 3D transport and \ce{H2O} microphysics, all play important roles.

Our results have possible ramifications for the early Earth, particularly palaeobiology and photogeochemistry, as our understanding of the surface UV environment evolves. The \ce{O3} column discrepancies are important because reduced \ce{O3} columns result in higher fluxes of UV radiation reaching the surface. Some examples at 1\% and 0.1\% PAL show that for an overhead Sun the integrated UV damage differs by up to a factor of 8. A high irradiance of UV may have provided a selection pressure during Earth's Proterozoic Eon, but this needs further confirmation with simulations that are more representative of the Proterozoic.

Beyond Earth, our results also have implications for exoplanets, particularly for the theoretical modelling of oxygenated atmospheres, biosignature detection and interpretation, and future mission designs. Generally, 1D photochemical models are benchmarked against some solar system exoplanets (e.g. Earth and Jupiter), which is a sensible step. Some model ``tuning'' is often required to get a composition consistent with observations, which may account for missing physical processes or atmospheric transport. These unknown errors can then propagate through to predictions of other planetary atmospheres, as we have demonstrated when we simulated the atmosphere of Proxima Centauri b. This is not to say that 3D models do not also have these issues, because model intercomparisions have shown diverse predictions in transport regimes. Nevertheless, it is clear that independent 1D models do not agree with each other or with a 3D model when equivalent boundary conditions and assumptions are used. Therefore, care must be taken when using these models towards designs for future telescopes such as HWO and LIFE, both of which will search for life on other worlds in the near future.

\section*{Data availability}

Simulation outputs for the 1D models can be found in the following Github repository - \href{https://github.com/Paleo-Exo-Atmospheres/1D-atmospheric-simulations-of-the-Early-Earth-and-Exoplanets}{https://github.com/Paleo-Exo-Atmospheres/1D-atmospheric-simulations-of-the-Early-Earth-and-Exoplanets}. 3D chemistry-climate model data can be found in \citet{cooke_kumar_2026}. We have also included a website application to explore our simulations interactively: \href{https://atmospheric-simulations-of-the-early-earth-and-exoplanets.streamlit.app/}{https://atmospheric-simulations-of-the-early-earth-and-exoplanets.streamlit.app}. More figures and data can be viewed there.

\section*{Software availability}

The GitHub containing the model setups and boundary conditions, as well as description for simulations, can be found at \href{https://github.com/Paleo-Exo-Atmospheres/1D-atmospheric-simulations-of-the-Early-Earth-and-Exoplanets}{https://github.com/Paleo-Exo-Atmospheres/1D-atmospheric-simulations-of-the-Early-Earth-and-Exoplanets}. 

\section*{Conflict of interest declaration}

We declare we have no competing interests.

\section*{Funding}

This work used the DiRAC Data Intensive service (CSD3—project dp410) at the University of Cambridge - see the Acknowledgments statement for more information.

\section*{Acknowledgments}

We would like to thank Aoshuang Ji for valuable assistance in running the \texttt{Kasting} model. 
Additionally, we thank Nicholas Wogan and Shawn Domagal-Goldman for creating video tutorials on the \texttt{Photochem} and \texttt{Atmos} models, respectively. We also express our sincere appreciation to all the research teams and individuals who generously make their code publicly available to the scientific community. Finally, Akhil thanks Oliver Shorttle for insightful discussions.

This work used the DiRAC Data Intensive service (CSD3—project dp410) at the University of Cambridge, managed by the University of Cambridge University Information Services on behalf of the STFC DiRAC HPC Facility (\href{www.dirac.ac.uk}{www.dirac.ac.uk}). The DiRAC component of CSD3 at Cambridge was funded by BEIS, UKRI and STFC capital funding and STFC operations grants. DiRAC is part of the UKRI Digital Research Infrastructure.

\clearpage
\newpage
\appendix

\section{Boundary Conditions}
\label{boundary conditions appendix}
To maintain a consistent baseline, we aligned the lower boundary conditions of our 1D models with the lower boundary conditions of our \texttt{WACCM6} runs (shown in table \ref{WACCM6_boundary_conditions}). In all models, we retained the native default deposition velocities ($v{dep}$) and emission fluxes inherent to each respective 1D model to preserve their independently validated settings.

In \texttt{Photochem}, we use  fluxes at altitude to capture a more realistic emission for volcanic sulphur dioxide (SO$_2$) since \texttt{WACCM6}  includes an altitude dependent emission profile. We derived a globally averaged vertical volcanic \ce{SO2} profile from CESM2 CMIP6 fluxes and calculated the flux-weighted centroid of this profile. We then derived a single effective injection height of 1.9km for this flux after accounting for surface emissions in the CMIP6 dataset. Furthermore, several species with lower boundary conditions in \texttt{WACCM6}  runs were not available in \texttt{Photochem} and were consequently omitted. These include CCl$_4$, CF$_2$ClBr, CFC-11eq, CH$_2$Br$_2$, CH$_3$Br, CH$_3$Cl, and CHBr$_3$. As these are primarily trace halogens or anthropogenic compounds, their exclusion are not expected to significantly impact the pre-industrial baseline. 

Like \texttt{Photochem}, \texttt{Atmos} also allows for fluxes at an altitude, which we leverage for sulphur dioxide \ce{SO2} as described in section \ref{Photochem simulation description}. Since the \texttt{Kasting} model allows for a distributed flux and a surface flux, we used this feature to reflect the vertical distribution of \ce{SO2} and decoupled the volcanic and surface \ce{SO2} sources. For continuous volcanic emissions, we replaced the model's default 10 km outgassing assumption and derived a physically representative 1D injection band from the CMIP6 profile. We isolate the vertical bounds of the volcanic emissions and distributed the flux between 0.24 km and 5.45 km. This prevents artificially inflating surface concentrations while realistically capturing the tropospheric distribution. Additionally, we incorporate the CMIP6 surface fluxes as an additional lower boundary condition. The default model also includes a distributed flux for \ce{H2S}, which we retain. 

\begin{table}[H]
\label{boundary conditions table}
\centering
\caption{Supplemental lower boundary conditions across \texttt{Photochem}, \texttt{Atmos}, \texttt{VULCAN}, and \texttt{Kasting} models. Default fluxes (molecules cm$^{-2}$ s$^{-1}$) and deposition velocities ($v_{dep}$, cm s$^{-1}$) are shown. Values not applicable or not specified for a model in the default configuration are indicated with a dash (-).  When VULCAN is run without sulphur, the boundary conditions in this table that involve sulphur are not included.}
\label{combined_boundary_conditions}
\small
\setlength{\tabcolsep}{4pt} 
\resizebox{\textwidth}{!}{%
\begin{tabular}{l cc cc cc cc}
\toprule
\multirow{2}{*}{\textbf{Species}} & \multicolumn{2}{c}{\textbf{Photochem}} & \multicolumn{2}{c}{\textbf{Atmos}} & \multicolumn{2}{c}{\textbf{VULCAN}} & \multicolumn{2}{c}{\textbf{Kasting}} \\
\cmidrule(lr){2-3} \cmidrule(lr){4-5} \cmidrule(lr){6-7} \cmidrule(lr){8-9}
& Flux & $v_{dep}$ & Flux & $v_{dep}$ & Flux & $v_{dep}$ & Flux & $v_{dep}$ \\
\midrule
\textbf{CO} & $4.98 \times 10^{10}$ & 0.03 & $4.98 \times 10^{10}$ & 0.03 & $4.98 \times 10^{10}$ & 0.03 & $4.98 \times 10^{10}$ & 0.0 \\
\textbf{SO$_2$} (Surface)$^a$ & $1.43 \times 10^{9}$ & 1.0 & $1.43 \times 10^{9}$ & 1.0 & $1.43 \times 10^{9}$ & 1.0 & $1.20 \times 10^{8}$ & 1.0 \\
\textbf{SO$_2$} (Distributed)$^b$ & - & - & - & - & - & - & $1.31 \times 10^{9}$ & - \\
\textbf{H$_2$S} (Surface) & $2.0 \times 10^{8}$ & 0.015 & $2.0 \times 10^{8}$ & 0.02 & $2.0 \times 10^{8}$ & 0.015 & 0 & 0.02 \\
\textbf{H$_2$S} (Distributed)$^c$ & - & - & - & - & - & - & $3.50 \times 10^{8}$ & - \\
\textbf{H$_2$SO$_4$} & $7.0 \times 10^{8}$ & 1.0 & $7.0 \times 10^{8}$ & 1.0 & $7.0 \times 10^{8}$ & 1.0 & 0 & 1.0 \\
\textbf{H$_2$SO$_4$(l)} & - & - & - & - & 0 & 0.2 & - & - \\
\textbf{O$_3$} & 0 & 0.1 & 0 & 0.1 & 0 & 0.1 & 0 & 0.1 \\
\textbf{H$_2$CO (CH$_2$O)} & $5.16 \times 10^{8}$ & - & $5.16 \times 10^{8}$ & 0.2 & $5.16 \times 10^{8}$ & - & $5.16 \times 10^{8}$ & 0.2 \\
\textbf{C$_2$H$_6$} & - & - & $9.0 \times 10^{8}$ & 0.0 & - & - & - & - \\
\textbf{CS$_2$} & - & - & $2.0 \times 10^{7}$ & 0.0 & - & - & - & - \\
\textbf{C$_2$H$_6$S (DMS)} & - & - & $1.62 \times 10^{9}$ & 0.0 & - & - & - & - \\
\textbf{OCS} & - & - & - & - & - & 0.003 & - & - \\
\textbf{O} & - & - & 0 & 1.0 & - & - & 0 & 1.0 \\
\textbf{HCO} & - & - & 0 & 1.0 & - & - & - & - \\
\textbf{HSO} & - & - & 0 & 1.0 & - & - & 0 & 1.0 \\
\textbf{HS} & - & - & - & - & - & - & 0 & 1.0 \\
\textbf{SO} & - & - & - & - & - & - & 0 & $3.0 \times 10^{-4}$ \\
\textbf{SO$_4$AER} & - & - & 0 & 0.01 & - & - & - & - \\
\textbf{S$_8$AER} & - & - & 0 & 0.01 & - & - & - & - \\
\midrule
\multicolumn{9}{c}{\textit{NO$_x$ and Nitrogen Species}} \\
\midrule
\textbf{NO} & $2.37 \times 10^{9}$ & 0.01 & $2.37 \times 10^{9}$ & 0.016 & $2.37 \times 10^{9}$ & 0.001 & $2.37 \times 10^{9}$ & 0.2 \\
\textbf{NO$_2$} & 0 & 0.05 & 0 & 0.003 & 0 & 0.01 & 0 & 0.2 \\
\textbf{NO$_3$} & 0 & 0.05 & - & - & 0 & 0.1 & 0 & 0.5 \\
\textbf{NH$_3$} & $1.5 \times 10^{10}$ & 1.0 & $1.50 \times 10^{10}$ & 0.0 & $1.5 \times 10^{9}$ & 1.0 & - & - \\
\textbf{HNO$_3$} & 0 & 4.0 & 0 & 0.2 & 0 & 4.0 & 0 & 0.5 \\
\textbf{HCN} & $1.7 \times 10^{8}$ & 0.13 & - & - & $1.7 \times 10^{8}$ & 0.13 & - & - \\
\textbf{CH$_3$CN} & $1.3 \times 10^{8}$ & 0.13 & - & - & $1.3 \times 10^{8}$ & 0.13 & - & - \\
\textbf{HO$_2$NO$_2$} & - & - & 0 & 0.2 & - & - & 0 & 0.5 \\
\textbf{HNO} & - & - & 0 & 1.0 & - & - & - & - \\
\textbf{N$_2$O} & - & - & - & - & 0 & $1 \times 10^{-4}$ & - & - \\
\textbf{HNO$_2$} & - & - & - & - & - & - & 0 & 0.5 \\
\textbf{N$_2$O$_5$} & - & - & - & - & - & - & 0 & 0.5 \\
\midrule
\multicolumn{9}{c}{\textit{HO$_x$, Hydrogen, and Halogen Species}} \\
\midrule
\textbf{H} & 0 & 1.0 & 0 & 1.0 & - & - & 0 & 1.0 \\
\textbf{OH} & 0 & 1.0 & 0 & 1.0 & - & - & 0 & 1.0 \\
\textbf{HO$_2$} & 0 & 1.0 & 0 & 1.0 & - & - & 0 & 1.0 \\
\textbf{H$_2$O$_2$} & 0 & 0.2 & 0 & 0.2 & - & - & 0 & 0.2 \\
\textbf{CH$_3$OOH} & - & - & - & - & - & - & 0 & 0.2 \\
\textbf{CH$_3$O$_2$} & - & - & - & - & - & - & 0 & 1.0 \\
\textbf{Cl$_2$O$_2$} & - & - & - & - & - & - & 0 & 0.5 \\
\textbf{HOCl} & - & - & - & - & - & - & 0 & 0.5 \\
\textbf{Cl} & - & - & - & - & - & - & 0 & 1.0 \\
\textbf{ClO} & - & - & - & - & - & - & 0 & 0.5 \\
\textbf{HCl} & - & - & - & - & - & - & 0 & 0.5 \\
\textbf{ClONO$_2$} & - & - & - & - & - & - & 0 & 0.5 \\
\bottomrule
\multicolumn{9}{p{\textwidth}}{\scriptsize{$^a$ \texttt{Photochem} and \texttt{Atmos} SO$_2$ fluxes are at an effective altitude of 1.9 km. \texttt{VULCAN} SO$_2$ flux represent total fluxes.}} \\
\multicolumn{9}{p{\textwidth}}{\scriptsize{$^b$ \texttt{Kasting} volcanic SO$_2$ flux is a vertically distributed continuous source injected between 0.24 km and 5.45 km.}} \\
\multicolumn{9}{p{\textwidth}}{\scriptsize{$^c$ \texttt{Kasting} volcanic H$_2$S flux is distributed across the lower 10 km.}} \\
\end{tabular}%
}
\end{table}

\newpage
\section{Additional Perturbation Experiments}
\label{Perturbation experiments appendix}

\subsection*{Surface Albedo}

\begin{figure}[h]
    \centering
    \includegraphics[width = 0.49 \linewidth]{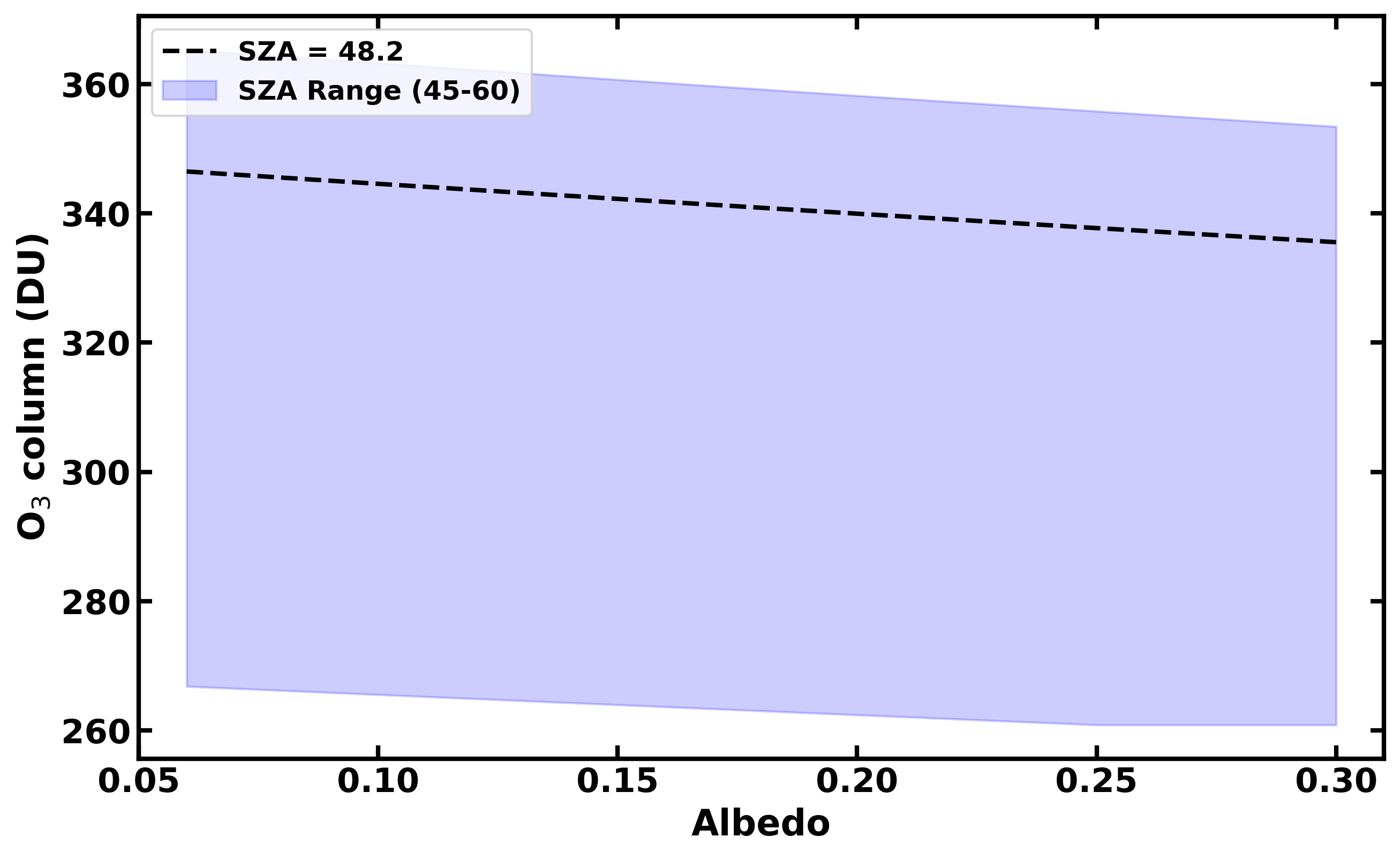}
    \includegraphics[width = 0.49 \linewidth]{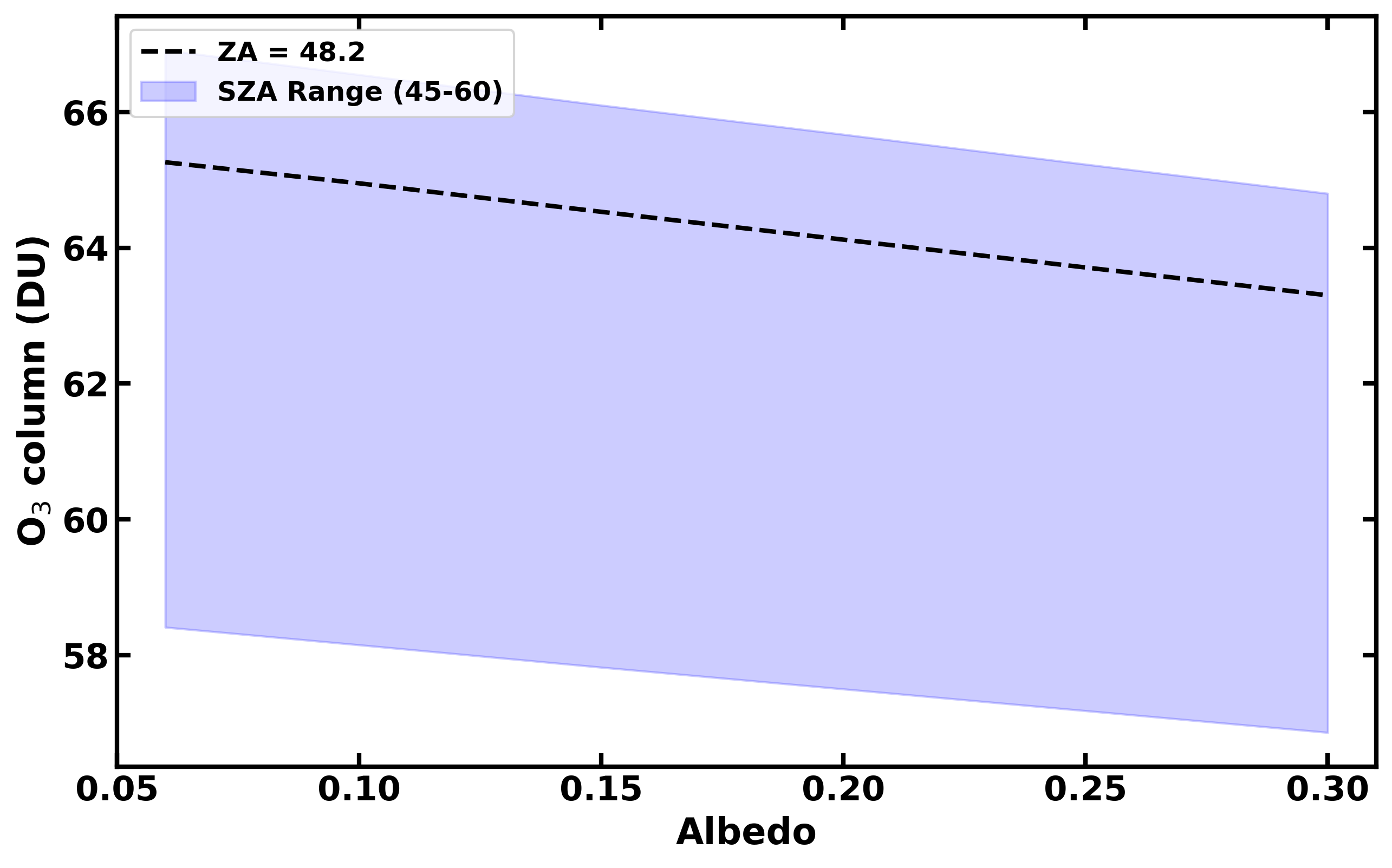}
    \caption{The \ce{O3} column depth is shown a function of surface albedo in \texttt{Photochem} for 100\% PAL conditions (left) and 0.1\% PAL conditions (right). The dotted line represents the reference case (SZA = 48.2\textdegree, DAF = 0.375), while the shaded regions span simulations with SZA 45\textdegree (DAF = 0.354) and SZA 60\textdegree  (DAF = 0.5). Increasing surface albedo results in a reduction in \ce{O3} column depth.}
    \label{albedo_photochem}
\end{figure}

In Fig.~\ref{albedo_photochem}, we show how increasing surface albedo from 0.06 to 0.30 in \texttt{Photochem} leads to a decreased \ce{O3} column for the 100\% PAL and the 0.1\% PAL atmospheres. The reflective surface in the high albedo case (0.3) increases the upwelling UV radiation compared to the low albedo (0.06) case. In both the 100\% PAL and 0.1\% PAL atmospheres, the \ce{O3} column decreases by an average of 2.9\%, with the primary driver of this discrepancy being catalytic destruction of \ce{O3} via NO\textsubscript{x} and HO\textsubscript{x} cycles. For 100\% PAL, when surface albedo is increased, the column integrated \ce{Ox} production is only increased by 0.5\%, while \ce{O3} destruction by NO\textsubscript{x} is higher by 6.6\% and \ce{O3} destruction by HO\textsubscript{x} species is higher by roughly 1.0\%. Similarly, for 0.1\% PAL, the transition to higher albedo increases the column integrated \ce{Ox} production by 0.5\%, while \ce{O3} destruction by NO\textsubscript{x} is higher by 1.9\% and \ce{O3} destruction by HO\textsubscript{x} species is higher by roughly 0.7\%.

\subsection*{Young Sun}
We tested the effect of the faint young sun on the \ce{O3} column depths. Variations in the uncoupled simulations are minor. 
In the 100 \% PAL case the \ce{O3} column decreases from 335.4 DU to 329.4 DU (1.8 \%). This minor decrease is a result of the interplay between \ce{O3} production and destruction, with decreases in \ce{Ox} production are offset by concurrent decreases in catalytic ozone destruction. While \ce{Ox} production decreases by 22.1\%, NO\textsubscript{x}  decreases by 26.0\% and HO\textsubscript{x} decreases by 15.3\%. Similarly, in the 0.1\% PAL simulation, the column decreases from 27.79 DU to 27.26 DU (1.9\%), with \ce{Ox} production decreasing by 13.2\% while NO\textsubscript{x} decreases 20.1\% and HO\textsubscript{x} decreases by 15.6\%. 

\begin{figure}[h]
    \centering
    \includegraphics[width = 0.49 \linewidth]{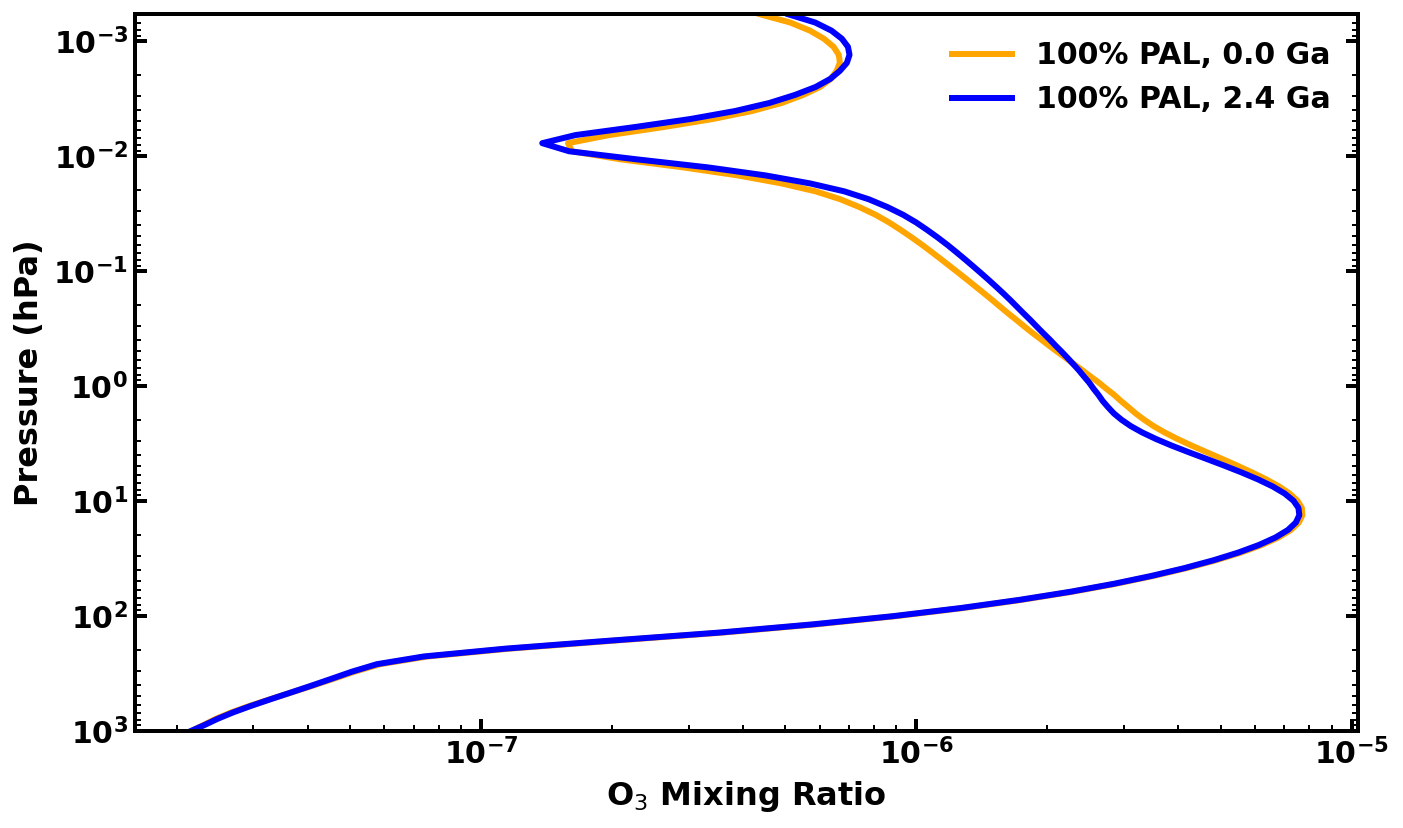}
    \includegraphics[width = 0.49 \linewidth]{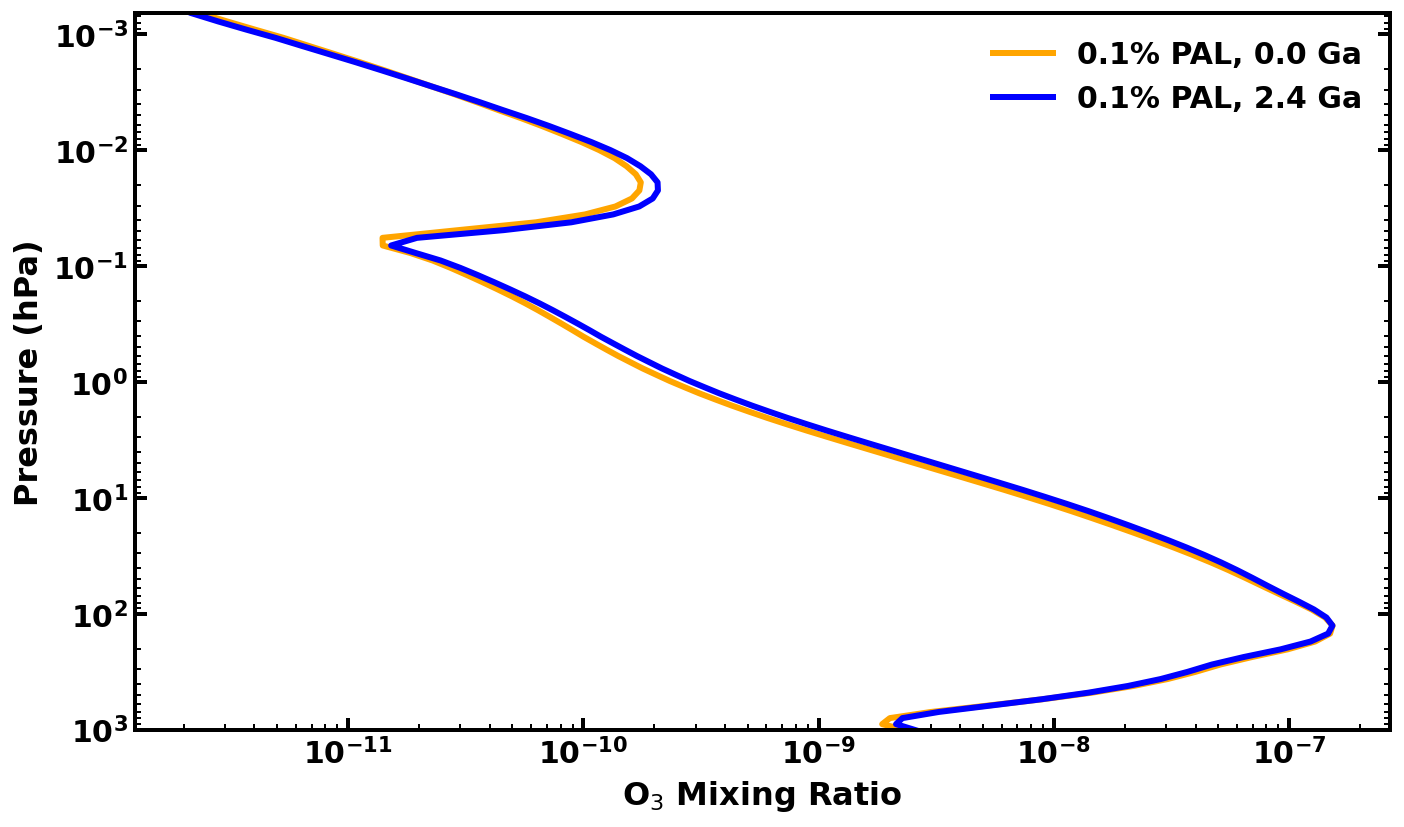}
    \caption{Uncoupled Ozone profiles for the 100\% PAL atmosphere (left) and the 0.1\% PAL atmosphere for both the present day sun (0.0 Ga, orange) and the young sun (2.4 Ga, blue) using \texttt{Photochem}. For each oxygenation state, the corresponding WACCM6 PT profile was used.}
    \label{ozone vs sun}
\end{figure}

\begin{figure}[h]
    \centering
    \includegraphics[width = 0.49 \linewidth]{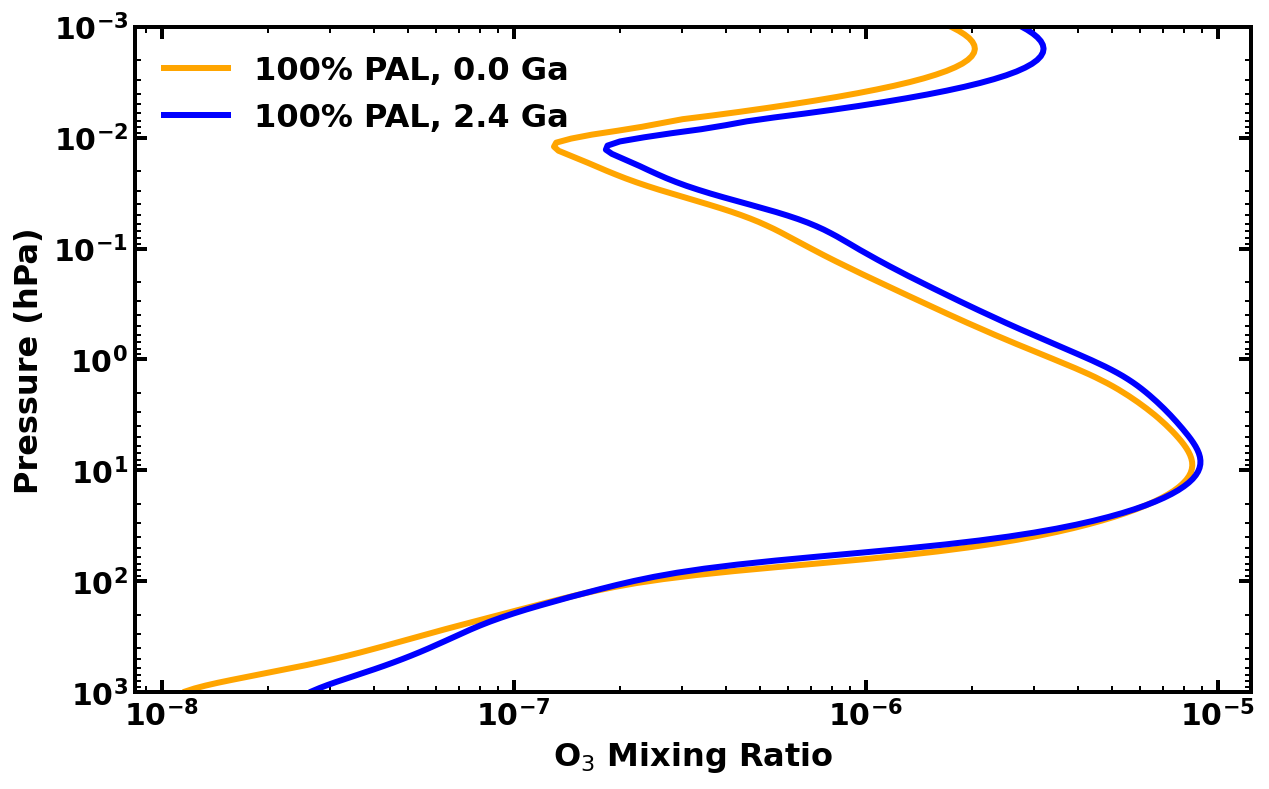}
    \includegraphics[width = 0.49 \linewidth]{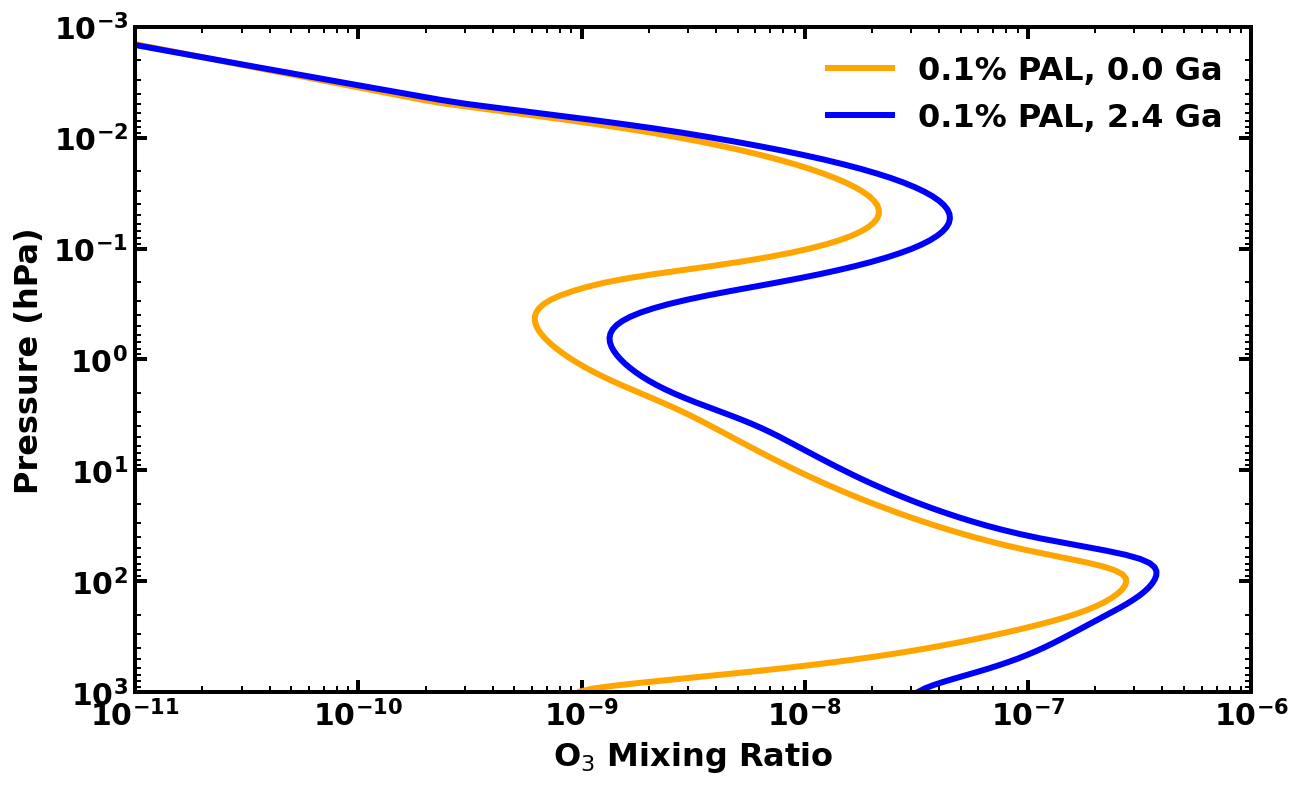}
    \caption{Coupled ozone profiles for the 100\% PAL atmosphere (left) and the 0.1\% PAL atmosphere for both the present day sun (0.0 Ga, orange) and the young sun (2.4 Ga, blue) using \texttt{Atmos}. For each oxygenation state and stellar flux, we use \texttt{Clima} to generate the PT profile and capture climate-chemistry feedbacks.}
    \label{ozone vs sun coupled}
\end{figure}

To test the effect of the young sun on atmospheric composition when climate is taken into account, we run coupled simulations of the atmosphere under a younger sun using \texttt{Atmos}. We find that changes in column depth are minor for the 100\% PAL case, with the column decreasing from 264.8 to 263.7 DU (0.4\%). However, the 0.1\% PAL atmosphere undergoes a dramatic increase, rising from 44.28 DU to 93.54 DU ( $\sim$111\%). This suggests that the effect of a younger sun is significantly amplified in a low-\ce{O2} atmosphere when temperature feedbacks are included (Fig.~\ref{ozone vs sun coupled}). This substantial accumulation of \ce{O3} is driven by differences in the amounts of HO\textsubscript{x} species due to climate-chemistry feedbacks. Under the fainter young sun, lower tropospheric temperatures reduce the saturation vapour pressure of \ce{H2O}, resulting in a drier lower atmosphere. This leads to a depletion in HO\textsubscript{x} mixing ratios in the troposphere, particularly at pressures greater than 100 hPa. Since HO\textsubscript{x} radicals are a primary catalytic sink for ozone, this reduction in HO\textsubscript{x} driven loss allows \ce{O3} to accumulate, driving the increase in the total \ce{O3} column. 

\subsection*{Methane}

\begin{figure}[t!]
    \centering
    \includegraphics[width = 0.9 \linewidth]{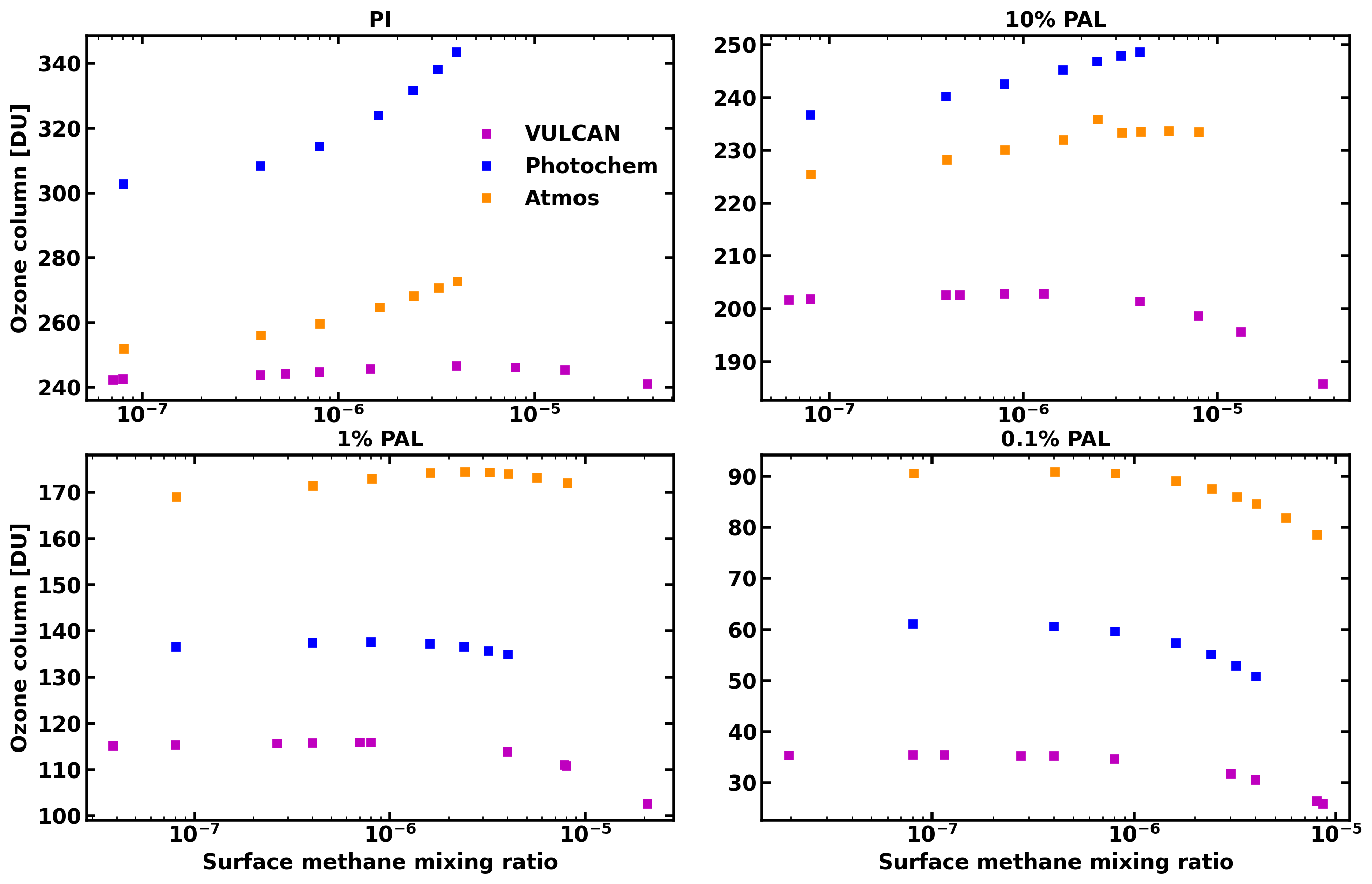}
    \caption{The relationship between \ce{O3} column depths and fixed \ce{CH4} mixing ratios is shown for the 100\% PAL (PI, top left), 10\% PAL (top right), 1\% PAL (bottom left) and 0.1\% PAL (bottom right) simulations in \texttt{VULCAN} (purple), \texttt{Photochem} (blue) and \texttt{Atmos} (orange). }
    \label{ozone vs methane}
\end{figure}

In Fig.~\ref{ozone vs methane}, we show the effect of increasing methane on atmospheric \ce{O3} column depths. As mentioned earlier, in high \ce{O2} simulations (100\% PAL and 10\% PAL), the \ce{O3} column are flat or increasing with increased \ce{CH4}, while in low \ce{O2} simulations (1\% PAL  and 0.1\% PAL), the \ce{O3} column curve is flat or slightly declines. These effects are largely consistent across models, with results largely being driven by the effects of \ce{CH4} on HO\textsubscript{x} and NO\textsubscript{x} cycles. 

In \texttt{Photochem} at 100\% PAL for example, increased \ce{CH4} raises the \ce{O3} column by 18.4\%, from 328.0 DU in the 0.1x case to 388.4 DU in the 10x case. The increased column is driven almost entirely by increased Chapman recombination (\ce{O + O2 + M -> O3 + M}), which rises by 11.3\%, despite \ce{Ox} production via \ce{O2} photolysis only increasing by \ce{1.1\%}. NO\textsubscript{x} loss reduces by 9.2\%, since peroxy radicals produced by the increased \ce{CH4} convert \ce{NO} to \ce{NO2} : 
\begin{align}
\ce{HO2 + NO &-> OH + NO2} \\
\ce{CH3O2 + NO &-> CH3O + NO2}.
\end{align}
Loss due to HO\textsubscript{x} also increases by 11\%. Despite this, the \ce{O3} column increase indicates that the suppression of the NO\textsubscript{x} cycle outweighs the increased HO\textsubscript{x} loss. In \texttt{Photochem} simulations at 0.1\% PAL, increased \ce{CH4} lowers the \ce{O3} 
column from 65.8 to 44.3DU (-32.7\%), driven by suppression of 
Chapman recombination by 16.6\% despite \ce{Ox} production from \ce{O2} photolysis remaining similar (+2.4\%). \ce{O3} loss due to HO\textsubscript{x} decreases by 34\%, while \ce{O3} loss due to NO\textsubscript{x} decreases by around 44\%. Despite this, the generation of species that consume atomic oxygen reduces the Chapman mechanism sufficiently to reduce \ce{O3} column depths. 

\clearpage
\newpage
\bibliographystyle{unsrtnat}

\bibliography{refs}

\end{document}